\pdfoutput=1
\documentclass[acmsmall, nonacm]{acmart}

\newcommand{\sun}[1]{{\textcolor{black}{#1}}}

\usepackage{enumitem}
\usepackage{tabularx} 
\usepackage[linesnumbered,vlined,ruled]{algorithm2e}
\usepackage{multirow}
\usepackage{graphicx}
\usepackage{caption}
\usepackage{subcaption}
\usepackage{booktabs} 
\usepackage{geometry} 
\usepackage{hyperref} 
\usepackage{pgfplots}
\usepackage{placeins}
\usepackage{afterpage}
\pgfplotsset{compat=1.17}
\usepgfplotslibrary{groupplots}
\setcopyright{rightsretained}
\acmYear{2025} \acmVolume{3} \acmNumber{1} \acmArticle{70} \acmMonth{2} \acmPrice{15.00}\acmDOI{XX.XX/XXX.XX}


\begin{document}


\title{Revisiting the Design of In-Memory Dynamic Graph Storage}

\author{Jixian Su}
\affiliation{
  \institution{Shanghai Jiao Tong University}
  \city{Shanghai}
  \country{China}
}
\email{sjx13623816973@sjtu.edu.cn}

\author{Chiyu Hao}
\affiliation{
  \institution{Shanghai Jiao Tong University}
  \city{Shanghai}
  \country{China}
}
\email{hcahoi11@sjtu.edu.cn}

\author{Shixuan Sun}
\affiliation{
  \institution{Shanghai Jiao Tong University}
  \city{Shanghai}
  \country{China}
}
\email{sunshixuan@sjtu.edu.cn}

\author{Hao Zhang}
\affiliation{
  \institution{Huawei Cloud}
  \city{Beijing}
  \country{China}
}
\email{zhanghao687@huawei.com}

\author{Sen Gao}
\affiliation{
  \institution{National University of Singapore}
  \country{Singapore}
}
\email{sen@u.nus.edu}

\author{Jiaxin Jiang}
\affiliation{
    \institution{National University of Singapore}
    \country{Singapore}
}
\email{jxjiang@nus.edu.sg}

\author{Yao Chen}
\affiliation{
    \institution{National University of Singapore}
    \country{Singapore}
}
\email{yaochen@nus.edu.sg}

\author{Chenyi Zhang}
\affiliation{
  \institution{Huawei Cloud}
  \city{Hangzhou}
  \country{China}
}
\email{zhangchenyi@huawei.com}

\author{Bingsheng He}
\affiliation{
    \institution{National University of Singapore}
    \country{Singapore}
}
\email{hebs@comp.nus.edu.sg}

\author{Minyi Guo}
\affiliation{
    \institution{Shanghai Jiao Tong University}
    \city{Shanghai}
    \country{China}
}
\email{guo-my@cs.sjtu.edu.cn}

\renewcommand{\shortauthors}{Jixian Su et al.}

\begin{abstract}

The effectiveness of in-memory dynamic graph storage (DGS) for supporting concurrent graph read and write queries is crucial for real-time graph analytics and updates. Various methods have been proposed, for example, LLAMA, Aspen, LiveGraph, Teseo, and Sortledton. These approaches differ significantly in their support for read and write operations, space overhead, and concurrency control. However, there has been no systematic study to explore the trade-offs among these dimensions. In this paper, we evaluate the effectiveness of individual techniques and identify the performance factors affecting these storage methods by proposing a common abstraction for DGS design and implementing a generic test framework based on this abstraction. Our findings highlight several key insights: 1) Existing DGS methods exhibit substantial space overhead. For example, Aspen consumes 3.3-10.8x more memory than CSR, while the optimal fine-grained methods consume 4.1-8.9x more memory than CSR, indicating a significant memory overhead.  2) Existing methods often overlook memory access impact of modern architectures, leading to performance degradation compared to continuous storage methods. 3) Fine-grained concurrency control methods, in particular, suffer from severe efficiency and space issues due to maintaining versions and performing checks for each neighbor. These methods also experience significant contention on high-degree vertices.  Our systematic study reveals these performance bottlenecks and outlines future directions to improve DGS for real-time graph analytics.

\end{abstract}



\maketitle

\section{Introduction} \label{sec:introduction}

\begin{figure}[t]\small
    \setlength{\abovecaptionskip}{3pt}
    \setlength{\belowcaptionskip}{0pt}
    \includegraphics[scale=0.75]{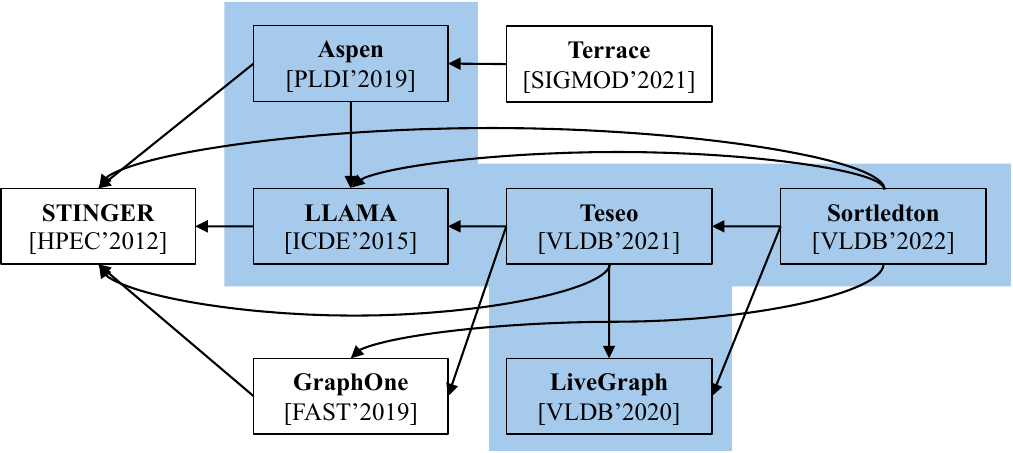}
    \centering
    \caption{Comparison of DGS methods from previous experiments. An edge from $x$ to $y$ indicates that $x$’s experiments include $y$. Shaded methods are transactional approaches.}
    \label{fig:method_relationship}
\end{figure}

Graph storage is the foundation for efficient in-memory graph data processing. As graph data often require frequent updates~\cite{sahu2017ubiquity,sakr2021future}, in-memory dynamic graph storage (DGS) is essential for supporting concurrent graph read and write queries, enabling real-time graph analytics and updates. To address this need, a variety of DGS methods have been proposed~\cite{ediger2012stinger,dhulipala2019low,macko2015llama,kumar2020graphone,pandey2021terrace,de2021teseo,fuchs2022sortledton,zhu2019livegraph}, as shown in Figure \ref{fig:method_relationship}. STINGER~\cite{ediger2012stinger}, GraphOne~\cite{kumar2020graphone}, and Terrace~\cite{pandey2021terrace} do not provide concurrency control, meaning read and write queries can only execute alternately. In contrast, the other approaches support concurrent read and write queries with serializability~\cite{ramakrishnan2002database}, i.e., the results of executing concurrent queries are equivalent to some serial execution order. \sun{LLAMA~\cite{macko2015llama} and Aspen~\cite{dhulipala2019low} primarily focus on batch updates with a single-writer model, while recent methods have shifted toward optimizing individual updates to support a broader range of applications, which has attracted significant research interest~\cite{zhu2019livegraph,de2021teseo,fuchs2022sortledton}. Despite their varied update strategies, these approaches all enable concurrent read and write queries, thus called \emph{transactional approaches} in this paper.}

These approaches differ significantly in their support for read and write operations, space overhead, and concurrency control. First, they employ different graph container designs to support efficient read and write operations. For instance, LLAMA~\cite{macko2015llama} and LiveGraph~\cite{zhu2019livegraph} store a vertex's neighbor set $N(u)$ in a dynamic array, enabling fast scan operations with continuous storage and simple append inserts, but at the cost of expensive search operations. Recent approaches like Aspen~\cite{dhulipala2019low}, Teseo~\cite{de2021teseo}, and Sortledton~\cite{fuchs2022sortledton} use a segmented strategy, dividing $N(u)$ into small blocks (e.g., $|B| = 256$), storing each block as a sorted array, and linking them with a block index (e.g., a skip list). This design balances insert, scan, and search efficiency. Moreover, these approaches propose additional optimizations, such as adaptive indexing to use different data structures for varying neighbor set sizes to further accelerate graph operations.

Second, although existing methods all use multi-version concurrency control (MVCC) to maximize parallelism among queries~\cite{wu2017empirical}, they differ in version granularity. \emph{Coarse-grained} methods like LLAMA and Aspen use the typical single-writer-multiple-reader scheme with copy-on-write (CoW)~\cite{lmdb}. For each update, they create a new graph snapshot, while read queries work on the snapshot at their start. Recent works such as LiveGraph, Teseo, and Sortledton use a \emph{fine-grained} strategy, maintaining version information for each neighbor and synchronizing read and write operations with concurrency control protocols like 2PL~\cite{ramakrishnan2002database}. This allows multiple writers to update different elements simultaneously. 

However, there has been no systematic study exploring the trade-offs among these dimensions. Particularly, we observe several issues in current DGS research. First, there is a lack of a common abstraction to model the problem and capture the design space. Second, although Sortledton~\cite{fuchs2022sortledton} covers most existing methods in the experiments, as shown in Figure \ref{fig:method_relationship}, existing methods compare different approaches as black boxes, while numerous factors can affect performance, making the effectiveness of individual techniques unclear. Additionally, existing experiments focus on evaluating the efficiency of proposed graph containers but lack assessments of concurrency performance. Third, the performance gap between DGS and CSR regarding read efficiency and space consumption is unclear, obscuring the overhead of using DGS for read queries.

\vspace{2pt}
\noindent\textbf{Our Work.} We propose to revisit the design of in-memory dynamic graph storage to evaluate the effectiveness of individual techniques and identify key performance factors. We first propose a common abstraction for DGS to capture the nature of graph queries and data, highlighting key performance factors and facilitating a systematic study of existing methods. Within this common abstraction, we compare key techniques in existing DGS methods, including graph containers, concurrency control, and specific optimizations. Additionally, we implement a generic test framework based on our abstraction and re-implement these techniques within the framework for fair empirical comparisons.

\sun{We evaluate five DGS methods and include two static graph storage methods, CSR and AdjLst, for comparison. We conduct detailed experiments with eight real-world graphs to assess graph container efficiency, the effectiveness of concurrency control techniques, and memory consumption across methods. Our results show that, while fine-grained methods improve write throughput, they also incur significant space and efficiency overhead. Fine-grained strategies face challenges with version maintenance and lock contention, especially when managing high-degree vertices. Although coarse-grained methods alleviate these issues, Aspen still suffers from inefficiencies in graph container design. Consequently, CSR consistently outperforms DGS methods, achieving 2.4–11.0x higher query performance and consuming 3.3–10.8x less memory. We summarize the relative performance of competing methods, key insights, and research opportunities in Section \ref{sec:conclusion}.  This work is open-sourced on GitHub at https://github.com/SJTU-Liquid/DynamicGraphStorage.git. In summary, this paper makes the following contributions.}

\begin{itemize} [leftmargin=*]
    \item \sun{We propose a simple yet effective abstraction for DGS, enabling a systematic study of existing methods.}
    \item \sun{Using this abstraction, we compare key techniques in DGS, e.g., graph containers and concurrency control.}
    \item \sun{We develop a generic testing framework based on this abstraction to systematically evaluate existing methods.}
    \item \sun{Our findings reveal the strengths of current approaches and highlight critical performance factors, providing valuable insights to guide the design of future DGS systems.}
\end{itemize}

\section{Preliminaries} \label{sec:preliminaries}

We focus on the directed graph $G = (V, E)$, where $V$ is the set of vertices and $E \subseteq V \times V$ is the set of edges. For a vertex $u \in V$, $N(u)$ denotes its neighbors and $d(u)$ is its degree, i.e., $|N(u)|$. An edge $e(u, v)$ is directed from $u$ to $v$. An undirected graph can be represented by storing edges in both directions, $e(u, v)$ and $e(v, u)$. Each vertex has a unique \emph{vertex ID}. Previous studies~\cite{aberger2017emptyheaded,han2018speeding,yang2024hero,wang2017experimental,gonzalez2012powergraph} have shown that using vertex IDs as integers in $[0, |V|)$ benefits both computation and storage due to compact ID representation. Existing works~\cite{zhu2019livegraph,macko2015llama,de2021teseo,fuchs2022sortledton,dhulipala2019low} either require input vertex IDs to belong to $[0, |V|)$ or use dictionary encoding techniques (e.g., concurrent hash maps) to map external vertex IDs to this range. Thus, in our case, each vertex ID $u \in V$ is an integer in $[0, |V|)$.

\vspace{2pt}
\noindent\textbf{Graph Storage.} A \emph{static graph} can be stored as an \emph{adjacency list} (AdjLst), which uses an array of arrays (or lists), where each array at index $u$ contains all of $u$‘s neighbors, sorted by vertex IDs. A common variant, the \emph{compressed sparse row} (CSR) format, stores all vertices’ neighbor sets in a single \emph{neighbor array} and uses an \emph{offset array} to record the start position of each vertex’s neighbor set in the neighbor array. Given that real-world graphs are often sparse, CSR is the most widely used storage method for static graphs due to its compact memory layout and data access efficiency. A \emph{dynamic graph} $G = (G_0, \Delta \mathcal{G})$ records the evolution of the graph over time. $G_0$ is the initial graph and $\Delta \mathcal{G} = (\Delta G_1, ..., \Delta G_i, ...)$ is a sequence of updates. $\Delta G_i = {(\oplus, v/e)}$ contains a set of updates, where $(\oplus = +/-)$ denotes an insertion/deletion of a vertex $v$ or an edge $e(u, v)$. Deleting a vertex also removes all its adjacent edges. $\Delta V$ denotes the vertices appeared in $\Delta G$. The graph $G_i$ is obtained by applying the first $i$ updates to $G_0$, i.e., $G_i = G_0 \oplus \Delta G_1 \oplus ... \oplus \Delta G_i$. If $\Delta G$ contains a single operation, it is called a \emph{single update}; otherwise, it is a \emph{batch update}. Static graph formats are inefficient for dynamic graphs because: 1) they cannot coordinate concurrent read and write operations, compromising the correctness of graph queries; and 2) they do not support fast updates. 

\small
\begin{table}[t]
\centering
\caption{The time and space complexity of involved data structures in our study.}
\label{tab:complexity_data_structure}
\begin{tabular}{|c|cccc|c|}
\hline
                                     & \multicolumn{4}{c|}{\textbf{Time}}                                                                                                 & \multirow{2}{*}{\textbf{Space}} \\ \cline{1-5}
                                     & \multicolumn{1}{c|}{\textsc{Insert}} & \multicolumn{1}{c|}{\textsc{Search}} & \multicolumn{1}{c|}{\textsc{Scan}} & \textsc{Resize} &                                 \\ \hline
\textbf{Dynamic Array(DA)}           & \multicolumn{1}{c|}{$O(1)$}          & \multicolumn{1}{c|}{$O(n)$}          & \multicolumn{1}{c|}{$O(n)$}        & $\Theta(n)$     & $O(n)$                          \\ \hline
\textbf{Sorted Dynamic Array(SDA)}   & \multicolumn{1}{c|}{$O(n)$}          & \multicolumn{1}{c|}{$O(\log n)$}     & \multicolumn{1}{c|}{$O(n)$}        & $\Theta(n)$     & $O(n)$                          \\ \hline
\textbf{Packed Memory Array(PMA)}    & \multicolumn{1}{c|}{$O(\log^2 n)$}   & \multicolumn{1}{c|}{$O(\log n)$}     & \multicolumn{1}{c|}{$O(n)$}        & $\Theta(n)$     & $O(n)$                          \\ \hline
\textbf{Skip List(SL)}               & \multicolumn{1}{c|}{$O(\log n)$}     & \multicolumn{1}{c|}{$O(\log n)$}     & \multicolumn{1}{c|}{$O(n)$}        & -               & $O(n\log n)$                    \\ \hline
\textbf{Hash Table(HT)}              & \multicolumn{1}{c|}{$O(1)$}          & \multicolumn{1}{c|}{$O(1)$}          & \multicolumn{1}{c|}{$O(n)$}        & -               & $O(n)$                          \\ \hline
\textbf{AVL Tree(AVL)}               & \multicolumn{1}{c|}{$O(\log n)$}     & \multicolumn{1}{c|}{$O(\log n)$}     & \multicolumn{1}{c|}{$O(n)$}        & -               & $O(n)$                          \\ \hline
\textbf{Parallel Augmented Map(PAM)} & \multicolumn{1}{c|}{$O(\log n)$}     & \multicolumn{1}{c|}{$O(\log n)$}     & \multicolumn{1}{c|}{$O(n)$}        & -               & $O(n)$                          \\ \hline
\end{tabular}%
\end{table}

\vspace{2pt}
\noindent\textbf{Data Structures for Vertex Sets.} In principle, both $V(G)$ and $N(u)$ (i.e., a subset of $V(G)$) are sets of integers. Various data structures can be used to maintain them. Table \ref{tab:complexity_data_structure} presents the time and space complexity of these data structures involved in this paper.

DA is a resizable array with append to perform insert. In contrast, SDA is a resizable sorted array that uses binary search to locate elements and insertion to keep the array sorted. PMA~\cite{bender2007adaptive,de2019packed} maintains elements in a partially filled array organized into blocks, facilitating efficient insertions while keeping elements sorted. To maintain balance, PMAs periodically rebalance elements within blocks to ensure even distribution and optimal space usage. Both SDA and PMA extend their capacity by allocating a larger block of memory, typically doubling the current size and copying the elements to the new block. The resize cost for $n$ elements is $\Theta (n)$. PAM~\cite{sun2018pam} is a data structure that supports parallel operations on ordered maps. It uses balanced binary trees and join-based algorithms to maintain order and balance efficiently. As SL~\cite{pugh1990skip}, HT, and AVL are widely used, we omit the details for brevity.

\vspace{2pt}
\noindent\textbf{Graph Queries.} The graph community~\cite{angles2014linked,besta2023demystifying} categorizes query workloads into three classes: \emph{interactive workloads} (IC)~\cite{erling2015ldbc}, \emph{analytics workloads} (AS)~\cite{iosup2016ldbc}, and \emph{business intelligence workloads} (BI)~\cite{szarnyas2022ldbc}. IC includes queries that involve simple vertex/edge insertion/deletion operations or start from specified vertices and access a small portion of the graph. AS consists of graph traversal algorithms like PageRank, which typically visit the entire graph. BI bridges the gap between IC and AS by introducing pattern queries. IC is referred to as OLTP, while AS and BI are considered OLAP~\cite{li2022bytegraph,besta2023graph}.


\textbf{In the dynamic graph model, these queries are categorized into \emph{read queries} $Q$, which are strictly read-only, and \emph{write queries} $\Delta G$, which include update operations ~\cite{fuchs2022sortledton,li2022bytegraph}.} Typically, graph queries are read-intensive, and both read and write queries exhibit distinct characteristics:

\begin{enumerate}[leftmargin=*]
\item Read queries can be complex and long-running, with an unpredictable \emph{read set} (i.e., the vertices and edges accessed).
\item Write queries are lightweight, involving a small, known \emph{write set} of vertex/edge insertion or deletion operations.
\item As graphs are widely used for connection analysis, scan operations are a common and important graph data access pattern~\cite{zhu2019livegraph,dhulipala2019low,de2021teseo,fuchs2022sortledton,ediger2012stinger,kumar2020graphone,macko2015llama,pandey2021terrace}.
\end{enumerate}

\noindent\textbf{MVCC.} Each query can be abstracted as a transaction with read and write operations on a database. MVCC is currently the most popular transaction management scheme~\cite{wu2017empirical}. The basic idea is to maintain multiple physical versions of each logical object to allow operations on the same object in parallel to maximize parallelism of concurrent queries without sacrificing serializability.

\emph{Strict two-phase locking} (S2PL) \cite{bernstein1987concurrency} is a concurrency control protocol that operates in two phases: 1) the growing phase, where locks can be acquired but not released; and 2) the shrinking phase, where no new locks can be acquired after the first lock is released. It holds exclusive locks until the transaction commits or aborts. To avoid deadlocks, an effective approach called the no-wait policy \cite{bernstein1981concurrency} is to immediately abort a transaction if it cannot acquire a lock \cite{yu2014staring}. S2PL is a classical approach to achieving serializability.

Unlike pessimistic protocols (e.g., S2PL) that assume queries will conflict and take preventive actions, \emph{optimistic concurrency control} (OCC)~\cite{larson2011high,neumann2015fast} assumes conflicts are rare and handles them at transaction commit. OCC typically has three phases: 1) in the read phase, the transaction performs read and write operations on local copies of the data; 2) in the validation phase, upon committing, OCC checks if the data read by the transaction has been modified by other transactions; and 3) in the write phase, if validation passes, the transaction's updates are applied to the database; otherwise, the transaction is aborted and its changes are discarded. To support validation and undo operations, OCC records accessed elements in a log. OCC does not require locking elements during the read and write phases but does track accessed elements to ensure consistency. For details of MVCC, please refer to this study~\cite{wu2017empirical}.

\vspace{2pt}
\noindent\textbf{Lock and Latch.} Locks are synchronization mechanisms used to regulate access to shared data objects during concurrent transaction processing. A DBMS employs a lock manager to handle lock requests and maintain the locks acquired by each transaction. There are two types of locks: shared locks, which allow reads from multiple transactions, and exclusive locks, which permit writes exclusively. Latches protect concurrent access to in-memory data structures by threads. A read latch allows multiple threads to access shared items simultaneously, whereas a write latch permits only one thread to access shared items at a time. Existing DGS methods~\cite{zhu2019livegraph,macko2015llama,de2021teseo,fuchs2022sortledton,dhulipala2019low} do not use a lock manager but instead rely on the locking mechanisms (e.g., mutex) provided by the OS to synchronize queries. Throughout the paper, we use the term "lock" without further distinction.

\section{A Common Abstraction for DGS}\label{sec:framework}

In this section, we propose a common abstraction for DGS to facilitate a systematical study of existing methods.

\subsection{Graph Query and Data Abstraction}\label{sec:data_abstraction}

\begin{figure}[h]\small
    \setlength{\abovecaptionskip}{3pt}
    \setlength{\belowcaptionskip}{0pt}
    \includegraphics[scale=0.75]{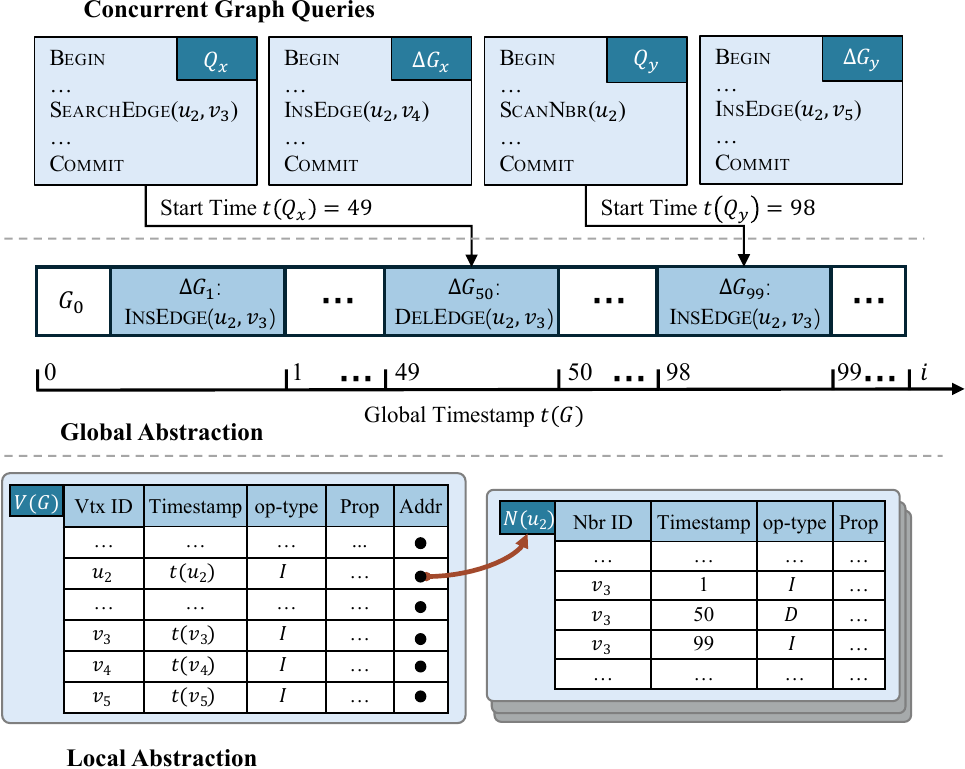}
    \centering
    \caption{The abstraction of graph query and data.}
    \label{fig:data_abstraction}
\end{figure}

In transaction management~\cite{ramakrishnan2002database}, a database is a set $\{x\}$ of tuples in tables. A transaction consists of a sequence of read and write operations on $\{x\}$, beginning with a \textsc{Begin} command and ending with either \textsc{Commit}, indicating successful execution, or \textsc{Abort}, indicating failure and reverting modifications. Building on this concept, we propose a simple yet effective multi-level abstraction for graph query and data that aims to: 1) reflect the characteristics of graph queries; 2) indicate the nature of graph data; and 3) capture graph data access patterns. Figure \ref{fig:data_abstraction} shows our abstraction.

\noindent\textbf{Global Abstraction.} Graph queries are categorized into write queries $\Delta G$ and read queries $Q$ as discussed in Section \ref{sec:preliminaries}. The global abstraction models the relationship among these queries by maintaining a global timestamp $t(G)$, initialized to 0 and incremented by 1 only when $\Delta G$ is committed. To ensure the serializability of graph queries, DGS requires that each committed write query be uniquely identified by its commit timestamp, denoted as $\Delta G_{i + 1}$ for the write query committed at $t(G) = i$. This abstraction effectively captures the construction of a dynamic graph $G = (G_0, \Delta \mathcal{G})$, where $\Delta \mathcal{G}$ is the serial execution order of committed write queries. A read query $Q$ starting at $t(G) = i$ has a local timestamp $t(Q)$.

\vspace{2pt}
\noindent\textbf{Local Abstraction.} In cooperation with the global abstraction, the local abstraction allows us to drill down to each data object and their primitive operations. The graph $G$ consists of vertices and edges. To indicate their differences and interconnections, $G$ is organized into a vertex table containing $V(G)$ and a set of neighbor tables, each corresponding to a neighbor set $N(u)$. Each entry in $V(G)$ contains the vertex ID $u$, the location of $N(u)$, and its properties, while each entry in $N(u)$ contains the neighbor ID $v$ and the properties associated with edge $e(u, v)$. Given a write query $\Delta G_i$, each operation creates a new version of a vertex or neighbor $u$ with the timestamp $t(u) = i$. Specifically, an insert (resp. delete) operation on $u$ creates a new version with the \textsc{op-type} as $I$ (resp. $D$). Updating an element is performed via an insert operation.

Lemma \ref{lemma:isolation} can be proven using the dependency graph \cite{fekete2005making} within the multi-level abstraction.  The lemma shows that by maintaining the serializability of write queries, DGS can achieve serializable isolation by ensuring $Q$ has a consistent view of $G_i$. This is done by allowing $Q$ to access only the latest version of vertices or neighbors $u$ such that $t(u) \leqslant t(Q)$. This approach allows us to coordinate $Q$'s data access based on timestamps without complex concurrency control protocols for read queries. Although existing DGS methods use this optimization, they do not explicitly and formally discuss it.

\begin{lemma} \label{lemma:isolation}
Suppose DGS maintains the serializability of write queries with the serial execution order $\Delta \mathcal{G}$. Given a read query $Q$ starting at timestamp $i$, ensuring $Q$ has a consistent view of $G_i = G_0 \oplus...\oplus \Delta G_i$ guarantees global serializable isolation for read and write queries.
\end{lemma}
\emph{Proof.} In the dependency graph, each node $Q_x$ represents a query, and an edge from $Q_x$ to $Q_y$ indicates that an operation $o_x \in Q_x$ conflicts with $o_y \in Q_y$, and $o_x$ precedes $o_y$ in execution. Operations $o_x$ and $o_y$ conflict if they act on the same object and at least one is a write. Queries are conflict serializable \emph{iff} the dependency graph is acyclic. Since DGS maintains the serializability of write queries, the dependency graph formed by them is acyclic. Ensuring $Q$ has a consistent view of $G_i$ guarantees that all operations in $Q$ depend only on operations from preceding write queries. Thus, the node representing $Q$ in the dependency graph has no incoming edges. Therefore, the combined dependency graph of read and write queries remains acyclic, proving the queries are conflict serializable.

\subsection{Graph Operations Abstraction}\label{sec:operation_abstraction}

We next analyze graph data access patterns based on the multi-level abstraction. From the perspective of DGS, a write (or read) query consists of a sequence of write (or read) operations on vertices and edges, while the computation logic and state of the query are beyond the scope of DGS. These \emph{graph operations} include:

\begin{itemize}[leftmargin=*]
\item \textsc{InsVtx($u$)}: inserting a vertex $u$ into $V(G)$;
\item \textsc{InsEdge($u, v$)}: inserting an edge $e(u, v)$ into $E(G)$;
\item \textsc{SearchVtx($u$)}: finding a vertex $u$ in $V(G)$;
\item \textsc{SearchEdge($u, v$)}: finding an edge $e(u, v)$ in $E(G)$;
\item \textsc{ScanVtx($G$)}: traversing the vertex set $V(G)$;
\item \textsc{ScanNbr($u$)}: traversing the neighbor set $N(u)$.
\end{itemize}

\begin{figure}[htbp]\small
    \setlength{\abovecaptionskip}{0pt}
    \setlength{\belowcaptionskip}{0pt}
    \includegraphics[scale=0.75]{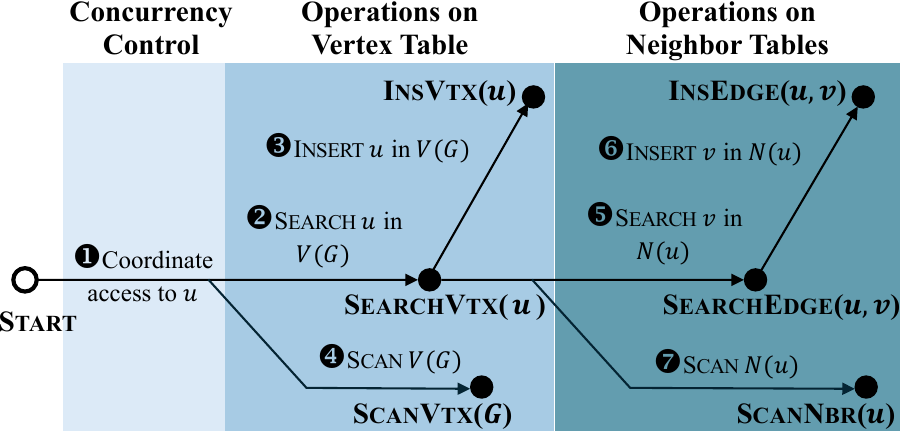}
    \centering
    \caption{The abstraction of graph operations.}
    \label{fig:primitive_opeartions}
\end{figure}

Figure \ref{fig:primitive_opeartions} presents the abstraction of graph operations, where each node represents a graph operation, and each edge denotes a primitive operator on vertex or neighbor tables. The path from the start node to an operation node illustrates the primitive operators required to perform the operation and shows the relationships between different operations (e.g., \textsc{InsEdge} invokes \textsc{SearchEdge}). In summary, this abstraction provides a unified execution routine on graph operations.


Let $P_V$ and $P_N$ denote the paths from the start node to the graph operation node in the vertex table and neighbor table operations, respectively, in Figure \ref{fig:primitive_opeartions}. Equation \ref{eq:cost} defines the cost $T$ of a graph operation, where $T_{CC}$ is the cost of coordinating access to the target vertex, and $T_p$ is the cost of the primitive operator $p$. Concurrency control requires the cooperation of underlying graph containers, such as checking a creation timestamp for each object. $\alpha_p$ is the overhead amplification ratio of concurrency control on the primitive operator $p$. An optimal value is 1, indicating no performance degradation due to concurrency control. This equation highlights the factors affecting DGS performance, serving as a tool to systematically study these performance factors rather than predicting the exact cost of a graph operation.

\begin{equation} \label{eq:cost}
T = T_{CC} + \sum_{p \in P_V} \alpha_p T_p + \sum_{p \in P_N} \alpha_p T_p.
\end{equation}

\noindent\textbf{Remark. } \sun{We exclude edge update and delete operations as they follow a similar process to insertions. Fine-grained methods like Sortledton, Teseo, and LiveGraph handle deletions by: 1) locating the target vertex, and 2) creating a new version marked as \emph{delete} or updating the end timestamp to indicate the deletion, as discussed above. Space is later reclaimed through garbage collection. Coarse-grained methods like Aspen, use a copy-on-write strategy, where deletion involves removing a vertex from a block rather than adding one, much like with insertions.} Graph query workloads generally focus on analyzing connections between vertices, with vertex updates, especially deletions, being rare \cite{zhu2019livegraph}. Consequently, existing DGS works typically focus on operations on neighbor tables. Therefore, this paper primarily focuses on $N(u)$ as well.

\section{DGS Methods Under Study}\label{sec:dgs_methods_under_study}

Existing DGS methods focus on optimizing two problems: 1) graph concurrency control, which includes version management and concurrency control protocols to coordinate the execution of concurrent graph queries, and minimizing overhead for each graph operation (i.e., $T_{CC}$ and $\alpha_p$ in Equation \ref{eq:cost}); and 2) graph containers, which include vertex indexes, neighbor indexes, and other optimizations to optimize graph data access $T_p$ for each operation in Equation \ref{eq:cost}. Table \ref{tab:summary_methods} summarizes the key design choices in these methods. In the following, we first briefly introduce these methods and then compare them.

\small
\begin{table}[t]
\centering
\caption{A summary of DGS methods under study.}
\label{tab:summary_methods}
\begin{tabular}{|c|clc|ccc|}
\hline
\textbf{}           & \multicolumn{3}{c|}{\textbf{Graph Concurrency Control}}                                                                                                       & \multicolumn{3}{c|}{\textbf{Graph Container}}                                                                                                                                                                                                                     \\ \hline
\textbf{Method}     & \multicolumn{2}{c|}{\textbf{\begin{tabular}[c]{@{}c@{}}Version\\ Management\end{tabular}}}          & \textbf{Protocol}                                       & \multicolumn{1}{c|}{\textbf{\begin{tabular}[c]{@{}c@{}}Vertex\\ Index\end{tabular}}} & \multicolumn{1}{c|}{\textbf{\begin{tabular}[c]{@{}c@{}}Neighbor\\ Index\end{tabular}}} & \textbf{\begin{tabular}[c]{@{}c@{}}Additional\\ Optimization\end{tabular}}          \\ \hline
\textbf{LiveGraph}  & \multicolumn{2}{c|}{\begin{tabular}[c]{@{}c@{}}Fine-Grained with\\ Continuous Version\end{tabular}} & S2PL                                                    & \multicolumn{1}{c|}{\begin{tabular}[c]{@{}c@{}}Dynamic\\ Array\end{tabular}}         & \multicolumn{1}{c|}{\begin{tabular}[c]{@{}c@{}}Dynamic\\ Array\end{tabular}}           & \begin{tabular}[c]{@{}c@{}}Bloom\\ Filter\end{tabular}                            \\ \hline
\textbf{Sortledton} & \multicolumn{2}{c|}{\begin{tabular}[c]{@{}c@{}}Fine-Grained with\\ Version Chain\end{tabular}}      & G2PL                                                    & \multicolumn{1}{c|}{\begin{tabular}[c]{@{}c@{}}Dynamic\\ Array\end{tabular}}         & \multicolumn{1}{c|}{\begin{tabular}[c]{@{}c@{}}Segmented\\ Skip List\end{tabular}}     & \begin{tabular}[c]{@{}c@{}}Adaptive\\ Indexing\end{tabular}                       \\ \hline
\textbf{Teseo}      & \multicolumn{2}{c|}{\begin{tabular}[c]{@{}c@{}}Fine-Grained with\\ Version Chain\end{tabular}}      & OCC                                                     & \multicolumn{1}{c|}{\begin{tabular}[c]{@{}c@{}}Hash\\ Table\end{tabular}}            & \multicolumn{1}{c|}{PMA}                                                               & \begin{tabular}[c]{@{}c@{}}Write-Optimized\\ Segment\end{tabular}                 \\ \hline
\textbf{Aspen}      & \multicolumn{2}{c|}{Coarse-Grained}                                                                 & \begin{tabular}[c]{@{}c@{}}Single\\ Writer\end{tabular} & \multicolumn{1}{c|}{\begin{tabular}[c]{@{}c@{}}AVL\\ Tree\end{tabular}}              & \multicolumn{1}{c|}{\begin{tabular}[c]{@{}c@{}}Segmented\\ PAM\end{tabular}}           & \begin{tabular}[c]{@{}c@{}}Vertex Index\\ Flatten \&\\ Data Encoding\end{tabular} \\ \hline
\textbf{LLAMA}      & \multicolumn{2}{c|}{Coarse-Grained}                                                                 & \begin{tabular}[c]{@{}c@{}}Single\\ Writer\end{tabular} & \multicolumn{1}{c|}{\begin{tabular}[c]{@{}c@{}}Dynamic\\ Array\end{tabular}}         & \multicolumn{1}{c|}{\begin{tabular}[c]{@{}c@{}}Dynamic\\ Array\end{tabular}}           & -                                                                                 \\ \hline
\end{tabular}%
\end{table}

\subsection{A Brief Introduction to DGS Methods}

\subsubsection{\textbf{LiveGraph}~\cite{zhu2019livegraph}}

\textbf{Graph Concurrency Control.} LiveGraph uses a lock for each vertex in $V(G)$ and adapts S2PL to synchronize data access. For a write query $\Delta G$, LiveGraph first obtains all exclusive locks on vertices in $\Delta V$ (vertices involved in $\Delta G$), performs the graph operations, and then releases the locks. To handle deadlocks, LiveGraph aborts $\Delta G$ if it cannot acquire a lock within a time limit.

\begin{figure}[h]\small
    \setlength{\abovecaptionskip}{3pt}
    \setlength{\belowcaptionskip}{0pt}
    \includegraphics[scale=0.75]{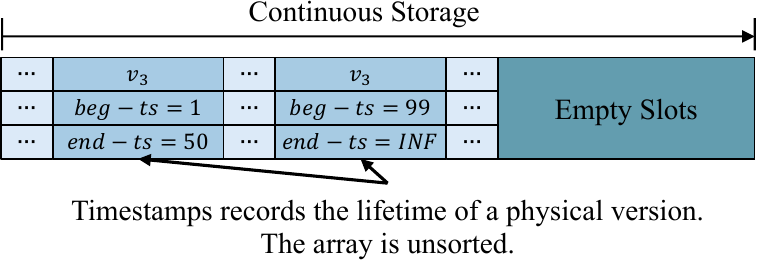}
    \centering
    \caption{The neighbor index of $N(u_2)$ in LiveGraph.}
    \label{fig:livegraph_index}
\end{figure}

For each version of a neighbor or vertex, LiveGraph keeps begin and end timestamps ($begin-ts$ and $end-ts$) to record its lifetime as shown in Figure \ref{fig:livegraph_index}. Given $\Delta G$ committed at timestamp $i$, \textsc{InsEdge($u, v$)} searches for the latest version of $v$ in $N(u)$. If found, it sets $end-ts$ to $i$ and creates a new version of $v$ with $begin-ts = i$ and $end-ts = INF$. Otherwise, it directly creates a new version of $v$. \textsc{DelEdge($u, v$)} searches for the latest version of $v$ and sets $end-ts$ to $i$. For read operations in $Q$, LiveGraph obtains a shared lock on the vertex and immediately releases it after the operation. For example, \textsc{ScanNbr($u$)} acquires the lock on $u$, accesses neighbors based on timestamps, and releases the lock immediately. Thus, $Q$ never leads to deadlocks because it never holds two locks simultaneously.

\noindent\textbf{Graph Container.} Given $u \in V(G)$, LiveGraph uses a dynamic array (DA) as the neighbor index of $N(u)$ where each element corresponds to a physical version of $v \in N(u)$. Graph operations in Figure \ref{fig:primitive_opeartions} are based on primitive operators of DA, the time complexity of which is listed in Table \ref{tab:complexity_data_structure}. As the storage of DA is continuous and no version chain exists, \textsc{Scan} is very fast. Moreover, LiveGraph executes \textsc{Scan} from the end to the beginning of DA since the latest element may be more frequently visited than the stale ones. However, \textsc{Search} is slow because DA is unsorted and uses \textsc{Scan} to perform the search. Consequently, \textsc{InsEdge} is slow because it depends on \textsc{SearchEdge} though adding an element only requires a simple append. To mitigate this issue, LiveGraph maintains a Bloom filter~\cite{mitzenmacher2001compressed} for each $N(u)$ to record whether an element exists in $N(u)$. LiveGraph uses DA as the vertex index of $V(G)$ as well. As the vertex ID is ranged in $[0, |V|)$, the element at the index $u$ is the vertex $u$. Therefore, the time complexity of \textsc{Search} on the vertex index is $O(1)$. As the implementation is simple, we omit the details.

\subsubsection{\textbf{Sortledton}~\cite{fuchs2022sortledton}}

\begin{figure}[h]\small
    \setlength{\abovecaptionskip}{3pt}
    \setlength{\belowcaptionskip}{0pt}
    \includegraphics[scale=0.75]{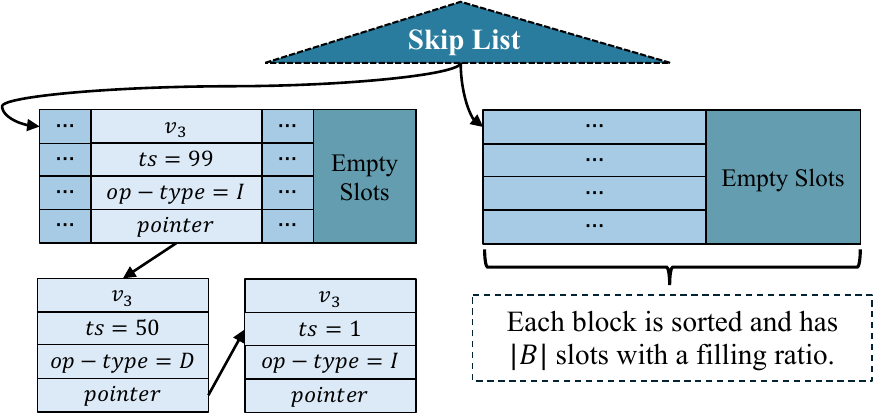}
    \centering
    \caption{The neighbor index of $N(u_2)$ in Sortledton.}
    \label{fig:sortledton_neighbor_index}
\end{figure}

\textbf{Graph Concurrency Control.} Sortledton also uses S2PL but optimizes the locking sequence: Sort the vertices in $\Delta V$ by ascending vertex IDs and acquire their exclusive locks in that order. This optimization prevents deadlocks by avoiding circular waiting among write queries~\cite{silberschatz1991operating}. Therefore, Sortledton does not need any mechanisms to handle deadlocks. For \textsc{InsEdge($u, v$)} (or \textsc{DelEdge($u, v$)}) in $\Delta G$ committed at timestamp $i$, Sortledton creates a new version of $v$ with timestamp $i$ and \textsc{op-type} as $I$ (or $D$) as shown in Figure \ref{fig:sortledton_neighbor_index}. Sortledton maintains a version chain for different versions of $v$, where the new version points to the old one. For read queries, Sortledton uses the same concurrency control strategy as LiveGraph.

\noindent\textbf{Graph Container.} Like LiveGraph, Sortledton uses a dynamic array as the vertex index. For the neighbor index, Sortledton splits $N(u)$ into blocks $B$, and uses the skip list as the block index linking them. The fill ratio of each block is maintained between 50\% and 100\%. When a block is full, Sortledton splits it into two blocks, equally distributing the neighbors. If the fill ratio drops below 50\%, Sortledton merges it with adjacent blocks. The first element of each block serves as its key in the skip list. We call this structure the \emph{segmented skip list}. A read operation traverses the version chain to find the target version. Since real-world graphs are sparse, Sortledton proposes the \emph{adaptive neighbor index}: If $|N(u)|$ is below a threshold (e.g., 256), it uses a sorted dynamic array as the neighbor index instead of the segmented skip list.

\subsubsection{\textbf{Teseo}~\cite{de2021teseo}}

\noindent\textbf{Graph Concurrency Control.} Teseo uses the same version management method as Sortledton but adopts optimistic concurrency control (OCC introduced in Section \ref{sec:preliminaries}) to coordinate concurrent write queries. Unlike LiveGraph and Sortledton, which keep a lock for each vertex, Teseo maintains a lock for each edge partition to synchronize concurrent data access. Specifically, Teseo logically divides $E(G)$ into equally sized partitions, each with a lock. Before accessing data in a partition, a write query acquires an exclusive lock (or a read query acquires a shared lock) on the partition and releases it immediately after access.

\begin{figure}[h]\small
    \includegraphics[scale=0.75]{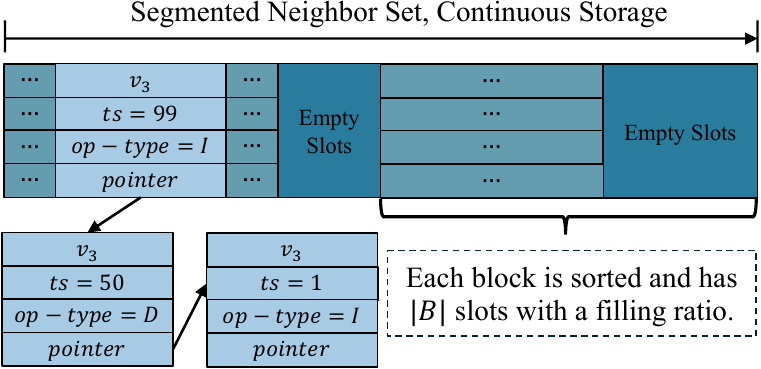}
    \centering
    \caption{The neighbor index of $N(u_2)$ in Teseo.}
    \label{fig:teseo_neighbor_index}
\end{figure}

\noindent\textbf{Graph Container.} Teseo employs the packed memory array (PMA)~\cite{bender2007adaptive,de2019packed} as the neighbor index as shown in Figure \ref{fig:teseo_neighbor_index}. But it stores all neighbor tables in a single PMA. If $N(u)$ is large, it spans multiple blocks in the PMA, while multiple small neighbor sets can share a single block. However, the global rebalance overhead is high for a PMA if $E(G)$ is stored together. To address this, Teseo divides the single PMA into multiple large leaves (several megabytes each) and indexes these leaves with an \emph{adaptive radix tree} (ART)~\cite{leis2013adaptive}, calling this data structure FAT. Additionally, Teseo uses a hash table as the vertex index to record the position of each vertex's neighbor index in FAT. By default, blocks in FAT are sorted, known as read-optimized segments. When the insertion rate is high, FAT switches to write-optimized segments (WOS), which handle updates by appending to the update log. For WOS, Teseo loops over the segment to find the target vertex. Teseo is built on top of HyPer~\cite{kemper2011hyper}.

\subsubsection{\textbf{Aspen}~\cite{dhulipala2019low}}

\begin{figure}[h]\small
    \setlength{\abovecaptionskip}{3pt}
    \setlength{\belowcaptionskip}{-5pt}
    \includegraphics[scale=0.75]{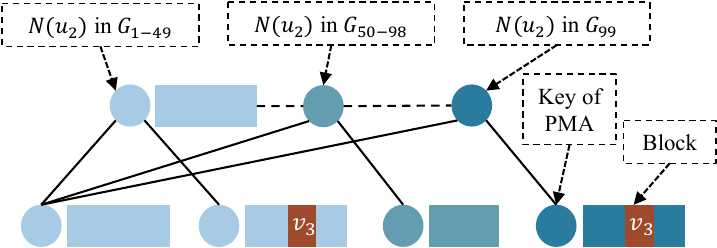}
    \centering
    \caption{The neighbor index of $N(u_2)$ in Aspen.}
    \label{fig:aspen_neighbor_index}
\end{figure}

\textbf{Graph Concurrency Control.} Aspen uses a coarse-grained strategy that maintains timestamps for each graph snapshot. Specifically, Aspen uses the \emph{single-writer-multiple-reader} scheme, which executes write queries serially and allows multiple readers to execute concurrently. A write query $\Delta G_i$ uses the \emph{copy-on-write} (CoW) method (also called shadow paging) to apply updates on a copy of the graph, creating a new snapshot $G_i$ as shown in Figure \ref{fig:aspen_neighbor_index}. $Q$ works on the latest version of the graph snapshot. Therefore, read and write queries never block each other, and multiple read queries can share the same graph snapshot.

\vspace{2pt}
\noindent\textbf{Graph Container.} Aspen partitions $N(u)$ into a set of sorted blocks and uses a parallel augmented map (PAM) to index these blocks. These blocks have no empty slots. \textsc{InsEdge($u, v$)} copies the block, inserts $v$, and then copies the path from the block to the root to create a new snapshot of $N(v)$. Given $N(u)$ and block size $B$, Aspen selects vertices such that $v \mod b = 0$ as heads, i.e., $heads = {v \in N(u) \mid v \mod b = 0}$. This approach ensures that updates to one block do not affect adjacent blocks. Aspen demonstrates that this segmentation ensures each block has $B$ elements with high probability. Moreover, Aspen uses an AVL tree as the vertex index and copies the path in the tree for each update operation.

Aspen proposes two optimization methods to enhance performance. First, for long-running read queries, Aspen can create an array storing the positions of each neighbor table based on the AVL tree to eliminate the overhead of \textsc{Search} in the AVL tree. Second, Aspen uses a difference encoding scheme to compress data in a block. For a block containing $(v0, v1, v2, \ldots)$, Aspen stores it as $(v0, v1 - v0, v2 - v0, \ldots)$ and compresses it with byte codes to accelerate set intersections~\cite{aberger2017emptyheaded}.

\subsubsection{\textbf{LLAMA}~\cite{macko2015llama}}

LLAMA, proposed in 2015, also employs a coarse-grained graph concurrency control mechanism similar to that in Aspen. Specifically, LLAMA divides the vertex table into partitions, each stored in a data page, and maintains an \emph{indirection table} to store the locations of these pages. Each write query must copy the indirection table to create a new graph snapshot, and this overhead limits update performance and graph scalability.

\subsection{Comparison of DGS Methods} \label{sec:discussion}

\noindent\textbf{Graph Containers.} We discuss the time and space cost of graph operations based on Table \ref{tab:complexity_data_structure}.

\emph{Time.} Given vertex IDs in the range $[0, |V|)$, using a dynamic array (DA) as vertex indexes allows \textsc{SearchVtx} and \textsc{InsVtx} in $O(1)$ time with simple memory accesses. Due to continuous storage, DA enables fast scan operations. In contrast, hash tables and AVL trees incur more overhead than DA, despite having the same time complexity for some operations.

Continuous storage, used in LiveGraph and LLAMA, stores $N(u)$ in DA, facilitating fast \textsc{ScanNbr} but resulting in slow \textsc{SearchEdge} due to the unsorted nature of the array. Since \textsc{InsEdge} depends on \textsc{SearchEdge} in DGS, its performance is also slow despite \textsc{Insert} in DA taking $O(1)$ time. For the segmented methods, scanning accesses data continuously within a block, while inserting typically moves only a few elements within a block. These blocks are linked by an index (e.g., PAM) to accelerate \textsc{SearchEdge}. Therefore, the cost of these operations includes the block index and the block itself. Increasing block sizes can improve scan performance due to continuous memory access but may degrade insert performance due to data movement within the block. Additionally, Aspen, using CoW, incurs more overhead for insertion than methods performing in-place updates because its insert operation copies the block as well as the root-to-leaf path.

\emph{Space.} Practical memory consumption is significantly affected by element size. Each element in Sortledton and Teseo consumes $3 \times w$ bytes, where $w$ is the word size: one for the vertex ID, one for the version, and one for the pointer. The \textsc{op-type} can use the high bit in the timestamp. Each element in LiveGraph also takes $3 \times w$ bytes. In contrast, each element in Aspen consumes only $w$ bytes due to its coarse-grained granularity. Additionally, Aspen’s neighbor index has no empty slots. Therefore, the coarse-grained method is more memory efficient than the fine-grained method.

\vspace{2pt}
\noindent\textbf{Graph Concurrency Control.} We first compare fine-grained and coarse-grained strategies, and then discuss fine-grained methods.

\emph{Fine-Grained vs. Coarse-Grained.} Fine-grained methods require lock operations on each graph operation, which can lead to lock contention and thus expensive $T_{CC}$. High-degree vertices are particularly prone to frequent access. Fine-grained methods necessitate version checks on each element, resulting in increased data loading from memory and more computation for version comparison, leading to a high $\alpha_p$ value. In contrast, coarse-grained methods avoid these issues. However, fine-grained methods allow multiple writers to update the graph simultaneously and perform in-place updates by simply inserting new elements, enhancing update performance.

Additionally, the fine-grained strategy does not place special requirements on the underlying graph containers, making it more generic. In contrast, the coarse-grained strategy requires support for fast snapshot creation. However, since the coarse-grained strategy does not maintain versions or perform version checks for each element, it can be effectively combined with data compression techniques~\cite{dhulipala2019low}, which is not feasible for the fine-grained approach.

\emph{Discussion on Fine-Grained Strategies.} First, the continuous version storage in LiveGraph can improve scanning efficiency by avoiding the traversal of a version chain. However, it may increase data access volume by including stale versions, negatively impacting search and insert efficiency. Second, G2PL in Sortledton is generally the optimal fine-grained concurrency control mechanism due to its effective deadlock avoidance optimization. Although OCC in Teseo does not require holding all mutexes of vertices in $\Delta V$, write queries are typically very short because $\Delta G$ generally contains a single update. Moreover, executing deadlock detection for write-write conflicts introduces significant overhead and implementation challenges. As such, our study uses G2PL for fine-grained methods because of its advantages.

\vspace{2pt}
\noindent\textbf{Empirical Evaluation Targets.} Following the above discussion, we will set up test frameworks and empirically evaluate these techniques by addressing the following five questions.

\begin{itemize} [leftmargin=*]
    \item \textbf{Graph Containers:} \emph{\textbf{Q1.} \sun{How effective are existing techniques in graph containers at efficiently performing key operations such as \textsc{SearchEdge}, \textsc{InsEdge}, and \textsc{ScanNbr}, as defined by $T_p$ in Equation \ref{eq:cost}?}} \emph{\textbf{Q2.} Which neighbor index performs the best, and what is the gap between it and CSR on read queries?}
    \item \textbf{Graph Concurrency Control:} \emph{\textbf{Q3.} What is the impact of graph concurrency control on graph operations?} \emph{\textbf{Q4.} How are the scalability and concurrency of competing methods?}
    \sun{
    \item \textbf{Batch Granularity:} \emph{\textbf{Q5.} How does the batch granularity affect the performance of competing methods?}
    }
    \item \textbf{Memory Consumption:} \emph{\textbf{Q6.} What is the impact of graph containers and version management in DGS on memory consumption, and what is the gap between DGS and CSR?}
\end{itemize}

\begin{figure*}[t]\small
    \setlength{\abovecaptionskip}{0pt}
    \includegraphics[scale=0.45]{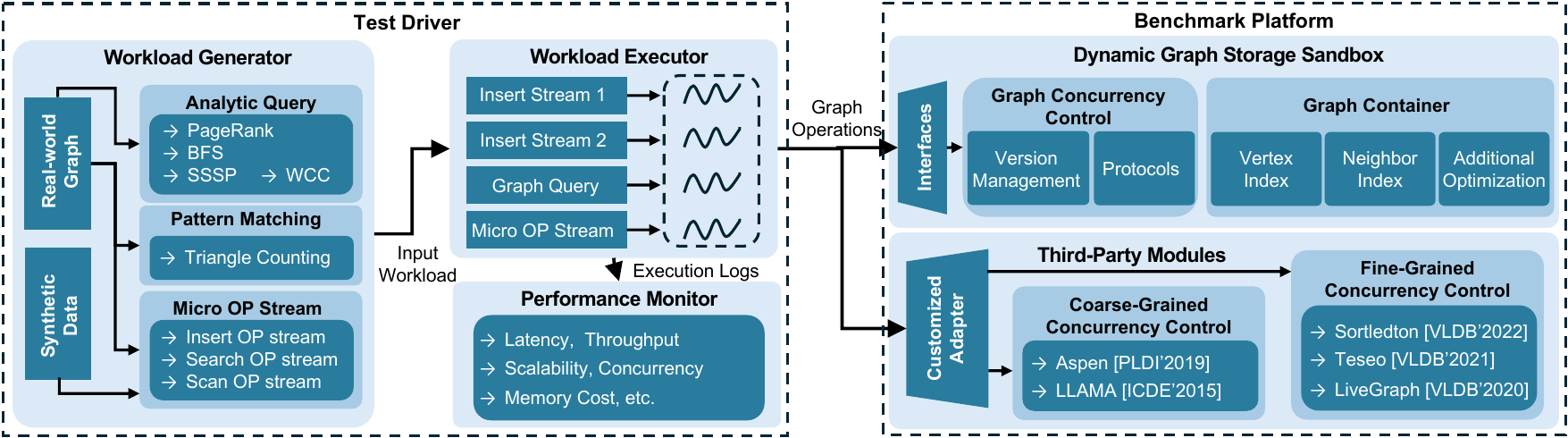}
    \centering
    \caption{An overview of the test framework.}
    \label{fig:overview_test_framework}
\end{figure*}

\section{Test Framework Setup} \label{sec:test_framework_setup}

Figure \ref{fig:overview_test_framework} provides an overview of this framework. It is implemented with around 7800 lines of optimized C++ code. The source code is compiled using g++ 10.5.0 with the -O3 optimization enabled. Experiments are conducted on a Linux server equipped with an AMD EPYC 7543 CPU (32 cores) and 128GB of memory.

\subsection{Benchmark Platform} \label{sec:benchmark_platform}

\vspace{2pt}
\noindent\textbf{Third-Party Modules.} This module integrates the original implementations of LLAMA~\cite{llamacodebase}, Aspen~\cite{aspencodebase}, LiveGraph~\cite{livegraphcodebase}, Teseo~\cite{teseocodebase}, and Sortledton~\cite{sortledtoncosebase}  from GitHub. Due to differences in their APIs, we implemented a customized adapter, a wrapper for each method, to standardize the evaluation of graph operations. The overhead of these adapters is negligible because they simply combine the interfaces to provide these functions, if not directly available. Each method is configured with its recommended settings. Specifically, the size of bloom filter in LiveGraph is set as $\frac{1}{16}$ of the block size. Sortledton, Teseo and Aspen set the block size to 256.

\vspace{2pt}
\noindent\textbf{DGS Sandbox.} In this module, we re-implement key techniques from competing methods within our abstraction for a fair and detailed investigation of individual techniques. These techniques can be composed differently to evaluate graph operations based on the unified execution routines shown in Figure \ref{fig:primitive_opeartions}. For fine-grained methods, we apply the following optimizations: 1) all methods use G2PL as the concurrency control protocol, and 2) all methods use a dynamic array as the vertex index by default. We implement a simple baseline DGS method by combining the static graph storage AdjLst with G2PL, naming it AdjLst. \sun{Specifically, AdjLst is implemented as an array where each element represents a vertex, and each vertex points to an array storing its neighbor set. When inserting a new element, a binary search is performed to find the correct position, after which the subsequent elements are shifted to make space for the new one. If the array is full, pre-allocated space is used, functioning like a dynamic array.} For comparison purposes, we also include CSR, the optimal baseline for static graphs.

\subsection{Test Driver}

\vspace{2pt}
\noindent\textbf{Workload Generator.} Table \ref{tab:datasets} presents the statistics of the real-world graphs used in the paper. They span six categories, with $|V|$ ranging from tens of thousands to tens of millions and $|E|$ scaling to hundreds of millions.  Both \emph{ldbc}~\cite{ldbcic} and \emph{nft}~\cite{Zhang2023LiveGL, livegraphlab} originally have timestamps to mark the insertion sequence of edges. \emph{ldbc} simulates actions in a social network. We set the scale factor to 10 to control the graph size. \emph{nft} records NFT transactions on Ethereum from 2017 to 2022. Other graphs are obtained from SNAP~\cite{snapnets} and do not include timestamps. These graphs are widely used in DGS research. We also considered larger graphs (e.g., \emph{friendster}) containing billions of edges but omitted them since existing DGS methods frequently run out of memory on these cases.
\small
\begin{table}[h]
\captionsetup{skip=0pt} 
\centering
\caption{The detailed information of the real-world graphs.}
\label{tab:datasets}
\begin{tabular}{|c|c|c|c|c|c|c|}
\hline
\textbf{Category}                & \textbf{Dataset}     & \textbf{Name} & \textbf{|\textit{V}|} & \textbf{|\textit{E}|} & \textit{\textbf{$d_{avg}$}} & $d_{max}$ \\ \hline
\multirow{4}{*}{\textbf{Social}} & \textbf{LiveJournal} & \emph{lj}       & 4.8M                  & 42.8M                 & 8.8                         & 20233     \\ \cline{2-7} 
                   & \textbf{LDBC}       & \emph{ldbc} & 30.0M  & 175.9M  & 5.9   & 4282595 \\ \cline{2-7} 
                   & \textbf{Twitter}    & \emph{tw}   & 21.3M & 265.0M & 12.4  & 698112 \\ \hline
\textbf{Game}      & \textbf{DotaLeague} & \emph{dl}   & 0.06M & 50.9M  & 831.6 & 17004   \\ \hline
\textbf{Web}       & \textbf{Wiki}       & \emph{wk}   & 14.0M & 59.0M  & 4.2   & 723404  \\ \hline
\textbf{Citation}  & \textbf{Cit}        & \emph{ct}   & 3.8M  & 16.5M  & 4.4   & 793     \\ \hline
\textbf{Synthetic} & \textbf{Graph500}   & \emph{g5}   & 8.9M  & 260.4M & 29.3  & 406416  \\ \hline
\textbf{Financial} & \textbf{NFT}   & \emph{nft}   & 29.6M  & 77.5M &  2.62 &  2290853 \\ \hline
\end{tabular}%
\end{table}

We generate three types of graph queries. First, the \emph{micro OP stream} contains a sequence of graph operations. For graphs with timestamps, we generate an \textsc{InsEdge} stream using the first 80\% of edges as the initial graph and the remaining 20\% as the insert edges. For graphs without timestamps, we shuffle the edges and generate the insert stream similarly, following previous works~\cite{zhu2019livegraph,de2021teseo,fuchs2022sortledton}. As these works focus on single updates, each operation corresponds to a write query $\Delta G$. For the \textsc{SearchEdge} stream, we randomly select 20\% of edges as the search targets. For the \textsc{ScanNbr} stream, we select 20\% of vertices based on their degrees. Each of these operations is a read query. The micro OP stream serves two purposes: 1) Investigating the performance of competing methods on basic graph operations, and 2) Studying the effectiveness of short queries (i.e., IC workloads).

Second, we integrate four representative \emph{graph analytic} algorithms from GAPBS~\cite{beamer2015gap}: PR (PageRank), BFS, SSSP, and WCC, which cover different graph data access patterns. PR sequentially accesses both vertices and neighbors. BFS and WCC visit neighbors sequentially while accessing vertices randomly. SSSP introduces a random access pattern to neighbors when retrieving weights. Third, we implement TC (triangle counting) as the representative \emph{graph pattern matching} query. This query requires DGS to support quick scanning in sorted order for fast set intersections. In summary, these two types of queries evaluate the effectiveness of complex, long-running graph queries (i.e., BI and IS workloads).

Real-world graphs following a power-law degree distribution complicate examining the performance of graph operations on different neighbor set sizes because accessing frequently used sets of elements can improve cache performance and lead to biased results. To address this, we design experiments using synthetic datasets. A \emph{synthetic dataset} consists of sets of elements with uniform sizes. Each element is a vertex with an ID ranging from $[0, 2^{22})$. Assuming each vertex ID is 8 bytes, to evaluate performance on a neighbor set with 512 elements, we generate $x = \frac{8 \text{GB}}{512 \times 8 \text{ bytes}}$ sets of the same size. These sets are labeled from $[0, x)$. We generate insert, scan, and search OP streams using the same strategy as for real-world graphs, treating set IDs as vertex IDs and the sets as neighbor sets. We default to using 8GB to prevent all sets from residing in the cache, thereby simulating random memory access in large graphs.

\vspace{2pt}
\noindent\textbf{Workload Executor.} Each thread executes a stream or a graph algorithm. Operations within the same stream execute sequentially by one thread, while multiple threads can execute concurrently. Although the graph algorithms in the test framework can run in parallel, we execute them in a single thread to focus on the DGS's capability to empower concurrent graph query execution and handle different queries.

\vspace{2pt}
\noindent\textbf{Performance Monitor.} We use \emph{throughput}, the number of edges processed (insert, search, scan) per second, to measure DGS efficiency. We use \emph{latency}, the time elapsed to complete one query, to examine service quality. Recording the latency of each operation incurs non-trivial overhead due to the short duration of single graph operations. Therefore, we record latency every one hundred micro-operations. \emph{Scalability} is measured by the throughput of micro-operations as the number of threads increases, while \emph{concurrency} is measured when multiple micro-operation streams (or graph algorithm queries) are mixed. \emph{Memory cost} tracks the memory consumption of competing storage methods.

\section{Experimental Results} \label{sec:experiments}

For brevity, we report the results for \emph{lj}, a sparse graph with no high-degree vertices, and \emph{g5}, a dense graph with high-degree vertices. Due to space limit, we have included the results on three graph analytics
queries: BFS, SSSP, and WCC, graph operation latency and the comparison with their original implementations in the complete version. We omit the results of LLAMA due to its memory consumption issues.  \sun{Guided by Equation \ref{eq:cost}, we conduct experiments according to the following roadmap.}


\sun{
\begin{itemize}[leftmargin=*]
\item \textbf{Graph Container Efficiency:} We measure the raw performance of graph containers on basic operations such as search, scan, and insert, as well as on graph analytics queries. These experiments are conducted in a single-threaded environment without concurrency control to isolate the baseline performance.
\item \textbf{Concurrency Control Effectiveness:} We assess the impact of concurrency control on query performance, scalability across varying workloads, and the efficiency of concurrent read and write operations.
\item \textbf{Batch Granularity:} We examine system performance under batch updates, focusing on behavior with different batch sizes.
\item \textbf{Memory Usage:} We compare the memory overhead of competing methods against static graph formats.
\end{itemize}
}

\subsection{Evaluation of Graph Containers} \label{sec:graph_container_efficiency}

To compare the absolute performance of graph containers, we execute graph queries bypassing concurrency control operations in Figure \ref{fig:primitive_opeartions}. The elements in both vertex and neighbor indexes contain only vertex IDs without version information.


\subsubsection{Efficiency of Vertex Indexes.} 
\begin{figure}[h]
	\setlength{\abovecaptionskip}{0pt}
	\setlength{\belowcaptionskip}{0pt}
		\captionsetup[subfigure]{aboveskip=0pt,belowskip=0pt}
	\centering
    \includegraphics[width=0.6\textwidth]{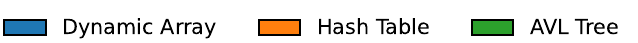}
    
    \begin{subfigure}[t]{0.20\textwidth}
    		\centering
    		\includegraphics[width=\textwidth]{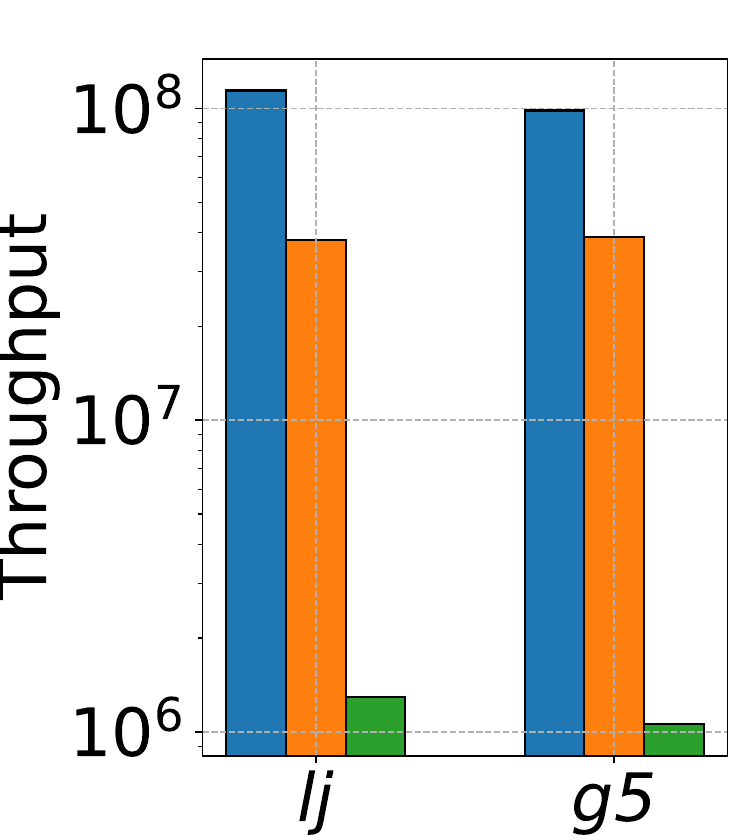}
    		\caption{\textsc{SearchVtx}.}
    		\label{fig:vertex_search_performance}
    \end{subfigure}  
    \begin{subfigure}[t]{0.20\textwidth}
    		\centering
    		\includegraphics[width=\textwidth]{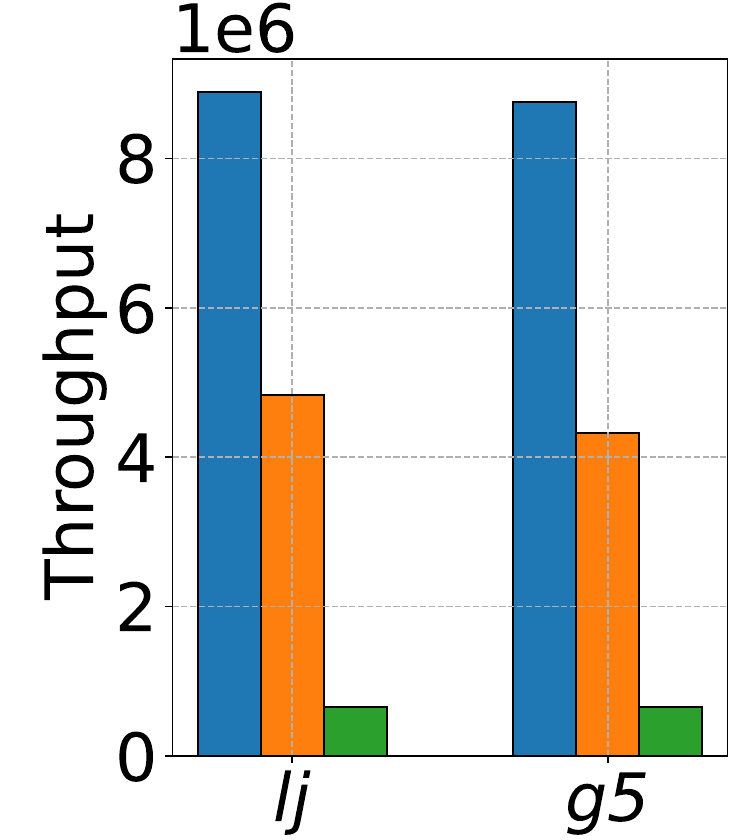}
    		\caption{\textsc{InsVtx}.}
    		\label{fig:vertex_insert_performance}
    \end{subfigure}
    \begin{subfigure}[t]{0.20\textwidth}
    		\centering
    		\includegraphics[width=\textwidth]{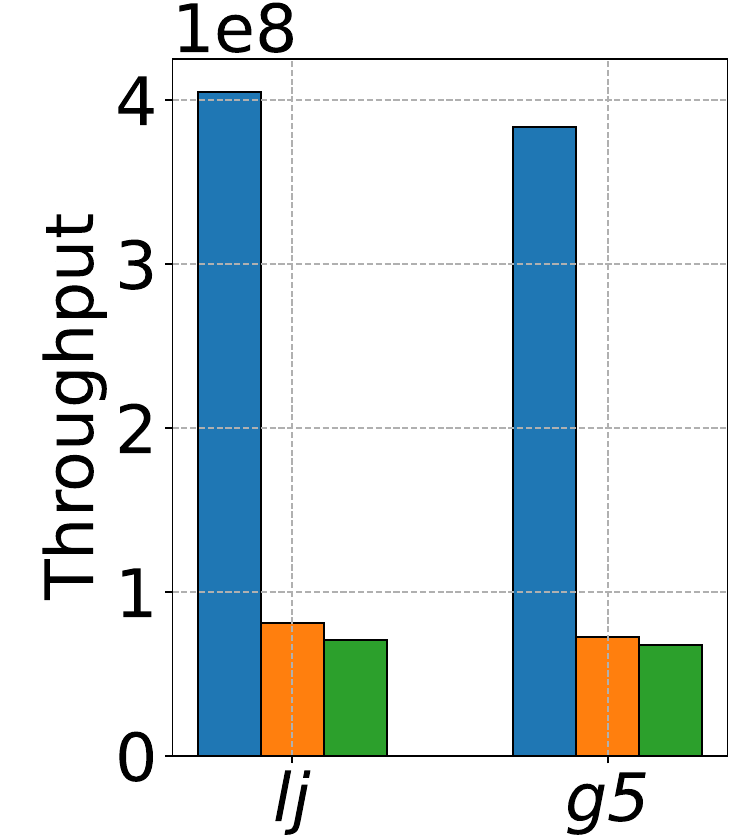}
    		\caption{\textsc{ScanVtx}.}
    		\label{fig:vertex_scan_performance}
    \end{subfigure}     
    \caption{Efficiency of vertex indexes.}
    \label{fig:vertex_index_performances}
\end{figure}

Figure \ref{fig:vertex_index_performances} presents the results for three vertex indexes listed in Table \ref{tab:summary_methods}. For search efficiency, the dynamic array achieves speedups of over 2.6x compared to the hash table and two orders of magnitude compared to the AVL tree. Since \textsc{SearchVtx} is a step in graph operations on edges, this difference can significantly affect edge operations. Vertices are inserted in vertex ID order since existing methods incrementally assign vertex IDs from 0, as discussed in Section \ref{sec:preliminaries}. The throughput of the dynamic array is approximately 2x and 8x higher than the hash table and AVL tree, respectively. The AVL tree is very slow because it copies the path for multiple graph snapshots. For scan operations, dynamic arrays are 4x faster than counterparts.


\subsubsection{Efficiency of Neighbor Indexes} \label{sec:effiency_of_neighbor_indexes}

To provide a performance reference for segmented PAM in Aspen, we implement O-Aspen, which uses a dynamic array as the vertex index and does not create vertex index snapshots for insert operations.

\begin{figure}[h!]
	\setlength{\abovecaptionskip}{0pt}
	\setlength{\belowcaptionskip}{-2pt}
		\captionsetup[subfigure]{aboveskip=0pt,belowskip=0pt}
	\centering
    \includegraphics[width=0.60\columnwidth]{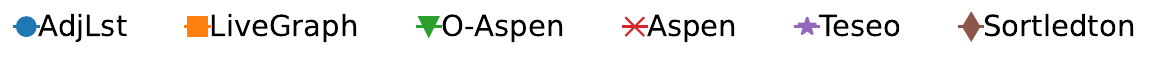}\\
	\begin{subfigure}[t]{0.30\textwidth}
		\centering
		\includegraphics[width=\textwidth]{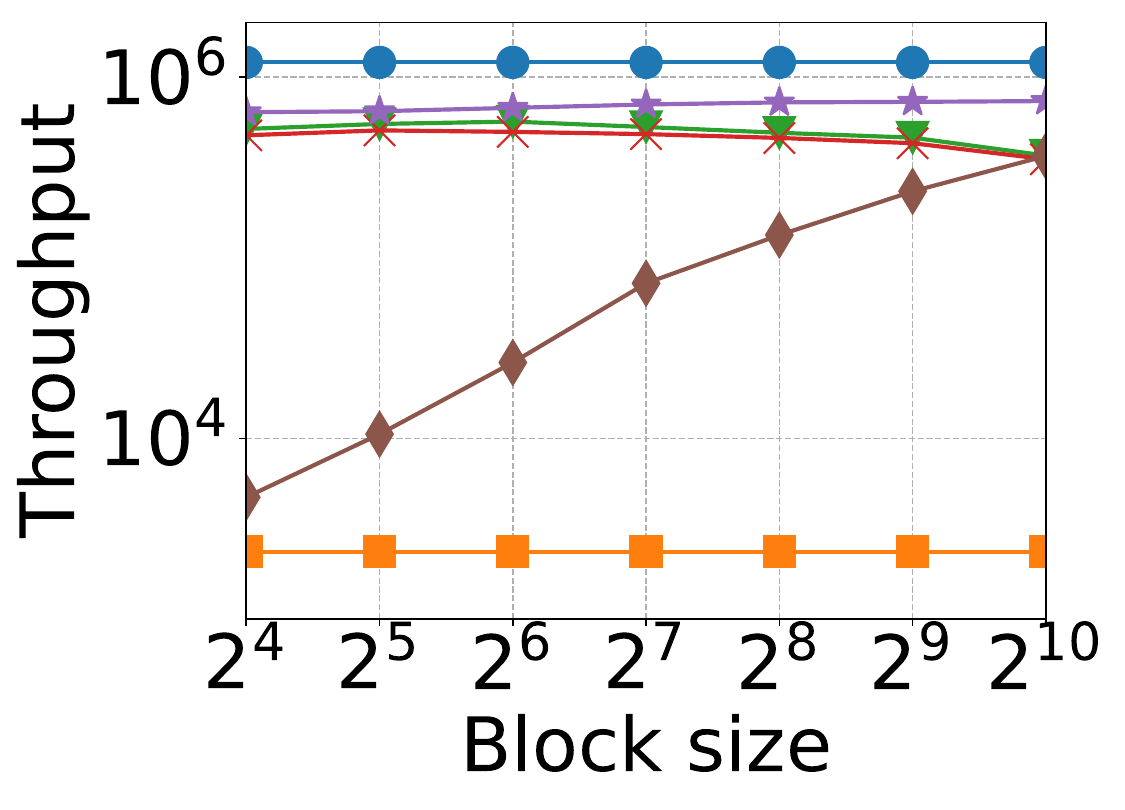}
		\caption{Vary $|B|$ when $|N(u)| = 2 ^ {20}$.}
		\label{fig:search_vary_block_size}
	\end{subfigure}
    \begin{subfigure}[t]{0.30\textwidth}
		\centering
		\includegraphics[width=\textwidth]{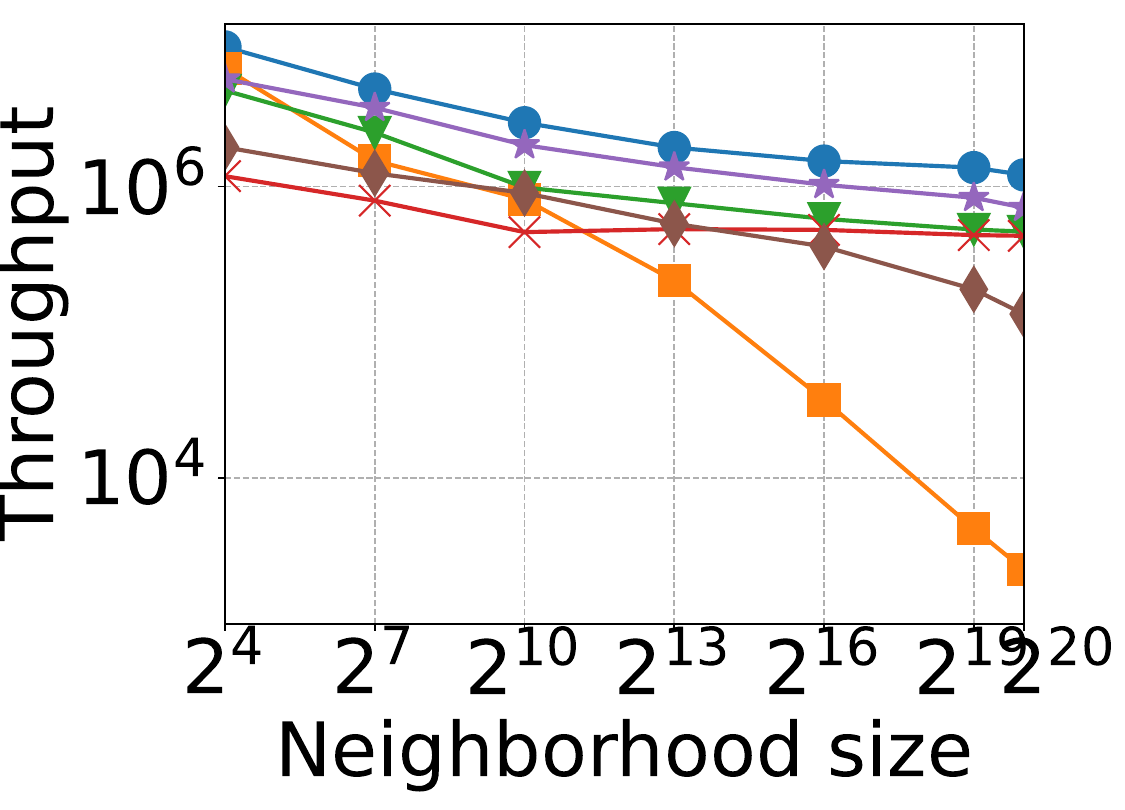}
		\caption{Vary $|N(u)|$ when $|B| = 256$.}
		\label{fig:search_vary_neighbor_set_size}
	\end{subfigure}
    \begin{subfigure}[t]{0.60\textwidth}
		\centering
		\includegraphics[width=\textwidth]{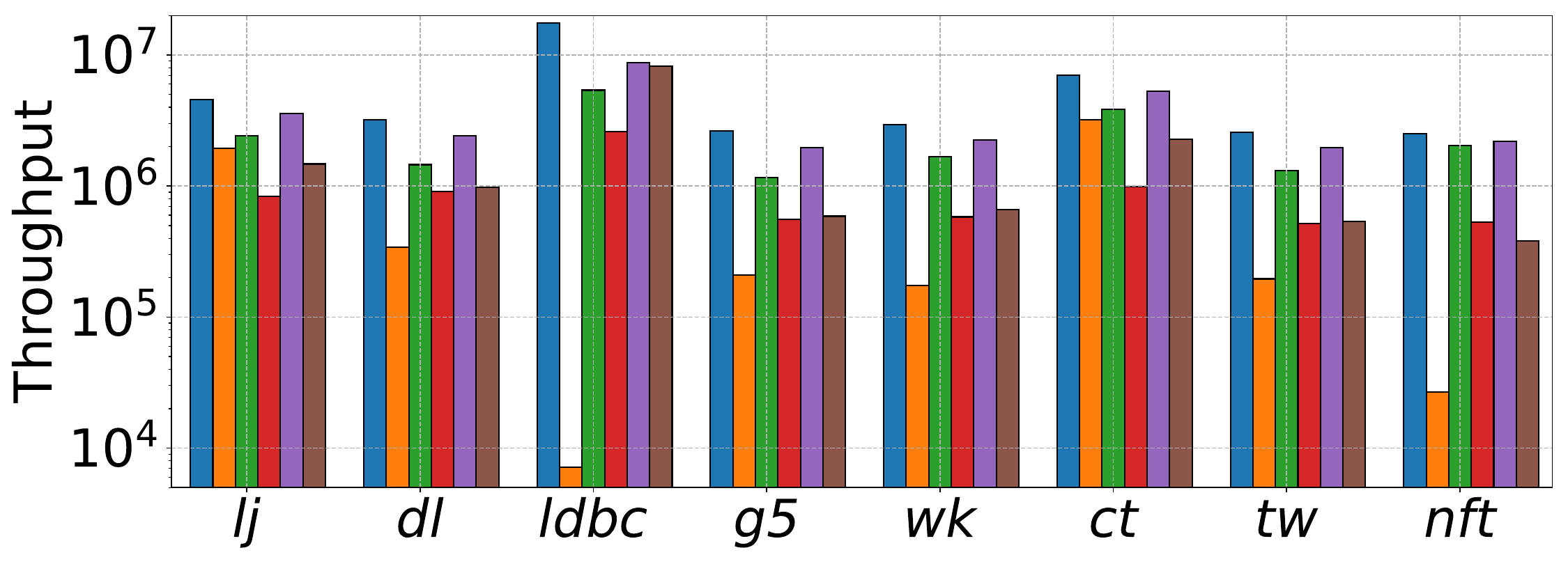}
		\caption{Vary real-world graphs when $|B| = 256$.}
		\label{fig:search_vary_graphs}
	\end{subfigure}
    
	\caption{Efficiency of \textsc{SearchEdge}.}
	\label{fig:search}
\end{figure}

\noindent\textbf{\textsc{SearchEdge}.} Figure \ref{fig:search} presents the results on search efficiency. LiveGraph and AdjLst keep $N(u)$ in a continuous array and are therefore unaffected by $|B|$ in Figure \ref{fig:search_vary_block_size}. LiveGraph runs slowly due to its unsorted neighbor index, while AdjLst performs the best, benefiting from efficient binary search on the continuous array. All segmented indexes perform better as $|B|$ increases, due to fewer blocks and lower block-index search overhead. O-Aspen, Aspen, and Teseo are much faster than Sortledton due to their efficient block indexing. Overall, AdjLst runs 1.2-5.8x times faster than O-Aspen and Teseo, and they are very close in performance on small neighbor sets in Figure \ref{fig:search_vary_neighbor_set_size}. Additionally, except for LiveGraph, search efficiency slightly decreases as $|N(u)|$ increases.

Figure \ref{fig:search_vary_graphs} illustrates the results on real-world graphs. These methods exhibit similar relative performance as shown in Figures \ref{fig:search_vary_block_size} and \ref{fig:search_vary_neighbor_set_size}, except for LiveGraph. LiveGraph is competitive with Sortledton on \emph{lj} and \emph{ct} but much slower on the other graphs. This is because most vertices in \emph{lj} and \emph{ct} have small degrees and no vertices with very high degrees.


\begin{figure}[h!]
	\setlength{\abovecaptionskip}{0pt}
	\setlength{\belowcaptionskip}{0pt}
		\captionsetup[subfigure]{aboveskip=0pt,belowskip=0pt}
	\centering
    \includegraphics[width=0.60\textwidth]{img/exp_figs/figure1/legend.pdf}\\
	\begin{subfigure}[t]{0.30\textwidth}
		\centering
		\includegraphics[width=\textwidth]{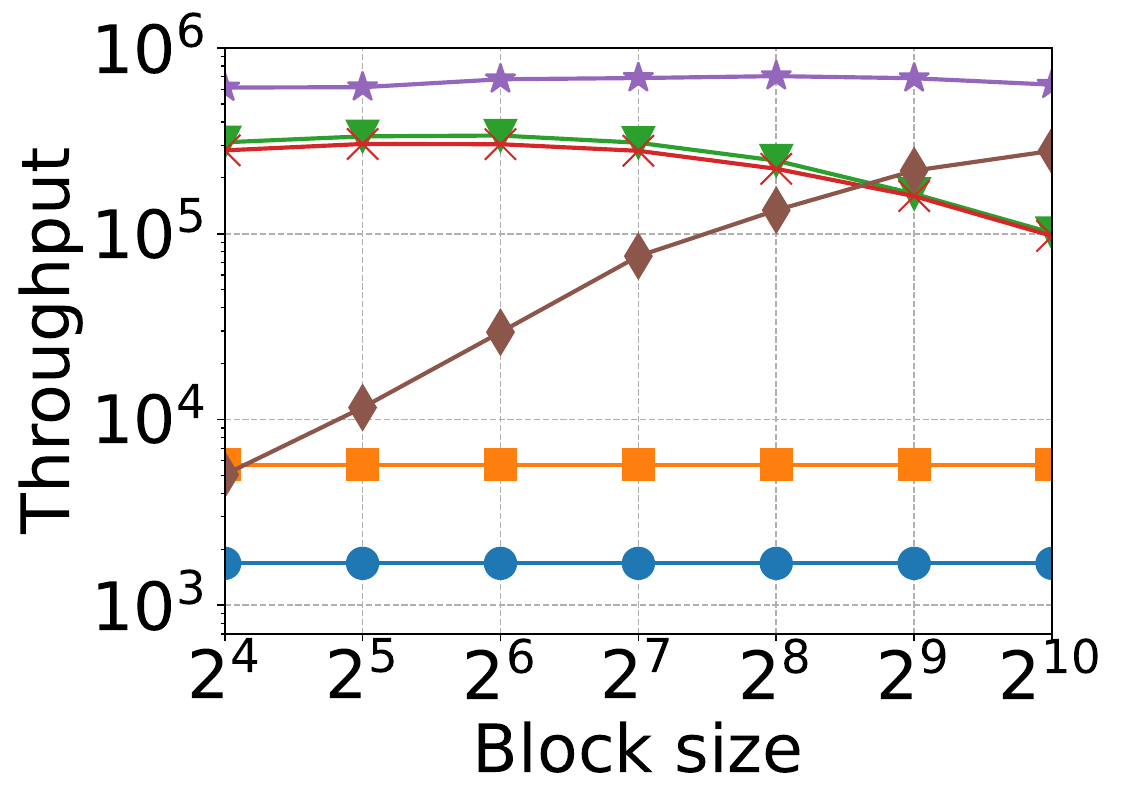}
		\caption{Vary $|B|$ when $|N(u)| = 2 ^ {20}$.}
		\label{fig:insert_vary_block_size}
	\end{subfigure}
        \begin{subfigure}[t]{0.30\textwidth}
		\centering
		\includegraphics[width=\textwidth]{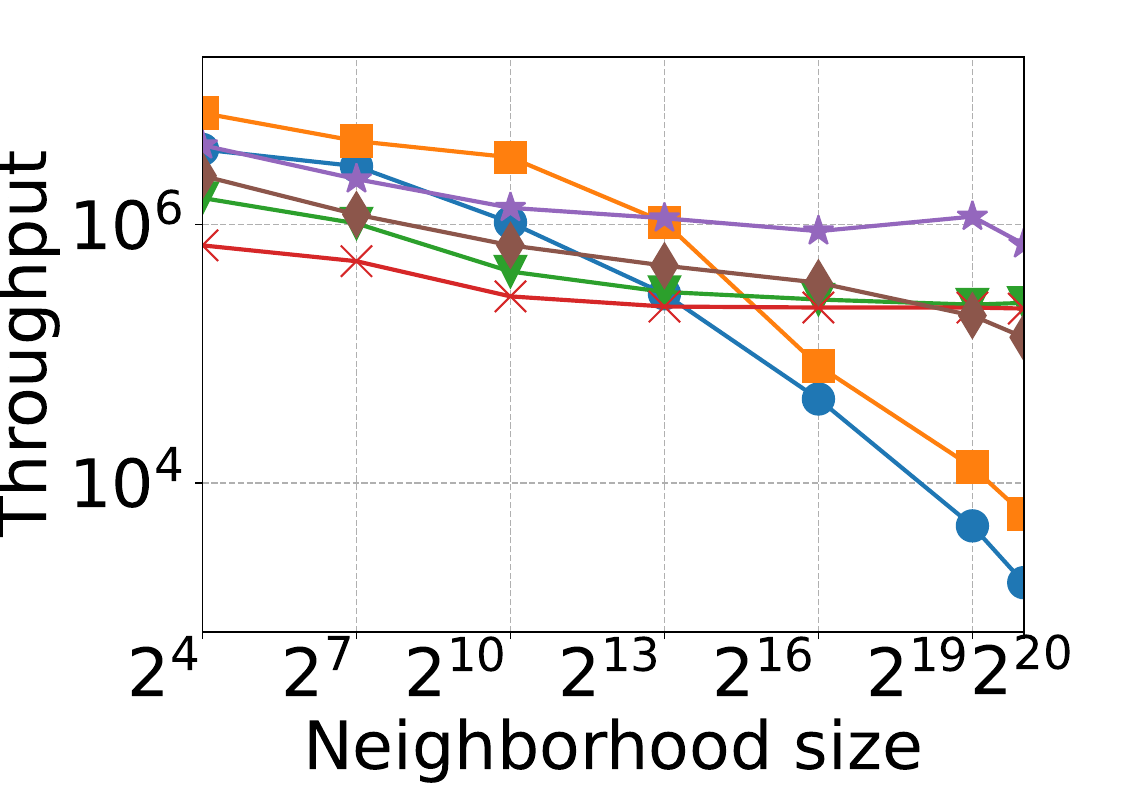}
		\caption{Vary $|N(u)|$ when $|B| = 256$.}
		\label{fig:insert_vary_neighbor_set_size}
	\end{subfigure}
        \begin{subfigure}[t]{0.60\textwidth}
		\centering
		\includegraphics[width=\textwidth]{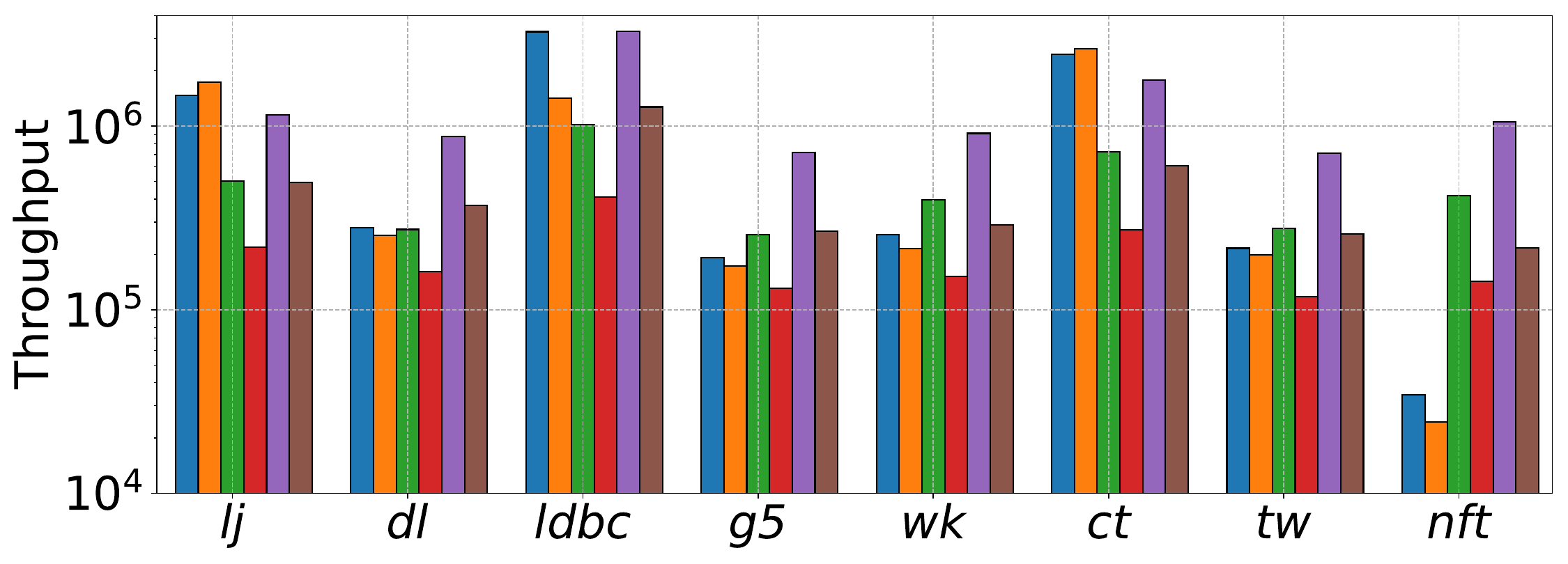}
		\caption{Vary real-world graphs when $|B| = 256$.}
		\label{fig:insert_vary_graphs}
        
	\end{subfigure}
    
	\caption{Efficiency of \textsc{InsEdge}.}
	\label{fig:insert}
\end{figure}

\vspace{2pt}
\noindent\textbf{\textsc{InsEdge}.} As shown in Figure \ref{fig:primitive_opeartions}, insert efficiency is closely related to search efficiency. Figures \ref{fig:insert_vary_block_size} and \ref{fig:insert_vary_neighbor_set_size} show that Teseo performs the best among segmented neighbor indexes. Although O-Aspen is faster than Teseo for search efficiency, it is slower for inserts due to the overhead of CoW. In Figure \ref{fig:insert_vary_block_size}, the performance of both Teseo and Aspen first increases and then decreases slightly because increasing block size reduces the time to locate the target block but increases the time required to copy data when inserting an element in the block.

While AdjLst performs the best for search, it is slow for inserts because it may need to move a large number of elements. Conversely, LiveGraph is slow for inserts because its search is slow. \sun{Specifically, LiveGraph checks whether the element to be inserted already exists, requiring a search operation. If the element exists, it updates the timestamp and appends a new version; otherwise, it directly appends the new version. However, since the neighbors are unsorted, LiveGraph must iterate through the entire neighbor set, leading to significant overhead. Although LiveGraph uses a bloom filter to reduce searches for non-existent edges, it struggles with existing edges, and false positives arise due to the filter’s limited memory space.} Figure \ref{fig:insert_vary_graphs} presents the results on the real-world graphs. LiveGraph and AdjLst outperform the counterparts on \emph{lj} and \emph{ct} because the two graphs are very sparse and have no vertices with high degree. Teseo generally works very well on all graphs.


\begin{figure}[h!]
	\setlength{\abovecaptionskip}{0pt}
	\setlength{\belowcaptionskip}{0pt}
		\captionsetup[subfigure]{aboveskip=0pt,belowskip=0pt}
	\centering
    \includegraphics[width=0.60\textwidth]{img/exp_figs/figure1/legend.pdf}\\
	\begin{subfigure}[t]{0.30\textwidth}
		\centering
		\includegraphics[width=\textwidth]{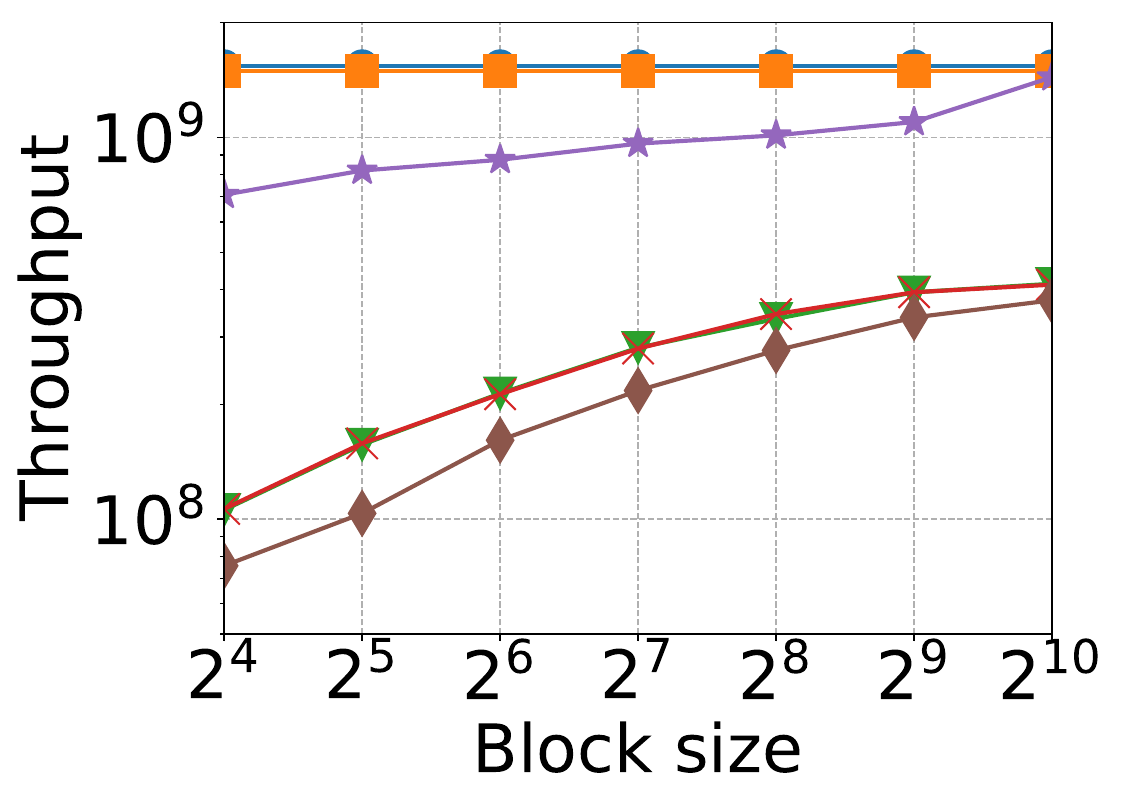}
		\caption{Vary $|B|$ when $|N(u)| = 2 ^ {20}$.}
		\label{fig:scan_vary_block_size}
	\end{subfigure}
        \begin{subfigure}[t]{0.30\textwidth}
		\centering
		\includegraphics[width=\textwidth]{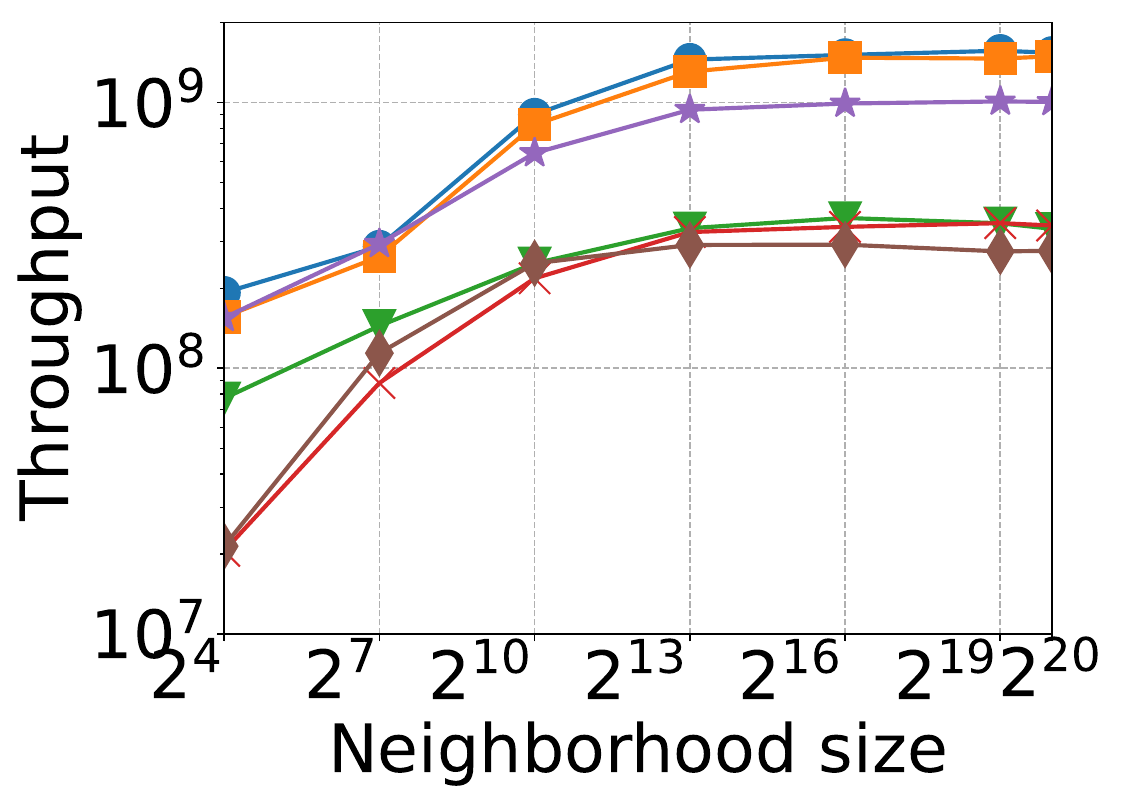}
		\caption{Vary $|N(u)|$ when $|B| = 256$.}
		\label{fig:scan_vary_neighbor_set_size}
	\end{subfigure}
        \begin{subfigure}[t]{0.60\textwidth}
		\centering
		\includegraphics[width=\textwidth]{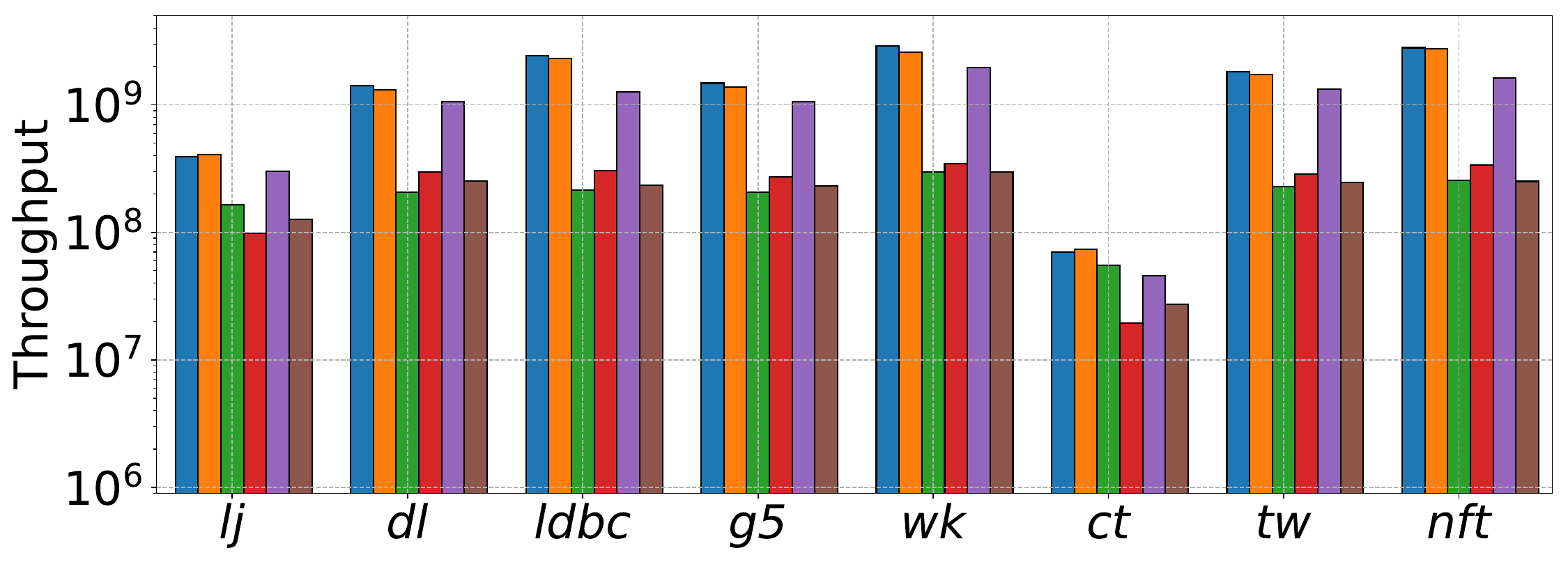}
		\caption{Vary real-world graphs when $|B| = 256$.}
		\label{fig:scan_vary_realworld_graphs}
	\end{subfigure}
	\caption{Efficiency of \textsc{ScanNbr}.}
	\label{fig:scan}
\end{figure}

\noindent\textbf{\textsc{ScanNbr}.} Figure \ref{fig:scan} presents the results on scan efficiency. LiveGraph and AdjLst perform the best due to their continuous storage. Among segmented methods, Teseo runs up to 10x faster than its counterparts due to its reduced branch misses and instruction overhead in PMA compared to the skip list and PAM. Notably, Teseo's performance is close to that of LiveGraph and AdjLst when $|B| = 1024$. Figure \ref{fig:scan_vary_realworld_graphs} illustrates the results on real-world graphs. Consistent with Figures \ref{fig:scan_vary_block_size} and \ref{fig:scan_vary_neighbor_set_size}, AdjLst and LiveGraph perform the best, with Teseo being the best among segmented methods. Specifically, Teseo achieves 52-77\% of AdjLst's performance.

\small
\begin{table}[ht]
\caption{\sun{Hardware metrics on \textsc{ScanNbr} with $|N(v)| = 2^{20}$ and $|B| = 256$. Lg, Ap, Ts, and Sl represent LiveGraph, Aspen, Teseo, and Sortledton, respectively.}}
\begin{tabular}{|c|cc|ccc|}
\hline
            & \multicolumn{2}{c|}{\textbf{\begin{tabular}[c]{@{}c@{}}Continuous Neighbor Index\end{tabular}}} & \multicolumn{3}{c|}{\textbf{\begin{tabular}[c]{@{}c@{}}Segmented Neighbor Index\end{tabular}}}             \\ \hline
\textbf{Metric} &
  \multicolumn{1}{c|}{\textbf{AdjLst}} &
  \textbf{Lg} &
  \multicolumn{1}{c|}{\textbf{Ap}} &
  \multicolumn{1}{c|}{\textbf{Ts}} &
  \textbf{Sl} \\ \hline
L1 miss     & \multicolumn{1}{c|}{2.57e8}           & 2.53e8          & \multicolumn{1}{c|}{2.38e8} & \multicolumn{1}{c|}{2.65e8} & 5.48e8 \\ \hline
L2 miss     & \multicolumn{1}{c|}{1.35e8}           & 2.38e8          & \multicolumn{1}{c|}{1.82e8} & \multicolumn{1}{c|}{2.15e8} & 2.86e8 \\ \hline
LLC miss    & \multicolumn{1}{c|}{1.34e8}           & 1.34e8          & \multicolumn{1}{c|}{1.84e8} & \multicolumn{1}{c|}{2.15e8} & 2.38e8 \\ \hline
DTLB miss   & \multicolumn{1}{c|}{1.19e4}           & 1.20e4          & \multicolumn{1}{c|}{1.94e6} & \multicolumn{1}{c|}{4.20e6} & 4.32e6 \\ \hline
Branch miss & \multicolumn{1}{c|}{1.20e3}           & 2.23e3                 & \multicolumn{1}{c|}{7.90e6} & \multicolumn{1}{c|}{1.50e3} & 6.24e6 \\ \hline
\begin{tabular}[c]{@{}c@{}}Instruction \\ Number\end{tabular} &
  \multicolumn{1}{c|}{1.07e9} &
  1.61e9 &
  \multicolumn{1}{c|}{7.57e9} &
  \multicolumn{1}{c|}{1.14e9} &
   1.18e10 \\ \hline
\end{tabular}%
\label{tab:hardware performance}
\end{table}

Table \ref{tab:hardware performance} presents hardware utilization for competing methods on \textsc{ScanNbr}. We specifically measure cache misses (L1, L2, LLC, and DTLB), branch mispredictions, and number of instructions. Segmented neighbor indexes incur more cache misses than continuous methods due to their non-contiguous memory layout. Particularly, benefiting from continuous storage, AdjLst and LiveGraph have much fewer DTLB misses than counterparts. Additionally, the number of instructions are higher for segmented methods, reflecting the complexity of managing multiple segments during traversal. Branch mispredictions are also more frequent in segmented indexes, leading to performance penalties from pipeline stalls. These factors together result in higher overhead for segmented neighbor indexes during scanning.

\subsubsection{Evaluation on Graph Analytical Queries}
\small
\begin{table}[h!]
\captionsetup{skip=0pt}
\setlength{\abovecaptionskip}{0pt}
\setlength{\belowcaptionskip}{0pt}
\caption{Performance of competing methods on graph analytical queries (measured in seconds). Lg, Ap, Ts and Sl denote LiveGraph, Aspen, Teseo and Sortledton, respectively.}
\label{tab:performance_comparison}


\centering
\begin{tabular}{|c|ccc|ccccc|}

\hline
 &
  \multicolumn{3}{c|}{\textbf{\begin{tabular}[c]{@{}c@{}}Continuous Neighbor Index\end{tabular}}} &
  \multicolumn{5}{c|}{\textbf{\begin{tabular}[c]{@{}c@{}}Segmented Neighbor Index\end{tabular}}} \\ \hline
\multirow{2}{*}{\textbf{Dataset}} &
  \multicolumn{1}{c|}{\multirow{2}{*}{\textbf{CSR}}} &
  \multicolumn{1}{c|}{\multirow{2}{*}{\textbf{AdjLst}}} &
  \multicolumn{1}{c|}{\multirow{2}{*}{\textbf{Lg}}} &
  \multicolumn{2}{c|}{\textbf{Ap}} &
  \multicolumn{1}{c|}{\multirow{2}{*}{\textbf{Ts}}} &
  \multicolumn{2}{c|}{\textbf{Sl}} \\ \cline{5-6} \cline{8-9} 
 &
  \multicolumn{1}{c|}{} &
  \multicolumn{1}{c|}{} &
   &
  \multicolumn{1}{c|}{w} &
  \multicolumn{1}{c|}{wo} &
  \multicolumn{1}{c|}{} &
  \multicolumn{1}{c|}{w} &
  wo \\ \hline
\textbf{PR} &
   &
   &
   &
   &
   &
   &
   &
   \\ \hline
\emph{lj} &
  \multicolumn{1}{c|}{\textbf{4.66}} &
  \multicolumn{1}{c|}{9.00} &
  11.88 &
  \multicolumn{1}{c|}{13.72} &
  \multicolumn{1}{c|}{24.61} &
  \multicolumn{1}{c|}{13.19} &
  \multicolumn{1}{c|}{\textbf{13.06}} &
  28.41 \\ \hline
\emph{dl} &
  \multicolumn{1}{c|}{\textbf{1.39}} &
  \multicolumn{1}{c|}{1.43} &
  1.92 &
  \multicolumn{1}{c|}{5.03} &
  \multicolumn{1}{c|}{5.22} &
  \multicolumn{1}{c|}{\textbf{2.01}} &
  \multicolumn{1}{c|}{5.53} &
  5.24 \\ \hline
\emph{ldbc} &
  \multicolumn{1}{c|}{\textbf{20.57}} &
  \multicolumn{1}{c|}{28.79} &
  34.19 &
  \multicolumn{1}{c|}{\textbf{44.28}} &
  \multicolumn{1}{c|}{113.36} &
  \multicolumn{1}{c|}{90.68} &
  \multicolumn{1}{c|}{48.09} &
  183.19 \\ \hline
\emph{g5} &
  \multicolumn{1}{c|}{\textbf{20.98}} &
  \multicolumn{1}{c|}{29.20} &
  55.73 &
  \multicolumn{1}{c|}{43.19} &
  \multicolumn{1}{c|}{80.35} &
  \multicolumn{1}{c|}{\textbf{38.51}} &
  \multicolumn{1}{c|}{48.30} &
  77.46 \\ \hline
\emph{wk} &
  \multicolumn{1}{c|}{\textbf{6.64}} &
  \multicolumn{1}{c|}{10.45} &
  14.67 &
  \multicolumn{1}{c|}{20.73} &
  \multicolumn{1}{c|}{53.06} &
  \multicolumn{1}{c|}{33.11} &
  \multicolumn{1}{c|}{\textbf{19.30}} &
  41.56 \\ \hline
\emph{ct} &
  \multicolumn{1}{c|}{\textbf{1.88}} &
  \multicolumn{1}{c|}{6.58} &
  7.25 &
  \multicolumn{1}{c|}{7.96} &
  \multicolumn{1}{c|}{15.84} &
  \multicolumn{1}{c|}{11.02} &
  \multicolumn{1}{c|}{\textbf{7.71}} &
  21.78 \\ \hline
\emph{tw} &
  \multicolumn{1}{c|}{\textbf{24.87}} &
  \multicolumn{1}{c|}{39.77} &
  43.89 &
  \multicolumn{1}{c|}{60.33} &
  \multicolumn{1}{c|}{105.96} &
  \multicolumn{1}{c|}{\textbf{55.58}} &
  \multicolumn{1}{c|}{65.24} &
  132.47 \\ \hline
\emph{nft} &
  \multicolumn{1}{c|}{\textbf{10.60}} &
  \multicolumn{1}{c|}{19.00} &
  24.52 &
  \multicolumn{1}{c|}{34.13} &
  \multicolumn{1}{c|}{103.19} &
  \multicolumn{1}{c|}{72.09} &
  \multicolumn{1}{c|}{\textbf{34.05}} &
  178.13 \\ \hline
\textbf{TC} &
   &
   &
   &
   &
   &
   &
   &
   \\ \hline
\emph{lj} &
  \multicolumn{1}{c|}{\textbf{15.36}} &
  \multicolumn{1}{c|}{23.57} &
  / &
  \multicolumn{1}{c|}{30.30} &
  \multicolumn{1}{c|}{52.95} &
  \multicolumn{1}{c|}{\textbf{29.95}} &
  \multicolumn{1}{c|}{27.73} &
  31.79 \\ \hline
\emph{dl} &
  \multicolumn{1}{c|}{\textbf{452.36}} &
  \multicolumn{1}{c|}{467.72} &
  / &
  \multicolumn{1}{c|}{620.72} &
  \multicolumn{1}{c|}{620.19} &
  \multicolumn{1}{c|}{\textbf{501.16}} &
  \multicolumn{1}{c|}{626.52} &
  571.12 \\ \hline
\emph{ldbc} &
  \multicolumn{1}{c|}{\textbf{89.24}} &
  \multicolumn{1}{c|}{115.70} &
  / &
  \multicolumn{1}{c|}{1404.68} &
  \multicolumn{1}{c|}{1750.89} &
  \multicolumn{1}{c|}{\textbf{142.19}} &
  \multicolumn{1}{c|}{172.66} &
  240.49 \\ \hline
\emph{g5} &
  \multicolumn{1}{c|}{\textbf{1042.68}} &
  \multicolumn{1}{c|}{1220.60} &
  / &
  \multicolumn{1}{c|}{1774.84} &
  \multicolumn{1}{c|}{1971.00} &
  \multicolumn{1}{c|}{\textbf{1395.10}} &
  \multicolumn{1}{c|}{1890.74} &
  1605.42 \\ \hline
\emph{wk} &
  \multicolumn{1}{c|}{\textbf{7.63}} &
  \multicolumn{1}{c|}{13.58} &
  / &
  \multicolumn{1}{c|}{42.43} &
  \multicolumn{1}{c|}{77.81} &
  \multicolumn{1}{c|}{\textbf{17.10}} &
  \multicolumn{1}{c|}{17.05} &
  18.91 \\ \hline
\emph{ct} &
  \multicolumn{1}{c|}{\textbf{3.17}} &
  \multicolumn{1}{c|}{4.74} &
  / &
  \multicolumn{1}{c|}{6.13} &
  \multicolumn{1}{c|}{16.78} &
  \multicolumn{1}{c|}{\textbf{5.91}} &
  \multicolumn{1}{c|}{5.31} &
  6.58 \\ \hline
\emph{tw} &
  \multicolumn{1}{c|}{\textbf{801.14}} &
  \multicolumn{1}{c|}{860.73} &
  / &
  \multicolumn{1}{c|}{1272.99} &
  \multicolumn{1}{c|}{1467.60} &
  \multicolumn{1}{c|}{\textbf{958.66}} &
  \multicolumn{1}{c|}{1295.46} &
  1092.51 \\ \hline
\emph{nft} &
  \multicolumn{1}{c|}{\textbf{218.23}} &
  \multicolumn{1}{c|}{346.04} &
  / &
  \multicolumn{1}{c|}{857.27} &
  \multicolumn{1}{c|}{1011.24} &
  \multicolumn{1}{c|}{\textbf{382.77}} &
  \multicolumn{1}{c|}{486.28} &
  482.18 \\ \hline
\end{tabular}

\end{table}


Table \ref{tab:performance_comparison} reports the results for PR and TC, while the results for BFS, SSSP, and WCC are included in the full version. PR requires fast scan efficiency, whereas TC relies on efficient set intersection, necessitating both efficient search and scan in sorted order. Consequently, LiveGraph cannot support TC. All segmented methods set $|B|$ to 256. Aspen-w enables the flatten optimization, and Sortledton-w enables adaptive indexing with the threshold set to 256.

First, continuous methods outperform segmented methods. CSR runs faster than AdjLst with all neighbor sets in a single array. Both AdjLst and LiveGraph store neighbor sets in separate arrays, but AdjLst runs 1.0-1.9x faster because LiveGraph scans from the end of the array, preventing automatic vectorization by the compiler using SIMD. Although manual vectorization is possible, it introduces significant engineering challenges. Second, Teseo generally performs the best among segmented methods without optimizations because it stores blocks of $N(u)$ continuously. Third, enabling adaptive indexing achieves a speedup of 1.0-5.2x. Enabling flatten optimization improves performance by 1.0-3.0x. Nevertheless, CSR achieves a significant speedup, ranging from 1.2-53.7x, over the best segmented methods in each case.


\begin{table}[h!]
\caption{\sun{Hardware metrics of competing methods on PR and TC (measured in billions). Lg, Ap, Ts and Sl denote LiveGraph, Aspen, Teseo and Sortledton, respectively.}}
\small
\begin{tabular}{|ccccc|ccccc|}
\hline
\multicolumn{2}{|c|}{}                                                                                             & \multicolumn{3}{c|}{\textbf{Continuous Neighbor Index}}                                                                                   & \multicolumn{5}{c|}{\textbf{Segmented Neighbor Index}}                                                                                                                 \\ \hline
\multicolumn{1}{|c|}{\multirow{2}{*}{\textbf{Metric}}}    & \multicolumn{1}{c|}{\multirow{2}{*}{\textbf{Dataset}}} & \multicolumn{1}{c|}{\multirow{2}{*}{\textbf{CSR}}} & \multicolumn{1}{c|}{\multirow{2}{*}{\textbf{AdjLst}}} & \multirow{2}{*}{\textbf{Lg}} & \multicolumn{2}{c|}{\textbf{Ap}}                                   & \multicolumn{1}{c|}{\multirow{2}{*}{\textbf{Ts}}} & \multicolumn{2}{c|}{\textbf{Sl}}              \\ \cline{6-7} \cline{9-10} 
\multicolumn{1}{|c|}{}                                    & \multicolumn{1}{c|}{}                                  & \multicolumn{1}{c|}{}                              & \multicolumn{1}{c|}{}                                 &                              & \multicolumn{1}{c|}{\textbf{w}} & \multicolumn{1}{c|}{\textbf{wo}} & \multicolumn{1}{c|}{}                             & \multicolumn{1}{c|}{\textbf{w}} & \textbf{wo} \\ \hline
\multicolumn{1}{|c|}{\textbf{PR}}                         &                                                        &                                                    &                                                       &                              &                                 &                                  &                                                   &                                 &             \\ \hline
\multicolumn{1}{|c|}{\multirow{2}{*}{L1 miss}}            & \multicolumn{1}{c|}{\emph{lj}}            & \multicolumn{1}{c|}{0.83}                          & \multicolumn{1}{c|}{0.86}                             & 0.97                         & \multicolumn{1}{c|}{0.89}       & \multicolumn{1}{c|}{0.95}        & \multicolumn{1}{c|}{0.92}                         & \multicolumn{1}{c|}{0.85}       & 0.93        \\ \cline{2-10} 
\multicolumn{1}{|c|}{}                                    & \multicolumn{1}{c|}{\emph{g5}}            & \multicolumn{1}{c|}{5.13}                          & \multicolumn{1}{c|}{5.23}                             & 5.47                         & \multicolumn{1}{c|}{5.28}       & \multicolumn{1}{c|}{5.46}        & \multicolumn{1}{c|}{5.37}                         & \multicolumn{1}{c|}{5.24}       & 5.37        \\ \hline
\multicolumn{1}{|c|}{\multirow{2}{*}{L2 miss}}            & \multicolumn{1}{c|}{\emph{lj}}            & \multicolumn{1}{c|}{0.80}                          & \multicolumn{1}{c|}{0.96}                             & 1.10                         & \multicolumn{1}{c|}{0.96}       & \multicolumn{1}{c|}{1.07}        & \multicolumn{1}{c|}{1.45}                         & \multicolumn{1}{c|}{0.94}       & 1.35        \\ \cline{2-10} 
\multicolumn{1}{|c|}{}                                    & \multicolumn{1}{c|}{\emph{g5}}            & \multicolumn{1}{c|}{4.22}                          & \multicolumn{1}{c|}{4.59}                             & 4.86                         & \multicolumn{1}{c|}{4.78}       & \multicolumn{1}{c|}{4.97}        & \multicolumn{1}{c|}{5.76}                         & \multicolumn{1}{c|}{4.96}       & 5.68        \\ \hline
\multicolumn{1}{|c|}{\multirow{2}{*}{LLC miss}}           & \multicolumn{1}{c|}{\emph{lj}}            & \multicolumn{1}{c|}{0.16}                          & \multicolumn{1}{c|}{0.32}                             & 0.46                         & \multicolumn{1}{c|}{0.30}       & \multicolumn{1}{c|}{0.43}        & \multicolumn{1}{c|}{0.81}                         & \multicolumn{1}{c|}{0.30}       & 0.70        \\ \cline{2-10} 
\multicolumn{1}{|c|}{}                                    & \multicolumn{1}{c|}{\emph{g5}}            & \multicolumn{1}{c|}{0.93}                          & \multicolumn{1}{c|}{1.26}                             & 1.53                         & \multicolumn{1}{c|}{1.27}       & \multicolumn{1}{c|}{1.53}        & \multicolumn{1}{c|}{2.38}                         & \multicolumn{1}{c|}{1.57}       & 2.28        \\ \hline
\multicolumn{1}{|c|}{\multirow{2}{*}{DTLB miss}}          & \multicolumn{1}{c|}{\emph{lj}}            & \multicolumn{1}{c|}{8.20}                          & \multicolumn{1}{c|}{8.74}                             & 8.91                         & \multicolumn{1}{c|}{8.12}       & \multicolumn{1}{c|}{8.29}        & \multicolumn{1}{c|}{10.96}                        & \multicolumn{1}{c|}{8.59}       & 9.89        \\ \cline{2-10} 
\multicolumn{1}{|c|}{}                                    & \multicolumn{1}{c|}{\emph{g5}}            & \multicolumn{1}{c|}{35.48}                         & \multicolumn{1}{c|}{38.34}                            & 37.34                        & \multicolumn{1}{c|}{35.96}      & \multicolumn{1}{c|}{37.57}       & \multicolumn{1}{c|}{39.59}                        & \multicolumn{1}{c|}{36.34}      & 38.15       \\ \hline
\multicolumn{1}{|c|}{\multirow{2}{*}{Branch miss}}        & \multicolumn{1}{c|}{\emph{lj}}            & \multicolumn{1}{c|}{0.05}                          & \multicolumn{1}{c|}{0.05}                             & 0.05                         & \multicolumn{1}{c|}{0.05}       & \multicolumn{1}{c|}{0.33}        & \multicolumn{1}{c|}{0.05}                         & \multicolumn{1}{c|}{0.05}       & 0.05        \\ \cline{2-10} 
\multicolumn{1}{|c|}{}                                    & \multicolumn{1}{c|}{\emph{g5}}            & \multicolumn{1}{c|}{0.09}                          & \multicolumn{1}{c|}{0.09}                             & 0.09                         & \multicolumn{1}{c|}{0.12}       & \multicolumn{1}{c|}{0.65}        & \multicolumn{1}{c|}{0.10}                         & \multicolumn{1}{c|}{0.09}       & 0.09        \\ \hline
\multicolumn{1}{|c|}{\multirow{2}{*}{\begin{tabular}[c]{@{}c@{}}Instruction \\ number\end{tabular}}} & \multicolumn{1}{c|}{\emph{lj}}            & \multicolumn{1}{c|}{8.27}                          & \multicolumn{1}{c|}{9.28}                             & 10.09                        & \multicolumn{1}{c|}{21.36}      & \multicolumn{1}{c|}{52.67}       & \multicolumn{1}{c|}{9.82}                         & \multicolumn{1}{c|}{10.05}      & 14.43       \\ \cline{2-10} 
\multicolumn{1}{|c|}{}                                    & \multicolumn{1}{c|}{\emph{g5}}            & \multicolumn{1}{c|}{44.25}                         & \multicolumn{1}{c|}{46.11}                            & 51.22                        & \multicolumn{1}{c|}{90.50}      & \multicolumn{1}{c|}{149.93}      & \multicolumn{1}{c|}{47.36}                        & \multicolumn{1}{c|}{69.83}      & 77.49       \\ \hline
\multicolumn{1}{|c|}{\textbf{TC}}                         &                                                        &                                                    &                                                       &                              &                                 &                                  &                                                   &                                 &             \\ \hline
\multicolumn{1}{|c|}{\multirow{2}{*}{L1 miss}}            & \multicolumn{1}{c|}{\emph{lj}}            & \multicolumn{1}{c|}{0.07}                          & \multicolumn{1}{c|}{0.14}                             & /                            & \multicolumn{1}{c|}{0.19}       & \multicolumn{1}{c|}{0.70}        & \multicolumn{1}{c|}{0.23}                         & \multicolumn{1}{c|}{0.16}       & 0.20        \\ \cline{2-10} 
\multicolumn{1}{|c|}{}                                    & \multicolumn{1}{c|}{\emph{g5}}            & \multicolumn{1}{c|}{2.35}                          & \multicolumn{1}{c|}{3.00}                             & /                            & \multicolumn{1}{c|}{24.44}      & \multicolumn{1}{c|}{19.93}       & \multicolumn{1}{c|}{7.78}                         & \multicolumn{1}{c|}{8.52}       & 8.61        \\ \hline
\multicolumn{1}{|c|}{\multirow{2}{*}{L2 miss}}            & \multicolumn{1}{c|}{\emph{lj}}            & \multicolumn{1}{c|}{0.93}                          & \multicolumn{1}{c|}{1.01}                             & /                            & \multicolumn{1}{c|}{1.28}       & \multicolumn{1}{c|}{1.86}        & \multicolumn{1}{c|}{1.37}                         & \multicolumn{1}{c|}{1.15}       & 1.27        \\ \cline{2-10} 
\multicolumn{1}{|c|}{}                                    & \multicolumn{1}{c|}{\emph{g5}}            & \multicolumn{1}{c|}{53.86}                         & \multicolumn{1}{c|}{56.33}                            & /                            & \multicolumn{1}{c|}{93.33}      & \multicolumn{1}{c|}{100.53}      & \multicolumn{1}{c|}{109.50}                       & \multicolumn{1}{c|}{111.48}     & 112.92      \\ \hline
\multicolumn{1}{|c|}{\multirow{2}{*}{LLC miss}}           & \multicolumn{1}{c|}{\emph{lj}}            & \multicolumn{1}{c|}{0.65}                          & \multicolumn{1}{c|}{0.74}                             & /                            & \multicolumn{1}{c|}{0.83}       & \multicolumn{1}{c|}{1.18}        & \multicolumn{1}{c|}{1.11}                         & \multicolumn{1}{c|}{0.76}       & 1.00        \\ \cline{2-10} 
\multicolumn{1}{|c|}{}                                    & \multicolumn{1}{c|}{\emph{g5}}            & \multicolumn{1}{c|}{31.63}                         & \multicolumn{1}{c|}{31.96}                            & /                            & \multicolumn{1}{c|}{62.87}      & \multicolumn{1}{c|}{66.15}       & \multicolumn{1}{c|}{68.99}                        & \multicolumn{1}{c|}{69.19}      & 71.77       \\ \hline
\multicolumn{1}{|c|}{\multirow{2}{*}{DTLB miss}}          & \multicolumn{1}{c|}{\emph{lj}}            & \multicolumn{1}{c|}{5.76}                          & \multicolumn{1}{c|}{10.67}                            & /                            & \multicolumn{1}{c|}{13.70}      & \multicolumn{1}{c|}{23.67}       & \multicolumn{1}{c|}{14.42}                        & \multicolumn{1}{c|}{12.45}      & 13.60       \\ \cline{2-10} 
\multicolumn{1}{|c|}{}                                    & \multicolumn{1}{c|}{\emph{g5}}            & \multicolumn{1}{c|}{97.34}                         & \multicolumn{1}{c|}{137.09}                           & /                            & \multicolumn{1}{c|}{336.05}     & \multicolumn{1}{c|}{441.16}      & \multicolumn{1}{c|}{211.64}                       & \multicolumn{1}{c|}{380.34}     & 237.88      \\ \hline
\multicolumn{1}{|c|}{\multirow{2}{*}{Branch miss}}        & \multicolumn{1}{c|}{\emph{lj}}            & \multicolumn{1}{c|}{0.85}                          & \multicolumn{1}{c|}{0.85}                             & /                            & \multicolumn{1}{c|}{0.84}       & \multicolumn{1}{c|}{1.27}        & \multicolumn{1}{c|}{0.82}                         & \multicolumn{1}{c|}{0.80}       & 0.79        \\ \cline{2-10} 
\multicolumn{1}{|c|}{}                                    & \multicolumn{1}{c|}{\emph{g5}}            & \multicolumn{1}{c|}{67.98}                         & \multicolumn{1}{c|}{63.01}                            & /                            & \multicolumn{1}{c|}{64.59}      & \multicolumn{1}{c|}{64.73}       & \multicolumn{1}{c|}{65.75}                        & \multicolumn{1}{c|}{63.33}      & 61.34       \\ \hline
\multicolumn{1}{|c|}{\multirow{2}{*}{\begin{tabular}[c]{@{}c@{}}Instruction \\ number\end{tabular}}} & \multicolumn{1}{c|}{\emph{lj}}            & \multicolumn{1}{c|}{31.44}                         & \multicolumn{1}{c|}{31.93}                            & /                            & \multicolumn{1}{c|}{60.00}      & \multicolumn{1}{c|}{75.00}       & \multicolumn{1}{c|}{32.62}                        & \multicolumn{1}{c|}{44.27}      & 56.05       \\ \cline{2-10} 
\multicolumn{1}{|c|}{}                                    & \multicolumn{1}{c|}{\emph{g5}}            & \multicolumn{1}{c|}{6929}                          & \multicolumn{1}{c|}{6932}                             & /                            & \multicolumn{1}{c|}{10778}      & \multicolumn{1}{c|}{10900}       & \multicolumn{1}{c|}{6976}                         & \multicolumn{1}{c|}{12020}      & 12075       \\ \hline
\end{tabular}

\label{tab:hardware pagerank}
\end{table}

\sun{Table \ref{tab:hardware pagerank} presents hardware metrics for PR and TC, including cache misses (L1, L2, LLC, and DTLB), branch mispredictions, and number of instructions. Although segmented neighbor indexes generally share the same big-O time complexity as continuous indexes, they result in higher cache miss rates (L1, L2, LLC), more branch mispredictions, and increased instruction counts, especially with the larger dataset \emph{g5}. Additionally, while both CSR and AdjLst store neighbor sets as continuous arrays, CSR exhibits lower cache miss rates due to its compact memory format. For page rank queries, adaptive indexing mechanisms reduce LLC misses by 57.1\% and 31.1\% on dataset \emph{lj} and \emph{g5}, respectively, while flatten optimizations lower LLC cache misses by  30.2\% and 17.0\%. For triangle counting queries, flatten optimizations lower LLC cache misses by 31.3\% and 49.2\% along with a decrease in DTLB misses by 42.1\% and 23.8\%, contributing to the performance improvements.}

\subsubsection{Summary of Findings} \textbf{Q1. How effective are existing techniques in graph containers?} The vertex index significantly influences operations on small neighbor sets, as well as search and insert efficiency. This aspect has been overlooked in some previous works, leading to biased conclusions about the efficiency of neighbor indexes. The dynamic array performs best, being orders of magnitude faster than its counterparts. We confirm that segmented strategies can significantly accelerate insert operations on large neighbor sets, while also providing considerable scan efficiency. Both search and scan benefit from large blocks, but there is an optimal block size ($|B| = 256$) for insert efficiency in our experiments. CoW is generally slower than in-place updates. Moreover, we reveal that memory layout and vectorization significantly impact scan and search performance.

For additional optimizations, adaptive indexing can improve performance on real-world graphs because most vertices have a small degree. Second, the bloom filter offers limited help for \textsc{SearchEdge} due to two reasons: 1) we still need to locate the neighbor if it exists, and 2) the bloom filter has a high false positive rate on large neighbor sets. Third, flattening optimization can enhance long-running query performance, but it is challenging to automatically determine when to apply this optimization for a given query.

\vspace{2pt}
\noindent\textbf{Q2. Which neighbor index performs the best, and what is the gap between it and CSR on read queries?} The simple sorted dynamic array performs best for search, insert, and scan operations on small neighbor sets ($|N(u)| \leqslant 256$). The neighbor indexes of Teseo and Aspen perform much better than Sortledton, and LiveGraph significantly outperforms it on scan efficiency. Sortledton's results are biased due to two factors: 1) using adaptive indexing improves overall performance, and 2) G2PL significantly simplifies concurrency control. Teseo performs the best for both read and write operations on large neighbor sets. Nevertheless, there is still a significant gap between Teseo and CSR on graph analytic queries.

\subsection{Evaluation of Graph Concurrency Control} \label{sec:evaluation_gcc}

We evaluate graph concurrency control (GCC) from three perspectives: overhead, scalability, and concurrency. All competing methods enable full-fledged GCC. Specifically, AdjLst, LiveGraph, Teseo, and Sortledton use the fine-grained concurrency control strategy with G2PL, while Aspen uses the coarse-grained strategy with CoW. For brevity, we refer to the two strategies as G2PL and CoW.

\subsubsection{Overhead Incurred By GCC}

\begin{figure}[t]
	\setlength{\abovecaptionskip}{0pt}
	\setlength{\belowcaptionskip}{0pt}
		\captionsetup[subfigure]{aboveskip=0pt,belowskip=0pt}
	\centering
    \includegraphics[width=0.70\textwidth]
    {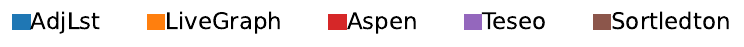}
    \\
    \begin{subfigure}[t]{0.25\textwidth}
    		\centering
    		\includegraphics[width=\textwidth]{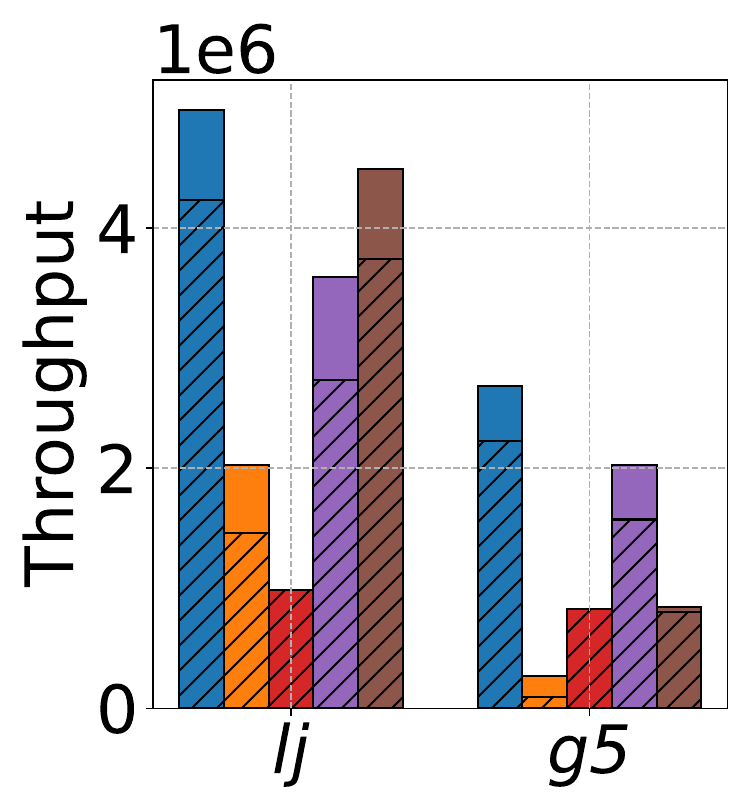}
    		\caption{\textsc{SearchEdge}}
    		\label{fig:slowdown_scan}
    \end{subfigure}
    \begin{subfigure}[t]{0.25\textwidth}
    		\centering
    		\includegraphics[width=\textwidth]{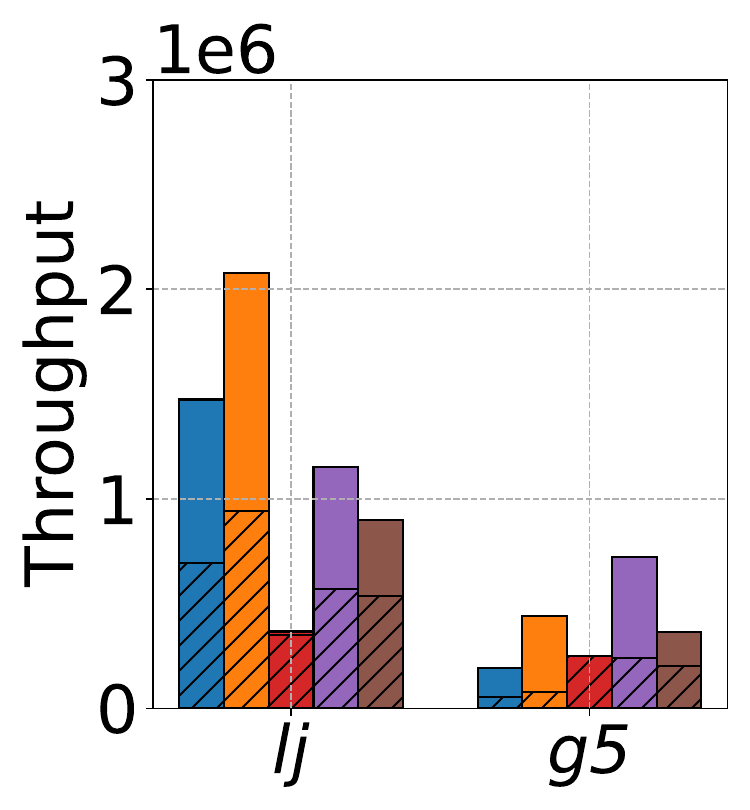}
    		\caption{InsertEdge}
    		\label{fig:slowdown_scan}
    \end{subfigure}
    \begin{subfigure}[t]{0.25\textwidth}
    		\centering
    		\includegraphics[width=\textwidth]{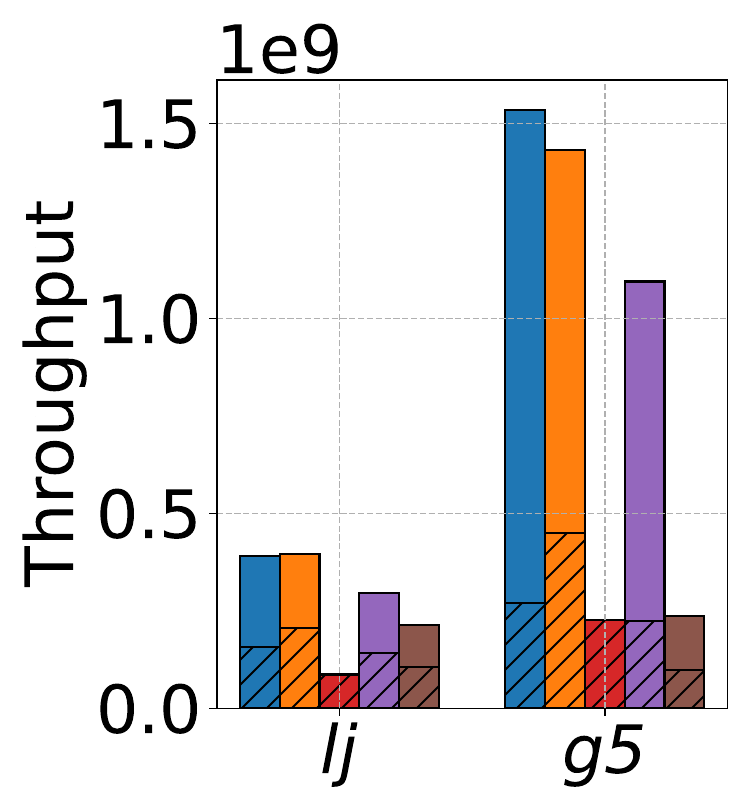}
    		\caption{ScanNbr}
    		\label{fig:slowdown_scan}
    \end{subfigure}
     
    \caption{Unshadowed denotes throughput without GCC, and shadowed denotes throughput with GCC.}
    \label{fig:overhead_of_cc}
\end{figure}

Figure \ref{fig:overhead_of_cc} presents the experimental results. As discussed in Section \ref{sec:discussion}, G2PL causes a 2.0-5.7x slowdown in scan operations, while Aspen’s performance remains unaffected. The slowdown on \emph{g5} is greater than on \emph{lj} because \emph{g5}’s vertices have a larger degree, making the cost of looping over neighbor sets dominate scan operations. We also compare the performance of graph operations with and without locking and find that the overhead of locking is negligible. Therefore, the slowdown is primarily caused by the fine-grained version information. In contrast, CoW incurs no additional overhead, as it employs a single-writer mechanism. As a result, the performance gap between Aspen and these fine-grained methods significantly reduces.

\begin{figure}[h]
	\setlength{\abovecaptionskip}{0pt}
	\setlength{\belowcaptionskip}{0pt}
    \captionsetup[subfigure]{aboveskip=0pt,belowskip=0pt}
	\centering
    \includegraphics[width=0.60\textwidth]{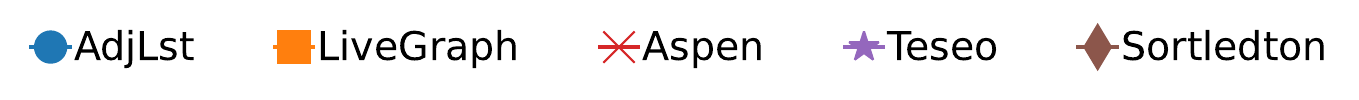}\\
    \begin{subfigure}[t]{0.30\textwidth}
        \centering
        \includegraphics[width=\textwidth]{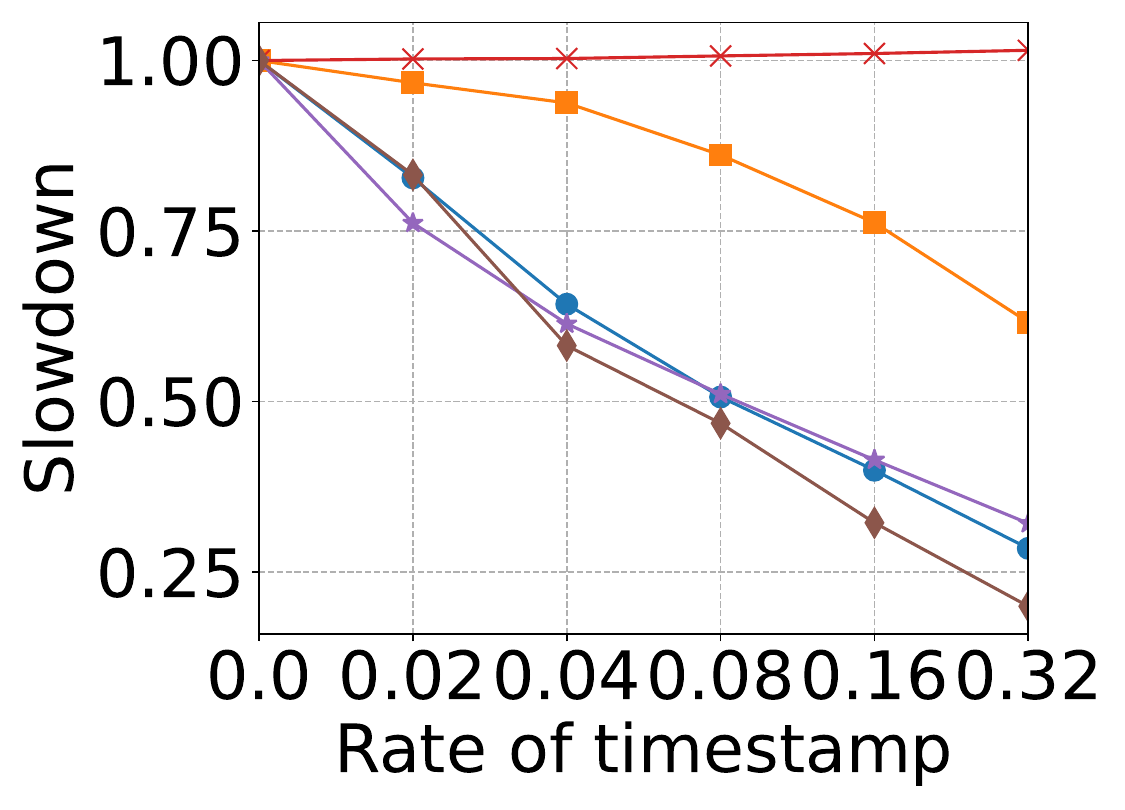}
        \caption{Slowdown of \textsc{ScanNbr}.}
        \label{fig:cc_efficiency_scan}
    \end{subfigure}
    \begin{subfigure}[t]{0.30\textwidth}
        \centering
        \includegraphics[width=\textwidth]{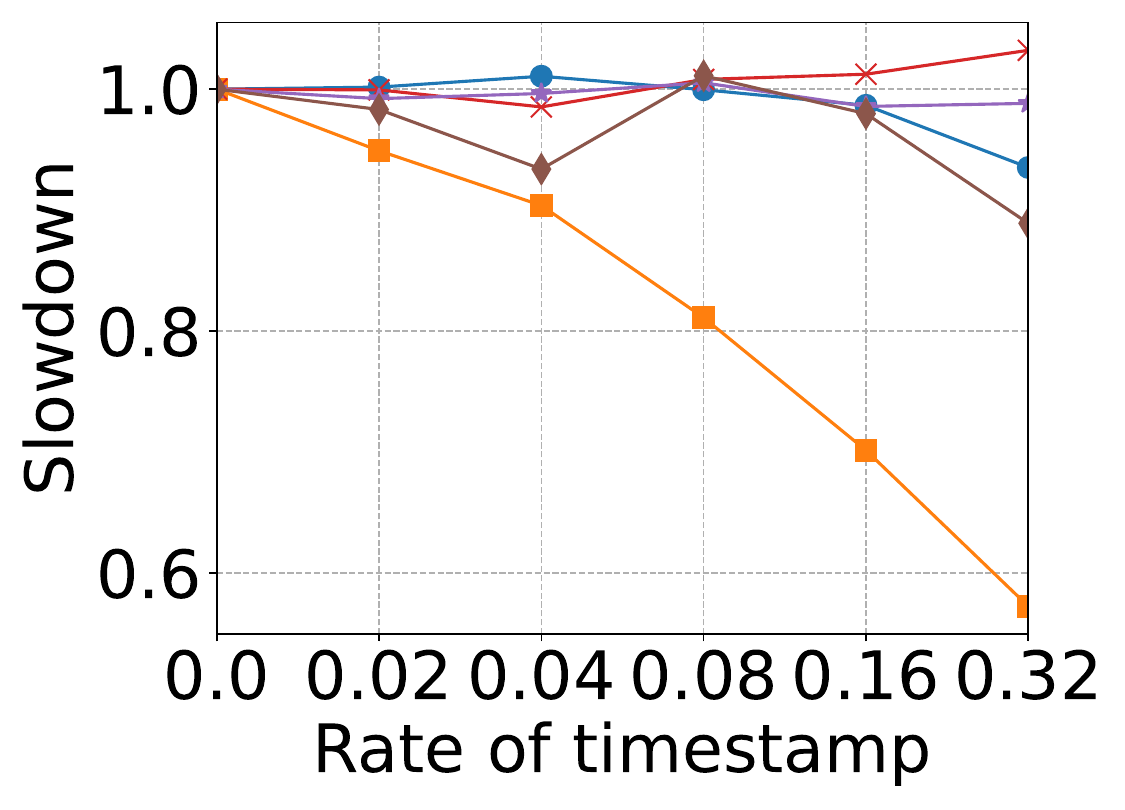}
        \caption{Slowdown of \textsc{SearchEdge}.}
        \label{fig:cc_efficiency_search}
    \end{subfigure} 
    \caption{Impact of varying the percentage of neighbors with more than one versions.}
    \label{fig:overhead_of_cc_varying_versions}
\end{figure}

\subsubsection{Scalability Evaluation} \label{sec:scalability_evaluation}

An element can have multiple versions stored as a version chain or continuous. We examine the impact of multiple versions on search and scan efficiency as follows: given $N(u)$, we randomly pick $x\%$ of vertices $v\in N(u)$ and set three different versions. We vary the ratio from 2\% to 32\%. Figure \ref{fig:overhead_of_cc_varying_versions} presents the results. The scan efficiency of methods using G2PL drops dramatically. The negative impact on AdjLst, Teseo, and Sortledton is greater than on LiveGraph due to the overhead of traversing the list and checking each element’s version. In Figure \ref{fig:cc_efficiency_search}, LiveGraph’s search efficiency drops significantly because it has to scan all versions to find the target, whereas AdjLst, Teseo, and Sortledton generally degrades by less than 10\%. In contrast, multiple versions do not affect Aspen since it maintains versions in coarse granularity. The trend in insert efficiency mirrors that of search, so we omit the results for \textsc{InsEdge}.

\begin{figure}[h]
    \setlength{\abovecaptionskip}{0pt}
    \setlength{\belowcaptionskip}{0pt}
    \captionsetup[subfigure]{aboveskip=0pt,belowskip=0pt}
    \centering
    \includegraphics[width=0.60\textwidth]{img/exp_figs/figure6/legend_line.pdf}\\
    \begin{subfigure}[t]{0.30\textwidth}
        \centering
        \includegraphics[width=\textwidth]{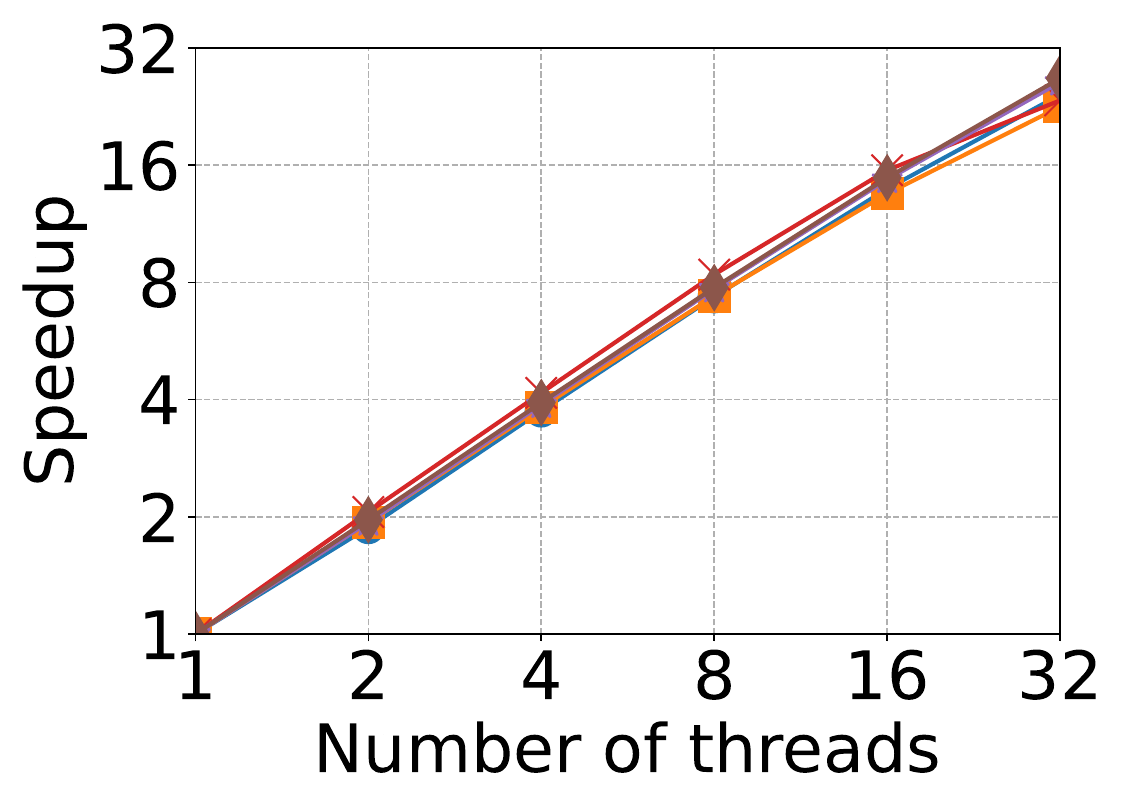}
        \caption{Search operations on \emph{lj}.}
        \label{fig:scalability_search_on_lj}
    \end{subfigure}
    \begin{subfigure}[t]{0.30\textwidth}
        \centering
        \includegraphics[width=\textwidth]{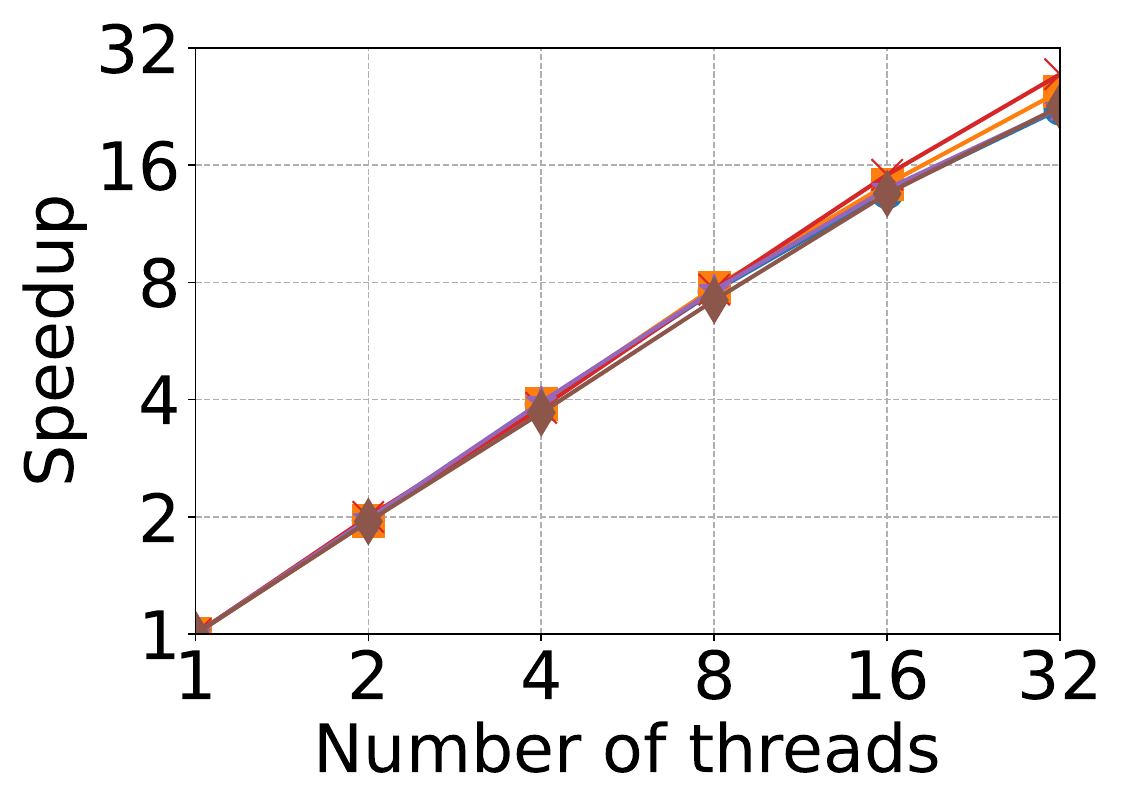}
        \caption{Scan operations on \emph{lj}.}
        \label{fig:scalability_scan_on_lj}
    \end{subfigure}
    \begin{subfigure}[t]{0.30\textwidth}
        \centering
        \includegraphics[width=\textwidth]{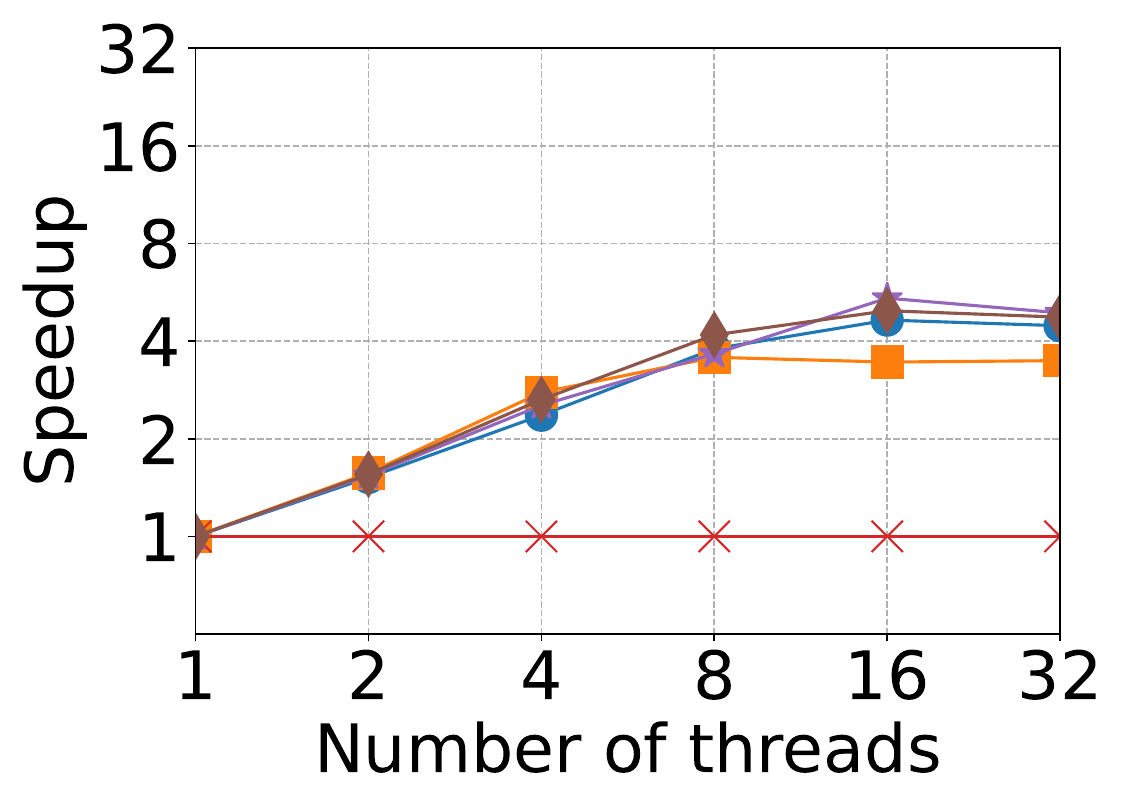}
        \caption{Insert operations on \emph{lj}.}
        \label{fig:scalability_insert_on_lj}
     \end{subfigure}
    \begin{subfigure}[t]{0.30\textwidth}
        \centering
        \includegraphics[width=\textwidth]{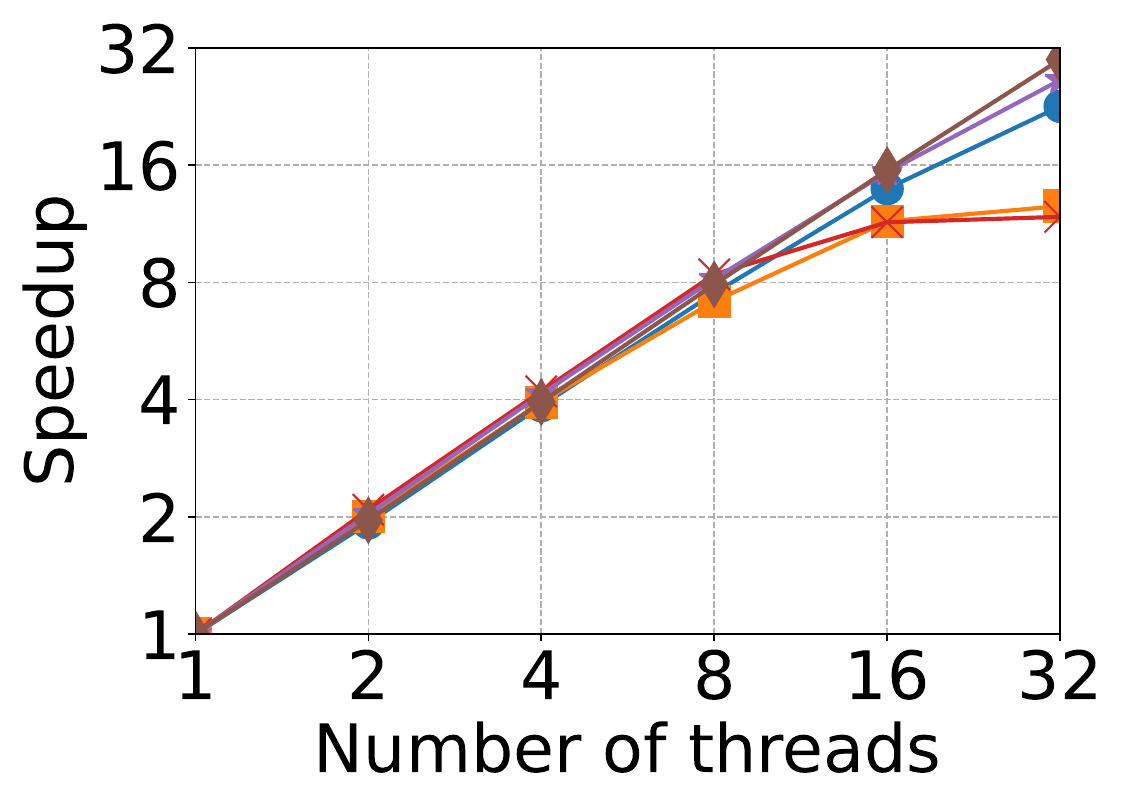}
        \caption{Search operations on \emph{g5}.}
        \label{fig:scalability_search_on_g5}
    \end{subfigure}
    \begin{subfigure}[t]{0.30\textwidth}
        \centering
        \includegraphics[width=\textwidth]{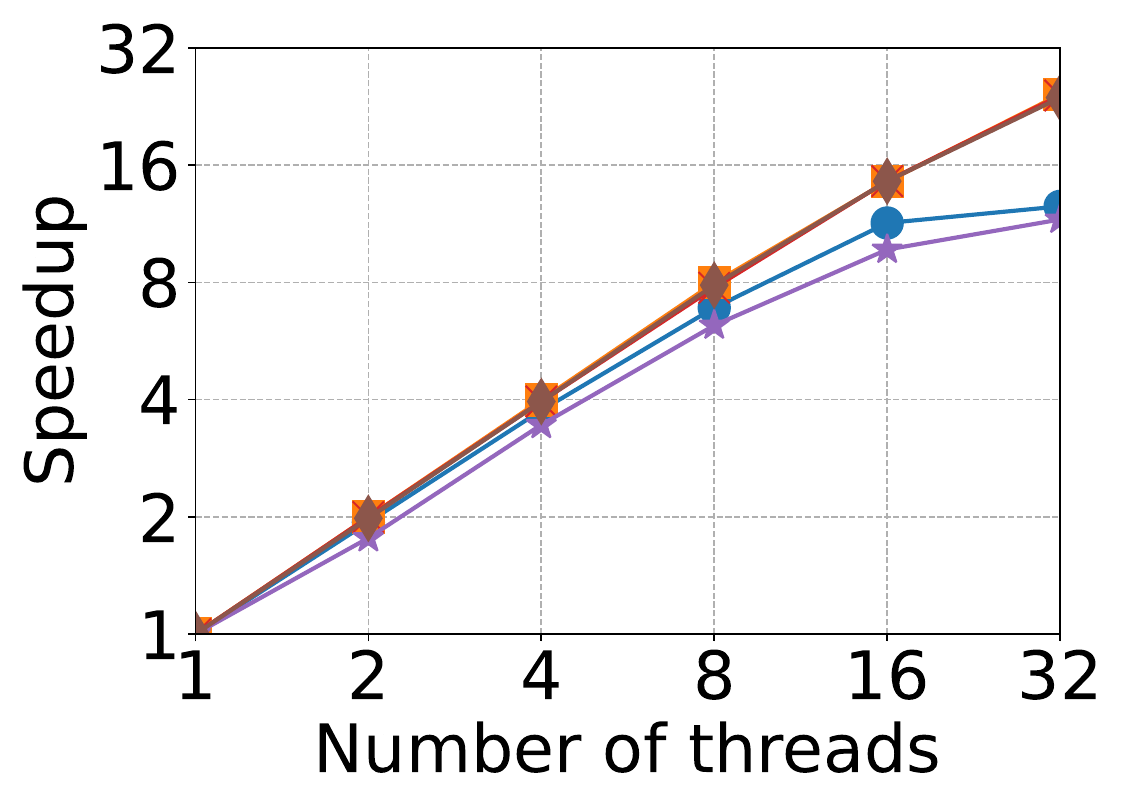}
        \caption{Scan operations on \emph{g5}.}
        \label{fig:scalability_scan_on_g5}
    \end{subfigure}
     \begin{subfigure}[t]{0.30\textwidth}
        \centering
        \includegraphics[width=\textwidth]{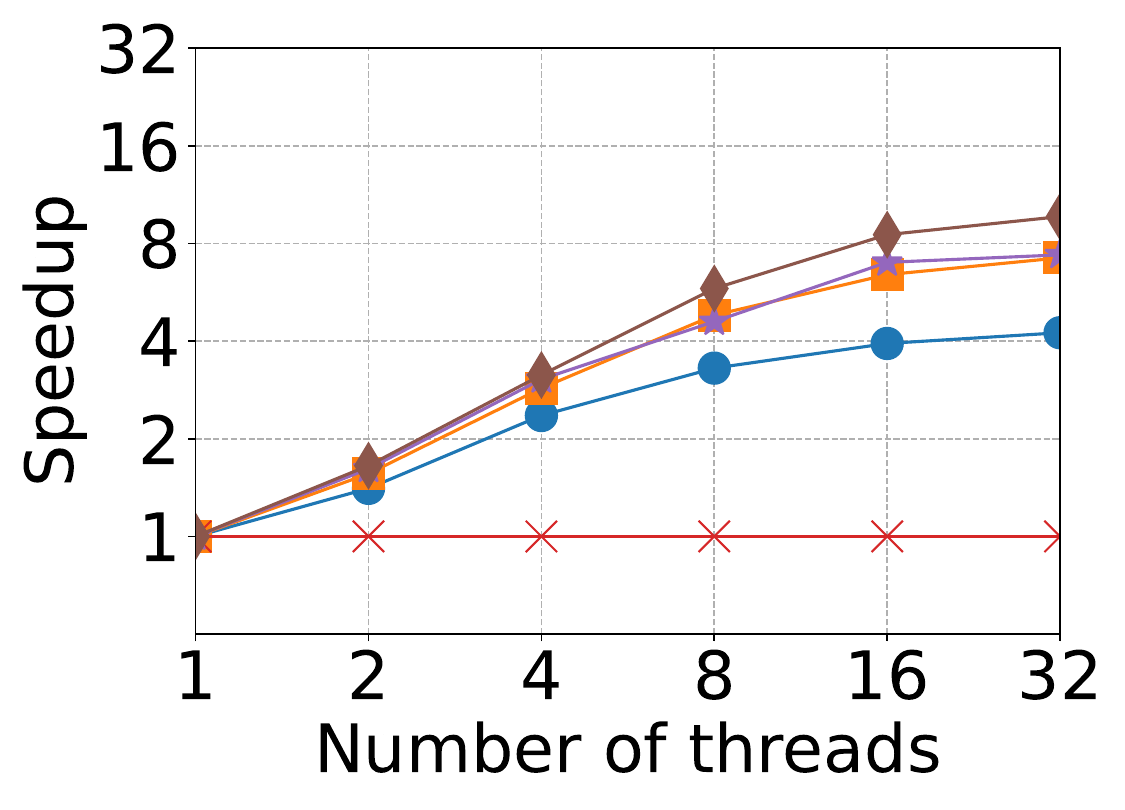}
        \caption{Insert operations on \emph{g5}.}
        \label{fig:scalability_insert_on_g5}
    \end{subfigure}
    
	\caption{Scalability with the number of threads varying.}
	\label{fig:scalability}
\end{figure}

We next evaluate the scalability of competing methods as the number of threads increases to 32, matching the number of physical cores in the testbed. Figure \ref{fig:scalability} presents the scalability results, and Table \ref{tab:performance_under_32_threads} shows the throughput with 32 threads. First, for search and scan efficiency on \emph{lj}, competing methods achieve 22.6-29.7x and 22.3-26.7x speedup, respectively. They fail to scale linearly because \emph{lj} is very sparse, and the irregular data accesses by many threads result in cache contention issues. As in the graph container evaluation, AdjLst significantly outperforms its counterparts on \emph{lj}, as shown in Table \ref{tab:performance_under_32_threads}.

\begin{table}[t]
\captionsetup{skip=0pt}
\setlength{\abovecaptionskip}{0pt}
\setlength{\belowcaptionskip}{0pt}
    \caption{Throughput comparison of competing methods using 32 threads (million edges per second).}
    \label{tab:performance_under_32_threads}
    \centering

\small
\begin{tabular}{|c|ccc|ccc|}
\hline
\textbf{Datasets}   & \multicolumn{3}{c|}{\textbf{\emph{lj}}}                               & \multicolumn{3}{c|}{\textbf{\emph{g5}}}                                        \\ \hline
Operations & \multicolumn{1}{c|}{search} & \multicolumn{1}{c|}{scan}    & insert & \multicolumn{1}{c|}{search} & \multicolumn{1}{c|}{scan}             & insert \\ \hline
AdjLst &
  \multicolumn{1}{c|}{\textbf{64.35}} &
  \multicolumn{1}{c|}{\textbf{3268.96}} &
  2.43 &
  \multicolumn{1}{c|}{\textbf{33.83}} &
  \multicolumn{1}{c|}{3955.47} &
  0.26 \\ \hline
LiveGraph  & \multicolumn{1}{c|}{19.49}  & \multicolumn{1}{c|}{1808.15} & 3.02   & \multicolumn{1}{c|}{0.55}   & \multicolumn{1}{c|}{2540.43}          & 1.24   \\ \hline
Aspen      & \multicolumn{1}{c|}{24.17}  & \multicolumn{1}{c|}{1669.10} & 0.20   & \multicolumn{1}{c|}{8.37}   & \multicolumn{1}{c|}{\textbf{4389.51}} & 0.32   \\ \hline
Teseo      & \multicolumn{1}{c|}{48.89}  & \multicolumn{1}{c|}{2822.68} & 3.40   & \multicolumn{1}{c|}{26.37}  & \multicolumn{1}{c|}{3021.91}          & 3.17   \\ \hline
Sortledton &
  \multicolumn{1}{c|}{40.20} &
  \multicolumn{1}{c|}{2335.66} &
  \textbf{3.45} &
  \multicolumn{1}{c|}{14.68} &
  \multicolumn{1}{c|}{2570.57} &
  \textbf{3.23} \\ \hline
\end{tabular}%
\end{table}

For search and scan efficiency on \emph{g5}, competing methods achieve 12.5-29.8x and 12.5-26.7x speedup, respectively. Specifically, LiveGraph and Aspen only achieve 12.5x and 14.7x speedup on search operations because \emph{g5} is large, and the AVL tree and bloom filter lead to severe cache contention issues. Interestingly, the speedup of AdjLst and Teseo is limited on scan but they achieve good throughput. \sun{To address this issue, we examine memory bandwidth utilization in Table \ref{tab:memory bandwidth}. At 8 threads, AdjLst and Teseo consume most of the available memory bandwidth, limiting further speedup even as more threads are added.} Notably, while AdjLst runs much faster than Aspen in the graph container evaluation, it is slower in Table \ref{tab:performance_under_32_threads} because AdjLst reads much more data due to fine-grained version management. \sun{As shown in Table \ref{tab:memory bandwidth}, Aspen uses less bandwidth than competing methods because it avoids reading timestamps for each edge.} Second, in Figures \ref{fig:scalability_insert_on_lj} and \ref{fig:scalability_insert_on_g5}, insert scalability is very limited because write queries require exclusive locks, and high-degree vertices frequently accessed amplify the lock contention issue.  Aspen is slower than Teseo and Sortledton because it has a single writer.

\begin{table}[h!]
\caption{\sun{Memory bandwidth utilization during concurrent scans for competing methods (max bandwidth: 60 GB/sec).}}
\small
\begin{tabular}{|c|ccccc|ccccc|}
\hline
\multicolumn{1}{|l|}{\multirow{2}{*}{\textbf{\begin{tabular}[c]{@{}l@{}}Num of\\ threads\end{tabular}}}} &
  \multicolumn{5}{c|}{\textbf{\emph{lj}}} &
  \multicolumn{5}{c|}{\textbf{\emph{g5}}} \\ \cline{2-11} 
\multicolumn{1}{|l|}{} &
  \multicolumn{1}{c|}{\textbf{AdjLst}} &
  \multicolumn{1}{c|}{\textbf{Lg}} &
  \multicolumn{1}{c|}{\textbf{Ap}} &
  \multicolumn{1}{c|}{\textbf{Ts}} &
  \textbf{Sl} &
  \multicolumn{1}{c|}{\textbf{AdjLst}} &
  \multicolumn{1}{c|}{\textbf{Lg}} &
  \multicolumn{1}{c|}{\textbf{Ap}} &
  \multicolumn{1}{c|}{\textbf{Ts}} &
  \textbf{Sl} \\ \hline
1 &
  \multicolumn{1}{c|}{3.66} &
  \multicolumn{1}{c|}{3.14} &
  \multicolumn{1}{c|}{1.35} &
  \multicolumn{1}{c|}{3.70} &
  2.80 &
  \multicolumn{1}{c|}{9.54} &
  \multicolumn{1}{c|}{8.07} &
  \multicolumn{1}{c|}{2.11} &
  \multicolumn{1}{c|}{10.19} &
  6.01 \\ \hline
2 &
  \multicolumn{1}{c|}{6.69} &
  \multicolumn{1}{c|}{5.97} &
  \multicolumn{1}{c|}{2.10} &
  \multicolumn{1}{c|}{6.87} &
  5.68 &
  \multicolumn{1}{c|}{18.37} &
  \multicolumn{1}{c|}{6.78} &
  \multicolumn{1}{c|}{5.35} &
  \multicolumn{1}{c|}{19.79} &
  12.54 \\ \hline
4 &
  \multicolumn{1}{c|}{11.59} &
  \multicolumn{1}{c|}{10.77} &
  \multicolumn{1}{c|}{4.50} &
  \multicolumn{1}{c|}{12.18} &
  10.39 &
  \multicolumn{1}{c|}{36.15} &
  \multicolumn{1}{c|}{29.05} &
  \multicolumn{1}{c|}{10.14} &
  \multicolumn{1}{c|}{36.49} &
  23.82 \\ \hline
8 &
  \multicolumn{1}{c|}{18.10} &
  \multicolumn{1}{c|}{17.44} &
  \multicolumn{1}{c|}{8.10} &
  \multicolumn{1}{c|}{19.34} &
  17.08 &
  \multicolumn{1}{c|}{51.03} &
  \multicolumn{1}{c|}{46.62} &
  \multicolumn{1}{c|}{19.63} &
  \multicolumn{1}{c|}{54.41} &
  41.07 \\ \hline
16 &
  \multicolumn{1}{c|}{23.96} &
  \multicolumn{1}{c|}{24.12} &
  \multicolumn{1}{c|}{12.68} &
  \multicolumn{1}{c|}{26.12} &
  24.14 &
  \multicolumn{1}{c|}{56.00} &
  \multicolumn{1}{c|}{54.51} &
  \multicolumn{1}{c|}{35.70} &
  \multicolumn{1}{c|}{57.33} &
  53.63 \\ \hline
32 &
  \multicolumn{1}{c|}{25.78} &
  \multicolumn{1}{c|}{26.61} &
  \multicolumn{1}{c|}{17.70} &
  \multicolumn{1}{c|}{28.76} &
  26.65 &
  \multicolumn{1}{c|}{57.76} &
  \multicolumn{1}{c|}{56.69} &
  \multicolumn{1}{c|}{49.07} &
  \multicolumn{1}{c|}{57.81} &
  55.62 \\ \hline
\end{tabular}%
\label{tab:memory bandwidth}
\end{table}

\begin{figure}[h]
    \setlength{\abovecaptionskip}{0pt}
    \setlength{\belowcaptionskip}{0pt}
    \captionsetup[subfigure]{aboveskip=0pt,belowskip=0pt}
    \centering
    \includegraphics[width=0.60\textwidth]{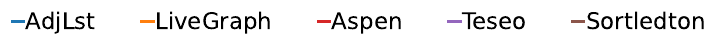}\\
    \begin{subfigure}[t]{0.32\textwidth}
        \centering
        \includegraphics[width=\textwidth]{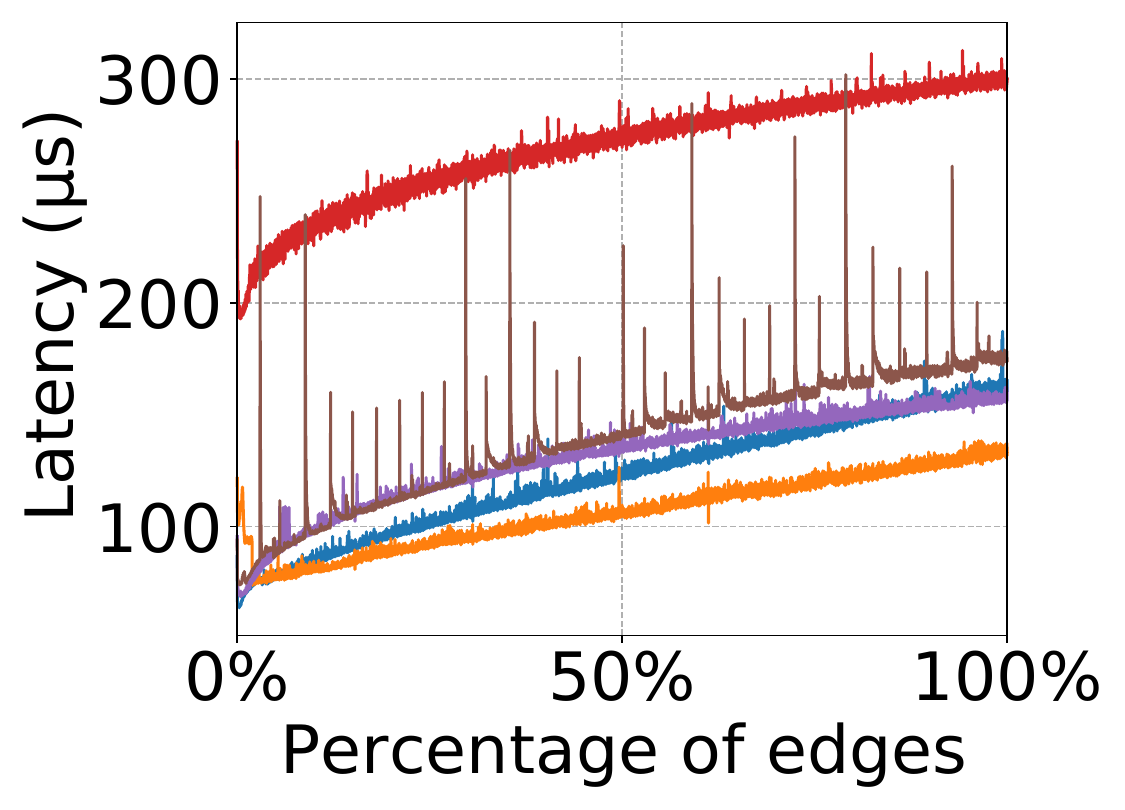}
        \caption{Insert latency on \emph{lj}.}
        \label{fig:insert_latency_lj}
    \end{subfigure}
    \begin{subfigure}[t]{0.32\textwidth}
        \centering
        \includegraphics[width=\textwidth]{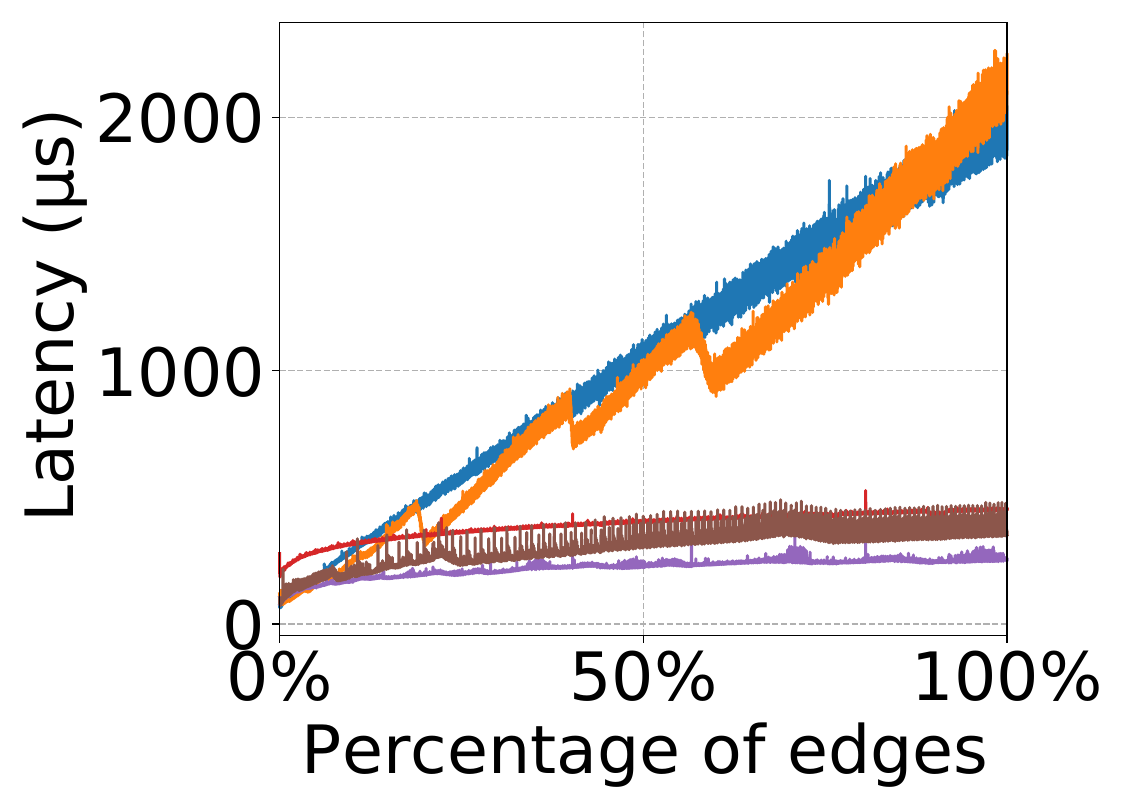}
        \caption{Insert latency on \emph{g5}.}
        \label{fig:insert_latency_g5}
    \end{subfigure}
    \caption{Insert latency for every 100 operations during the edge insertion process.}
    \label{fig:insert_latency_edge_insertion}
\end{figure}

We further compare performance variations during the insertion process. Figure \ref{fig:insert_latency_edge_insertion} shows the insert latency along the edge insertion process. Since \emph{lj} is sparse, AdjLst and LiveGraph generally outperform segmented methods, while they are slower on \emph{g5}, a dense graph. Aspen’s latency fluctuation is much smaller than that of other methods because 1) it uses a single writer and has no lock contention, and 2) the overhead of CoW on blocks is steady. The higher fluctuation in other methods is due to the expensive overhead of copying blocks and inserting into the skip list.

\subsubsection{Concurrency Evaluation} 

\begin{figure}[h!]
    \setlength{\abovecaptionskip}{0pt}
    \setlength{\belowcaptionskip}{0pt}
    \captionsetup[subfigure]{aboveskip=0pt,belowskip=0pt}
    \centering
    \includegraphics[width=0.60\textwidth]{img/exp_figs/figure6/legend_line.pdf}\\
    \begin{subfigure}[t]{0.30\textwidth}
        \centering
        \includegraphics[width=\textwidth]{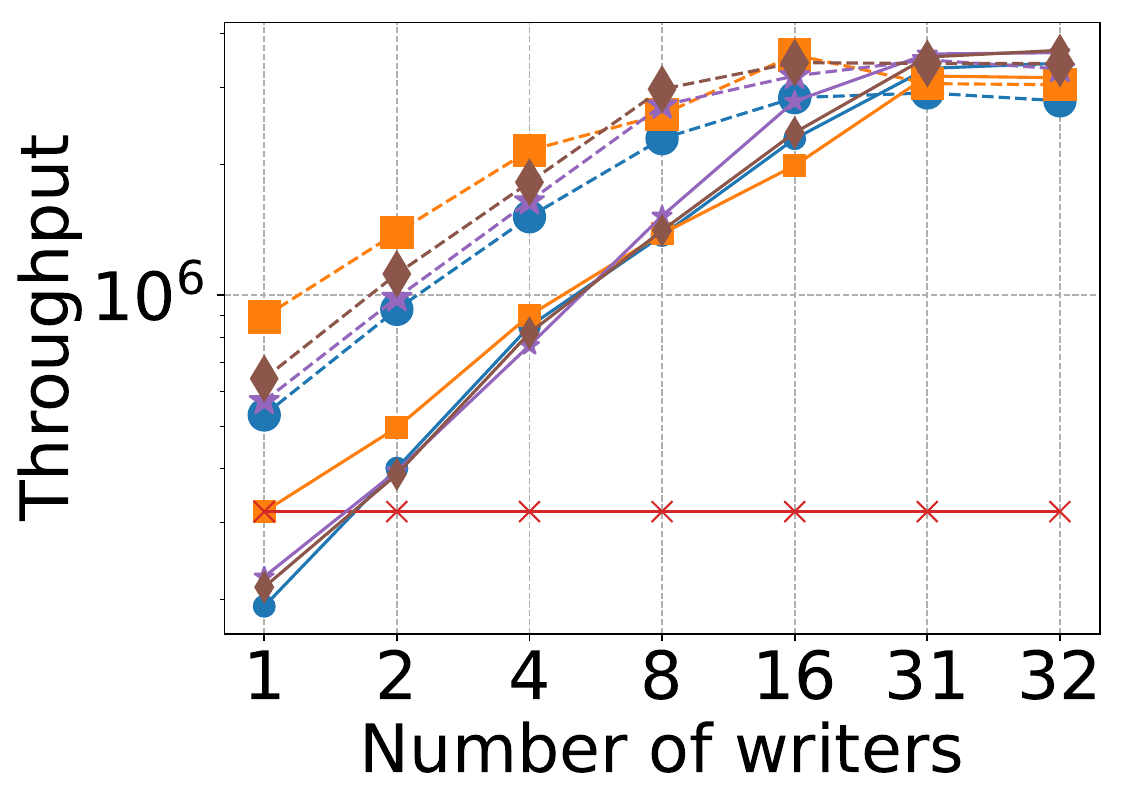}
        \caption{Insert operations on \emph{lj}.}
        \label{fig:concurrency_insert_lj}
    \end{subfigure}
    \begin{subfigure}[t]{0.30\textwidth}
        \centering
        \includegraphics[width=\textwidth]{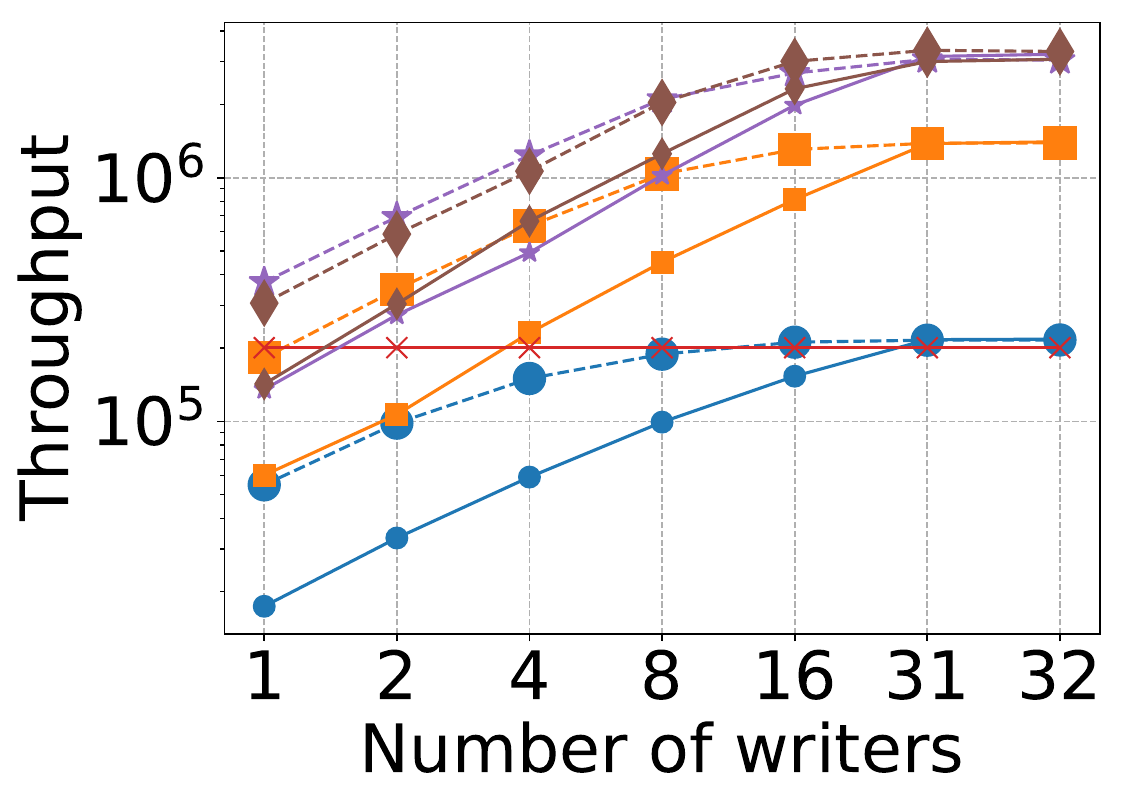}
        \caption{Insert operations on \emph{g5}.}
        \label{fig:concurrency_insert_g5}
    \end{subfigure}
    \caption{Write efficiency when mixing readers and writers.}
    \label{fig:concurrency_write}
\end{figure}

We evaluate the concurrency of competing methods by mixing readers and writers with their total count equal to 32, the number of cores. Each writer executes an insertion stream, whereas a reader runs a PR query. Figure \ref{fig:concurrency_write} presents the experiment results on write efficiency. The dashed line denotes write efficiency without readers executing, while the solid line represents write efficiency with readers. Introducing readers significantly degrades the performance of methods using G2PL due to vertex access contention. With 31 readers and 1 writer, the throughput is only 31.75-46.58\% of that without any readers. In contrast, Aspen’s throughput is only slightly affected by readers because they access the graph snapshot without locking data.

\begin{figure}[h!]
    \setlength{\abovecaptionskip}{0pt}
    \setlength{\belowcaptionskip}{0pt}
    \captionsetup[subfigure]{aboveskip=0pt,belowskip=0pt}
    \centering
     \includegraphics[width=0.60\textwidth]{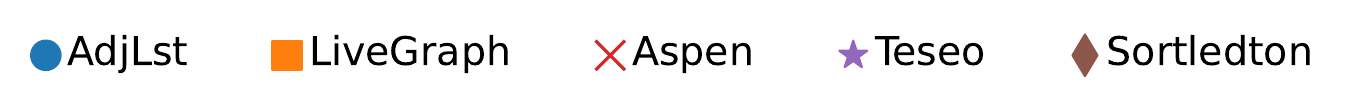}\\
    \begin{subfigure}[t]{0.30\textwidth}
        \centering
        \includegraphics[width=\textwidth]{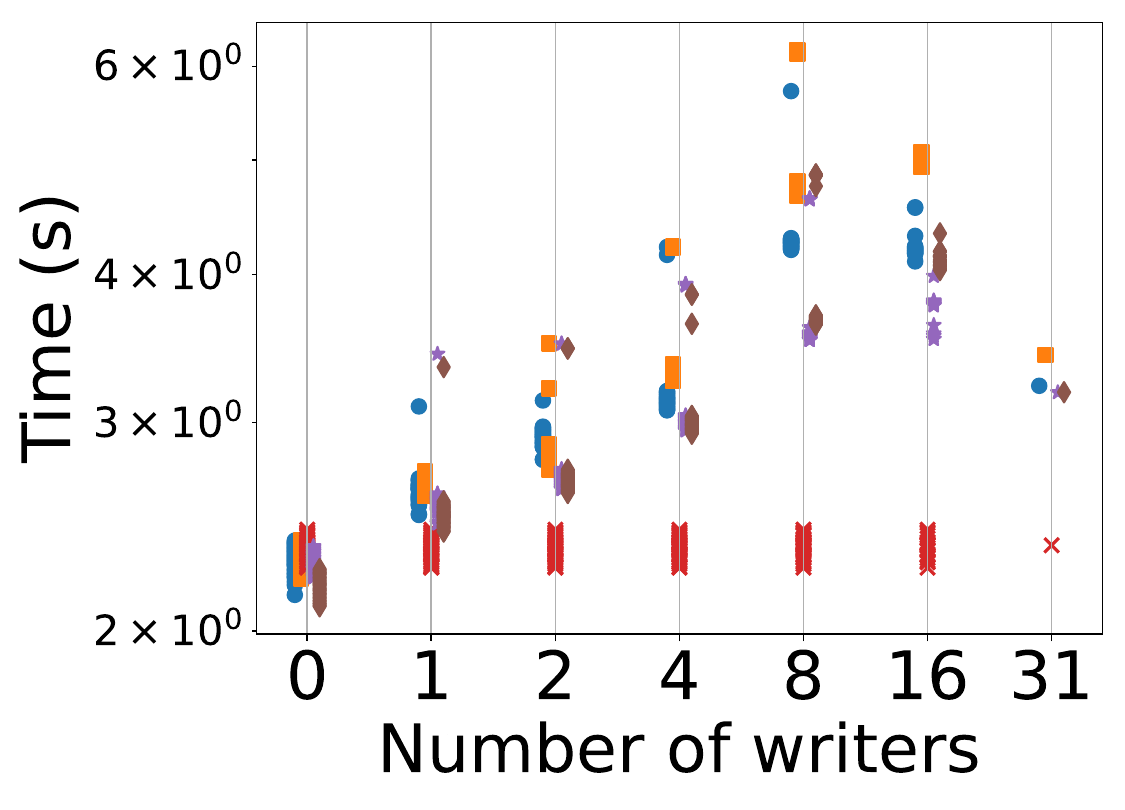}
        \caption{PR on \emph{lj}.}
        \label{fig:concurrency_pr_lj}
    \end{subfigure}
    \begin{subfigure}[t]{0.30\textwidth}
        \centering
        \includegraphics[width=\textwidth]{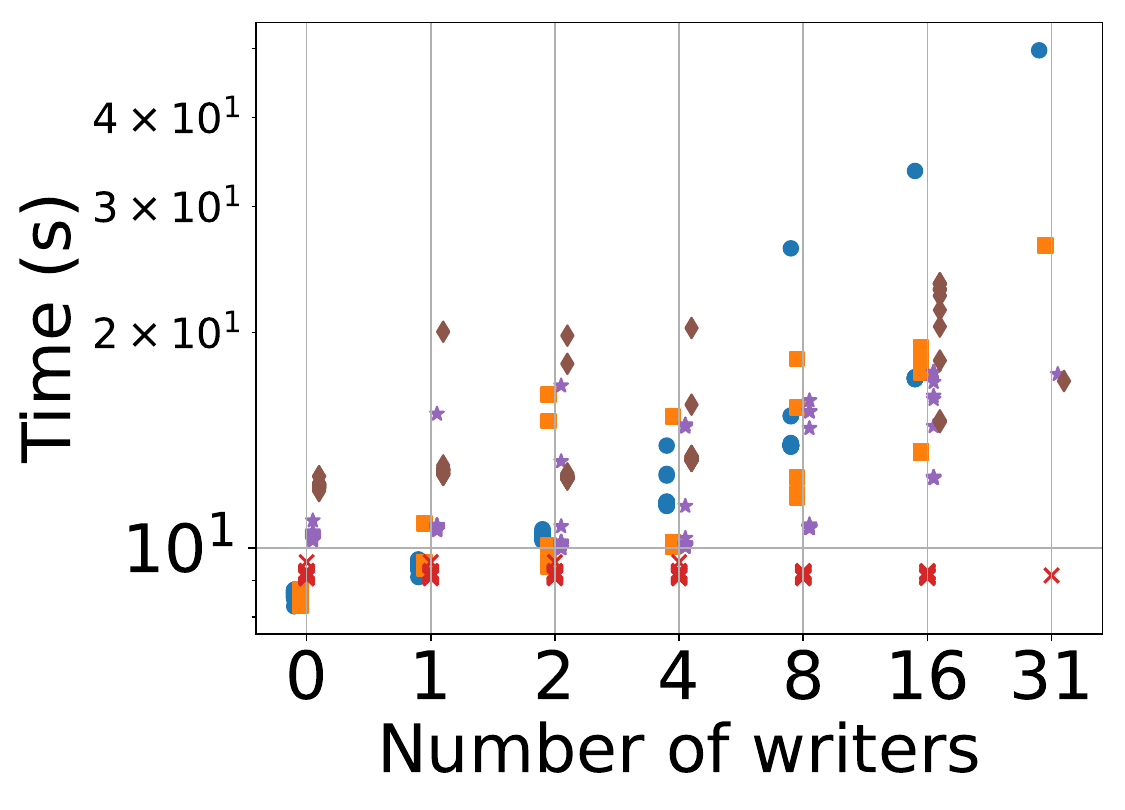}
        \caption{PR on \emph{g5}.}
        \label{fig:concurrency_pr_g5}
    \end{subfigure}
   
    \caption{Read efficiency when mixing readers and writers.}
    \label{fig:concurrency_pr}
\end{figure}

Figure \ref{fig:concurrency_pr} presents the experiment results on read efficiency with varying numbers of writers. Each dot represents the average time for a reader to perform one iteration of PR. Methods using G2PL show significant performance decreases as the number of writers increases due to lock contention. Methods with poor insert efficiency (e.g., LiveGraph and AdjLst) are more affected than Teseo and Sortledton because they must hold exclusive locks for longer periods. The performance variance among different readers increases. In contrast, Aspen maintains steady performance. Specifically, Aspen performs worse than its counterparts when there are no writers but outperforms them when there are multiple writers.

\subsubsection{Summary of Findings} \textbf{Q3. What is the impact of graph concurrency control on graph operations?} For fine-grained methods, maintaining and checking versions for each element incurs significant overhead due to the need to load more data and perform extra computations. Traversing the version list can cause more overhead for scans, while continuous version storage can be costlier for searches. In contrast, the coarse-grained method (i.e., Aspen) avoids these issues, resulting in a smaller performance gap.

\vspace{2pt}
\noindent\textbf{Q4. How is the scalability and concurrency of competing methods?} Overall, all competing methods generally exhibit good scalability in search and scan. Search scalability can be affected by cache contention, while scan scalability can be limited by memory bandwidth. Aspen can outperform the fine-grained methods despite its less efficient graph container. However, existing works overlook the problem of hardware utilization. In contrast, insert operations have very limited scalability due to heavy lock contention from frequent access to high-degree vertices.

Despite using MVCC, in practice, readers and writers of fine-grained methods interfere significantly on graphs because high-degree vertices are frequently accessed. While Aspen’s read performance can outperform its counterparts, it suffers from slow insert speeds due to its inefficient CoW strategy. In a word, both fine-grained and coarse-grained methods face severe performance issues in a multi-threaded context.

\begin{figure}[t]
    \setlength{\abovecaptionskip}{0pt}
    \setlength{\belowcaptionskip}{0pt}
    \captionsetup[subfigure]{aboveskip=0pt,belowskip=0pt}
    \centering
    \includegraphics[width=0.60\textwidth]{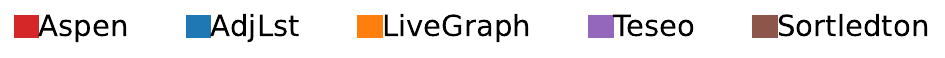}\\
    \begin{subfigure}[t]{0.30\textwidth}
        \centering
        \includegraphics[width=\textwidth]{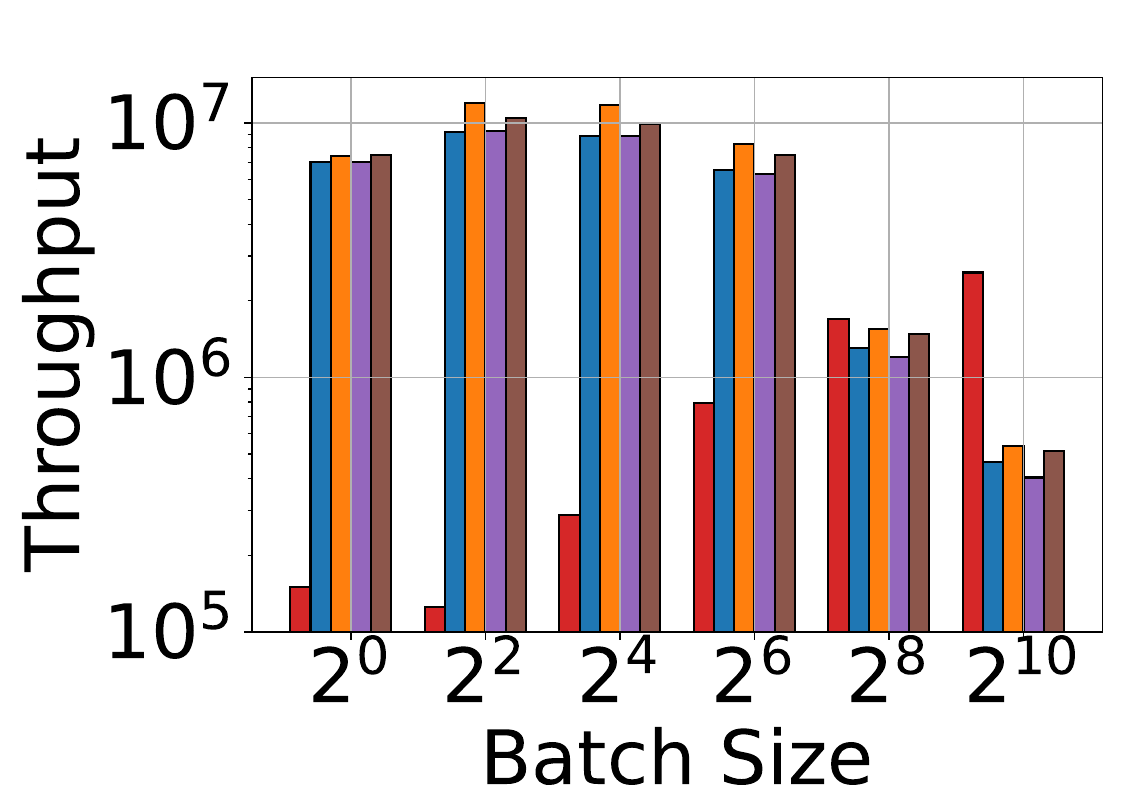}
        \caption{Batch update on \emph{lj}.}
        \label{fig:batch_update_on_lj}
    \end{subfigure}
    \begin{subfigure}[t]{0.30\textwidth}
        \centering
        \includegraphics[width=\textwidth]{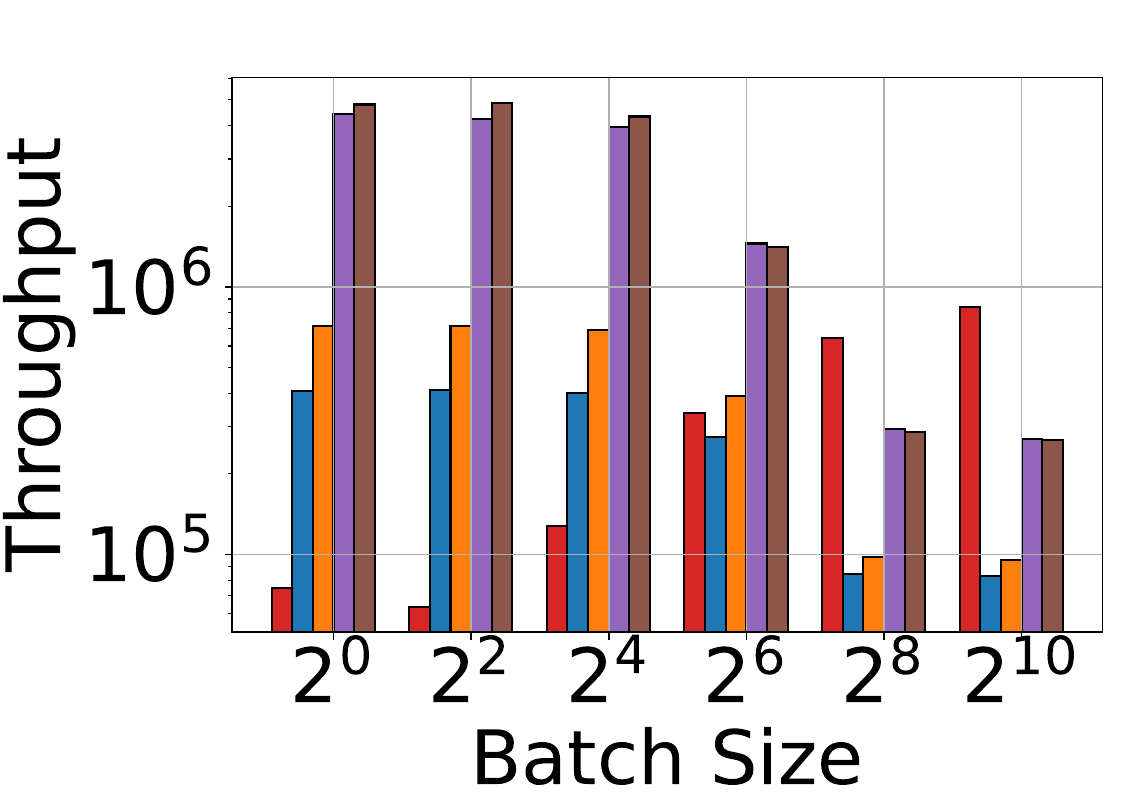}
        \caption{Batch update on \emph{g5}.}
        \label{figbatch_update_on_g5}
    \end{subfigure}
    
	\caption{\sun{Write efficiency with batch size increasing.}}
	\label{fig:batch_update_throughput}
\end{figure}

\subsection{\sun{Evaluation of Update Granularity}}

\sun{In the above experiments, each write query consisted of a single update. We now evaluate how batch size affects performance. Since our test system has 32 CPU cores, Aspen uses a single writer and 32 threads to process one batch in parallel, while the other methods use 32 writers, each independently processing its own batch.}

\sun{As shown in Figure \ref{fig:batch_update_throughput}, Aspen outperforms the other methods when the batch size exceeds $2^{8}$, with throughput steadily increasing as the batch size grows from $2^{0}$ to $2^{10}$. This is because Aspen avoids lock contention with a single writer, creates one snapshot for multiple updates, and leverages parallelism across update operations. In contrast, methods using G2PL exhibit a significant performance drop as batch size increases, especially beyond $2^8$, due to the need to hold more locks per query and the longer execution time required to process larger batches, which increases lock contention.}

\noindent\textbf{\sun{Q5. How does the batch granularity affect the performance of competing methods?}} \sun{Coarse-grained methods perform poorly with small batch sizes since parallelism within a write query is limited, and a snapshot must be created for each update. However, their performance improves significantly as batch size increases. In contrast, fine-grained methods excel with small batch sizes but face severe performance degradation due to lock contention among queries as the batch size grows.}

\subsection{Evaluation of Memory Consumption}

We evaluate the memory consumption of competing methods by measuring the \emph{resident set size} as reported by the OS. Table \ref{tab:memory_consumption} presents the results. \emph{wo} and \emph{w} denote the storage without and with version information, respectively. Aspen uses coarse-grained version management, resulting in negligible memory consumption differences (less than 10MB) between \emph{wo} and \emph{w}. Therefore, we report its \emph{w} implementation for reference.

\vspace{2pt}
\noindent\textbf{Q6. What is the impact of graph containers and version management in DGS on memory consumption, and what is the gap between DGS and CSR?} Overall, Aspen consumes 3.3-11.0x times more memory than CSR. Among the fine-grained methods, the most optimal consumes 0.9-1.3x and 4.1-8.9x times more memory than Aspen and CSR, respectively. This is primarily because: 1) they maintain version information for each element, and 2) they leave empty slots in neighbor indexes. These results indicate that fine-grained methods can cause severe memory issues, limiting the scalability of DGS on large graphs.

\small
\begin{table}[htbp]
\captionsetup{skip=0pt}
\setlength{\abovecaptionskip}{0pt}
\setlength{\belowcaptionskip}{0pt}
    \caption{Memory consumption of competing methods (GB).}
    \label{tab:memory_consumption}
    \begin{tabular}{|c|c|c|cc|cc|cc|cc|}
\hline
\multirow{2}{*}{\textbf{Dataset}} &
  \multirow{2}{*}{\textbf{CSR}} &
  \multirow{2}{*}{\textbf{Aspen}} &
  \multicolumn{2}{c|}{\textbf{AdjLst}} &
  \multicolumn{2}{c|}{\textbf{LiveGraph}} &
  \multicolumn{2}{c|}{\textbf{Teseo}} &
  \multicolumn{2}{c|}{\textbf{Sortledton}} \\ \cline{4-11} 
          &      &       & \multicolumn{1}{c|}{wo}   & w     & \multicolumn{1}{c|}{wo}   & w     & \multicolumn{1}{c|}{wo}    & w     & \multicolumn{1}{c|}{wo}    & w     \\ \hline
\emph{lj}   & 0.67 & 3.58  & \multicolumn{1}{c|}{0.98} & 2.74  & \multicolumn{1}{c|}{1.66} & 6.24  & \multicolumn{1}{c|}{9.67}  & 30.98 & \multicolumn{1}{c|}{2.62}  & 4.31  \\ \hline
\emph{dl}   & 0.76 & 2.52  & \multicolumn{1}{c|}{0.99} & 2.89  & \multicolumn{1}{c|}{1.21} & 3.71  & \multicolumn{1}{c|}{1.33}  & 6.87  & \multicolumn{1}{c|}{3.00}  & 3.11  \\ \hline
\emph{ldbc}   & 2.84 & 17.96 & \multicolumn{1}{c|}{3.99} & 10.52 & \multicolumn{1}{c|}{7.61} & 28.95 & \multicolumn{1}{c|}{61.16} & OOM   & \multicolumn{1}{c|}{15.30} & 20.19 \\ \hline
\emph{g5} & 3.95 & 15.33 & \multicolumn{1}{c|}{5.42} & 15.41 & \multicolumn{1}{c|}{7.17} & 24.62 & \multicolumn{1}{c|}{21.57} & 78.42 & \multicolumn{1}{c|}{15.25} & 19.55 \\ \hline
\emph{wk}   & 0.98 & 8.18  & \multicolumn{1}{c|}{1.41} & 3.49  & \multicolumn{1}{c|}{3.49} & 13.09 & \multicolumn{1}{c|}{28.11} & 86.08 & \multicolumn{1}{c|}{5.38}  & 7.12  \\ \hline
\emph{ct}   & 0.27 & 2.06  & \multicolumn{1}{c|}{0.42} & 1.11  & \multicolumn{1}{c|}{0.84} & 3.32  & \multicolumn{1}{c|}{7.42}  & 22.81 & \multicolumn{1}{c|}{1.11}  & 2.14  \\ \hline
\emph{tw}   & 4.11 & 20.20 & \multicolumn{1}{c|}{5.49} & 15.09 & \multicolumn{1}{c|}{9.02} & 31.98 & \multicolumn{1}{c|}{45.89} & OOM   & \multicolumn{1}{c|}{17.48} & 22.04 \\ \hline
\emph{nft}  & 1.38 & 14.94 & \multicolumn{1}{c|}{2.06} & 4.99  & \multicolumn{1}{c|}{5.93} & 23.01 & \multicolumn{1}{c|}{58.76} & OOM   & \multicolumn{1}{c|}{9.07}  & 12.27 \\ \hline
\end{tabular}%
\end{table}

\section{Related Work} \label{sec:related_work}

In this paper, we study in-memory dynamic graph storage~\cite{macko2015llama, zhu2019livegraph, de2021teseo, dhulipala2019low, fuchs2022sortledton} that supports concurrent read and write queries, particularly single updates. Below, we discuss the related work.

\noindent\textbf{Graph Databases and Benchmarks.} Graph databases such as Neo4J, Virtuoso, and K`uzu~\cite{kuzu:cidr} also support transactions. They typically operate on external storage and focus on supporting labeled property graphs. Some, like Neo4J, use the read-committed isolation level~\cite{ramakrishnan2002database} to improve performance and simplify transaction management. Additionally, there are graph databases designed for distributed environments~\cite{li2022bytegraph,zhang2024bg3,besta2023graph,carter2019nanosecond}, which focus on optimizing communication and distributed storage. For graph benchmarks, LDBC provides a variety of workloads~\cite{erling2015ldbc,szarnyas2022ldbc,iosup2016ldbc} to evaluate the performance of graph databases. These workloads also use labeled property graphs with various operations on labels or properties. Our paper focuses on in-memory DGS, optimizing operations on graph topology. These existing DGS methods do not consider labels and properties, and thus cannot be directly applied. However, as the labeled property graph model is widely used, researching DGS on this model is an interesting direction, for example, combining DGS with well-designed columnar graph storage~\cite{mhedhbi2021a+}.

\noindent\textbf{Graph Processing Frameworks.} Many frameworks have been proposed for parallel analysis of static graphs, such as Ligra~\cite{shun2013ligra} and Ligra+\cite{shun2015smaller} for single-machine environments, and Pregel\cite{malewicz2010pregel}, Giraph~\cite{grover2015hadoop}, Gemini~\cite{zhu2016gemini}, and Grapes~\cite{fan2018parallelizing} for distributed environments. These frameworks typically use CSR for in-memory storage and focus on parallelization strategies that exploit inter-query parallelism for graph queries like BFS, SSSP, and PR.

\noindent\textbf{Dynamic Graph Processing Frameworks.} Dynamic graph processing frameworks target frequently updated graphs. They focus on designing novel approaches to maintain intermediate results for graph queries, reducing the overhead of re-computation when the graphs are modified. For example, KickStarter~\cite{vora2017kickstarter} minimizes unnecessary computations by obtaining a trimmed approximate subset of the nodes affected by the update. RisGraph~\cite{feng2021risgraph} uses a variant of adjacency lists and sparse arrays as its storage structure and can switch between vertex parallelism and edge parallelism. To achieve streaming processing, it employs a tree-based classification of safe and unsafe updates. GraphZeppelin~\cite{tench2022graphzeppelin} uses new linear sketching data structures to solve the streaming connected components problem. These studies are orthogonal to our research.

\vspace{-10pt} 
\section{Conclusion} \label{sec:conclusion}

\sun{We summarize our findings on the relative performance of computing methods, key technical insights, and opportunities for future research and optimization below.}

\begin{figure}[t]
	\setlength{\abovecaptionskip}{0pt}
	\setlength{\belowcaptionskip}{0pt}
		\captionsetup[subfigure]{aboveskip=0pt,belowskip=0pt}
	\centering

		\includegraphics[width=0.60\textwidth]{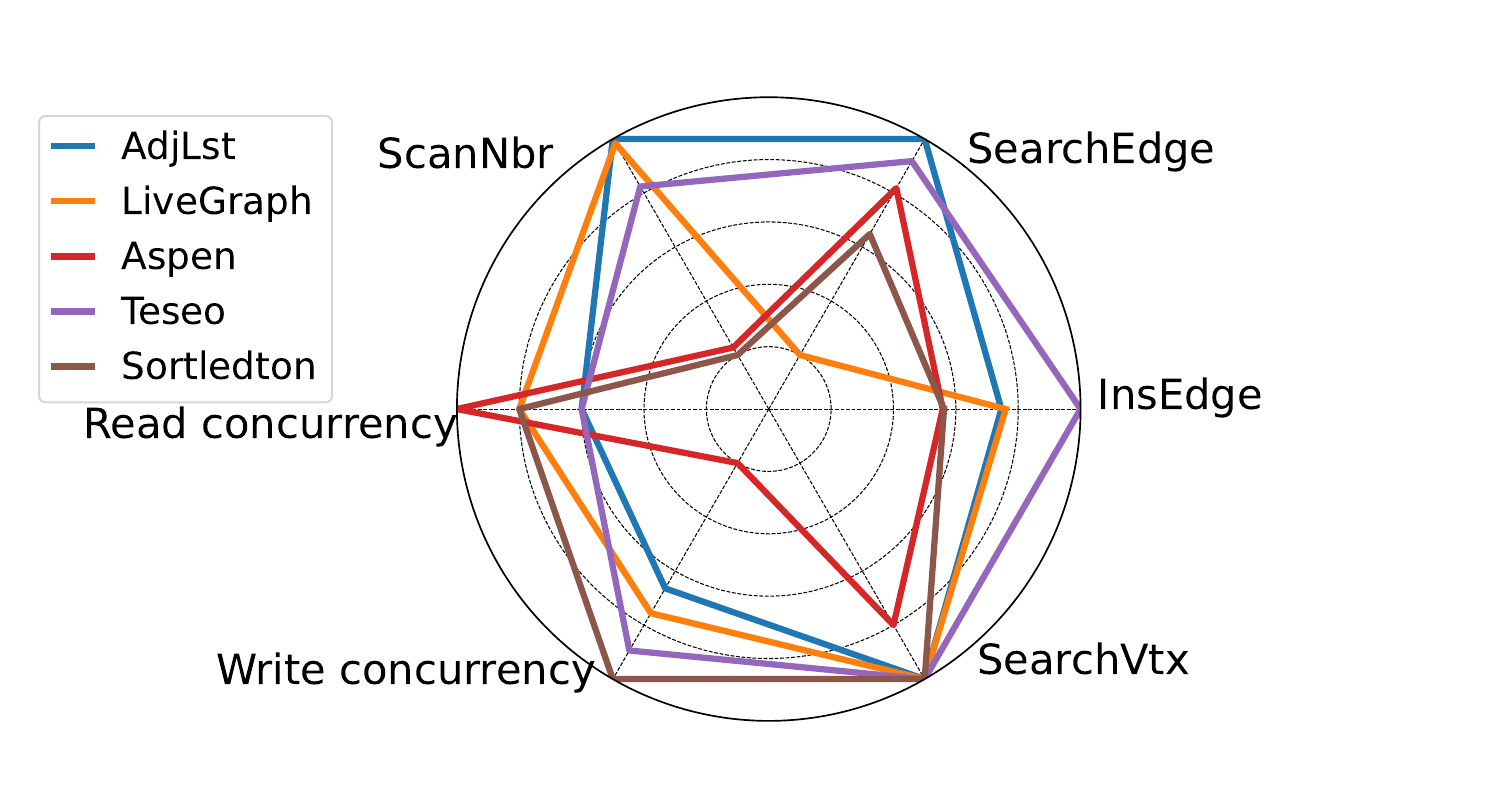}

	\caption{\sun{Relative performance of existing methods for factors in  Equation \ref{eq:cost}.}}
	\label{fig:radar}
\end{figure}

\noindent\sun{\textbf{Relative Performance.} Guided by Equation \ref{eq:cost}, we summarize the efficiency $T_p$ of graph access operations and the concurrency control overhead amplification ratio $\alpha_p$ across competing methods. We focus on four key operators: \textsc{SearchVtx}, \textsc{SearchEdge}, \textsc{ScanNbr}, and \textsc{InsEdge}. The efficiency and overhead amplification ratios are based on experimental results in Sections \ref{sec:effiency_of_neighbor_indexes} and \ref{sec:scalability_evaluation}. Relative performance is calculated by first computing the geometric mean of throughput across datasets, then normalizing the results on a scale of 1 to 5, with 5 being the best. Figure \ref{fig:radar} visualizes the relative performance of the methods across different dimensions.}

\sun{For read efficiency, including \textsc{ScanNbr} and \textsc{SearchEdge}, AdjLst performs the best. Teseo is slower on \textsc{ScanNbr} and \textsc{SearchNbr}, but offers a balanced read and write performance. For \textsc{SearchVtx}, dynamic array storage for vertices achieves the best results. Aspen excels in read concurrency due to its coarse-grained version management but performs poorly in write concurrency because it only supports a single-writer.}

\noindent\sun{\textbf{Key Technical Insights.} Based on our extensive experimental results, we confirm the following key findings from previous work: 1) segmenting neighbor sets into blocks effectively balances read and write performance; 2) in the single update setting, fine-grained methods improve write throughput and outperform coarse-grained methods; 3) adaptive indexing significantly enhances performance by leveraging the sparsity of real-world graphs, reducing LLC misses; and 4) converting tree indexes to arrays improves the performance of long-running queries in coarse-grained methods.}

\sun{\noindent However, we find that current design choices in concurrency control and graph containers lead to significant performance issues, which diverge from modern trends and present new challenges for the community.}

\begin{itemize}[leftmargin=*]
    \item \sun{\textbf{The fine-grained strategy, widely adopted in recent works, introduces severe performance issues.} First, performing version checks for each neighbor imposes substantial computation overhead and consumes excessive memory bandwidth. Second, maintaining versions for each neighbor requires a large amount of memory, limiting scalability. Third, readers and writers interfere heavily, and insert operations face limited scalability due to lock contention, especially when accessing high-degree vertices.}
    
    \item \sun{\textbf{Existing research often focuses on selecting various block indexes but neglects hardware-level considerations.} This creates a performance gap compared to continuous storage methods. Fragmented memory layouts increase cache misses (L1, L2, LLC, and DTLB). Additionally, the complexity of segmented methods introduces high instruction overhead and frequent branch mispredictions.}
\end{itemize}

\noindent\sun{\textbf{Research and Optimization Opportunities.} This study identifies key areas for future research and optimization. In graph concurrency control, new version management techniques are needed to reduce the space and efficiency overhead of fine-grained versions, improving scalability. One research opportunity is to manage versions in a more coarse-grained manner, potentially eliminating the overhead of version checks for each edge. Additionally, more efficient concurrency control mechanisms are required to minimize interference among concurrent queries. Given that graph queries are typically read-intensive, another research direction is to explore relaxed locking requirements for read queries, thereby reducing interference from write operations.}

\sun{For graph containers, segmenting neighbor sets and using adaptive indexing can enhance graph access efficiency. However, further optimization of memory layout and management strategies is necessary to reduce cache misses and instruction overhead, improving hardware utilization. Developing graph containers that balance read and write efficiency while integrating improved concurrency control methods is essential, as a significant read efficiency gap remains between dynamic and static graph storage formats.}

\begin{acks}
Jixian Su and Chiyu Hao contributed equally to this work. Shixuan Sun is the corresponding author.
\end{acks}
\bibliographystyle{ACM-Reference-Format}
\bibliography{reference}

\appendix
\clearpage

\section{Supplement Results}

\noindent\textbf{Graph Analytics Query.} We present the experiment results of the graph containers on three graph analytics queries: BFS, SSSP, and WCC. Aspen-w enables the flatten optimization, and Sortledton-w enables adaptive indexing with the threshold set to 256.
\small
\begin{table}[h]
\caption{Comparison of analytical queries (in seconds). Lg, Ap, Ts and Sl denote LiveGraph, Aspen, Teseo and Sortledton, respectively.}
\label{tab:performance_comparison_containers_appendix}
\centering
\begin{tabular}{|c|ccc|ccccc|}
\hline
 &
  \multicolumn{3}{c|}{\textbf{Continuous Neighbor Index}} &
  \multicolumn{5}{c|}{\textbf{Segmented Neighbor Index}} \\ \hline
\multirow{2}{*}{\textbf{Dataset}} &
  \multicolumn{1}{c|}{\multirow{2}{*}{\textbf{CSR}}} &
  \multicolumn{1}{c|}{\multirow{2}{*}{\textbf{AdjLst}}} &
  \multirow{2}{*}{\textbf{Lg}} &
  \multicolumn{2}{c|}{\textbf{Ap}} &
  \multicolumn{1}{c|}{\multirow{2}{*}{\textbf{Ts}}} &
  \multicolumn{2}{c|}{\textbf{Sl}} \\ \cline{5-6} \cline{8-9} 
 &
  \multicolumn{1}{c|}{} &
  \multicolumn{1}{c|}{} &
   &
  \multicolumn{1}{c|}{w} &
  \multicolumn{1}{c|}{wo} &
  \multicolumn{1}{c|}{} &
  \multicolumn{1}{c|}{w} &
  wo \\ \hline
\textbf{BFS} &
  \multicolumn{3}{c|}{} &
  \multicolumn{5}{c|}{} \\ \hline
\emph{lj} &
  \multicolumn{1}{c|}{\textbf{0.56}} &
  \multicolumn{1}{c|}{0.82} &
  1.57 &
  \multicolumn{1}{c|}{\textbf{1.43}} &
  \multicolumn{1}{c|}{3.15} &
  \multicolumn{1}{c|}{2.03} &
  \multicolumn{1}{c|}{1.89} &
  2.94 \\ \hline
\emph{dl} &
  \multicolumn{1}{c|}{\textbf{0.01}} &
  \multicolumn{1}{c|}{\textbf{0.01}} &
  0.02 &
  \multicolumn{1}{c|}{0.03} &
  \multicolumn{1}{c|}{0.04} &
  \multicolumn{1}{c|}{\textbf{0.02}} &
  \multicolumn{1}{c|}{0.03} &
  0.03 \\ \hline
\emph{ldbc} &
  \multicolumn{1}{c|}{\textbf{2.68}} &
  \multicolumn{1}{c|}{4.18} &
  5.41 &
  \multicolumn{1}{c|}{\textbf{6.60}} &
  \multicolumn{1}{c|}{14.87} &
  \multicolumn{1}{c|}{11.30} &
  \multicolumn{1}{c|}{6.61} &
  16.62 \\ \hline
\emph{g5} &
  \multicolumn{1}{c|}{\textbf{2.00}} &
  \multicolumn{1}{c|}{3.06} &
  5.68 &
  \multicolumn{1}{c|}{3.76} &
  \multicolumn{1}{c|}{7.57} &
  \multicolumn{1}{c|}{\textbf{3.33}} &
  \multicolumn{1}{c|}{4.07} &
  6.33 \\ \hline
\emph{wk} &
  \multicolumn{1}{c|}{\textbf{1.19}} &
  \multicolumn{1}{c|}{1.61} &
  2.21 &
  \multicolumn{1}{c|}{\textbf{2.46}} &
  \multicolumn{1}{c|}{6.08} &
  \multicolumn{1}{c|}{4.03} &
  \multicolumn{1}{c|}{2.71} &
  4.96 \\ \hline
\emph{ct} &
  \multicolumn{1}{c|}{\textbf{0.44}} &
  \multicolumn{1}{c|}{0.85} &
  1.62 &
  \multicolumn{1}{c|}{\textbf{1.54}} &
  \multicolumn{1}{c|}{3.17} &
  \multicolumn{1}{c|}{2.37} &
  \multicolumn{1}{c|}{1.57} &
  4.55 \\ \hline
\emph{tw} &
  \multicolumn{1}{c|}{\textbf{2.92}} &
  \multicolumn{1}{c|}{4.40} &
  5.92 &
  \multicolumn{1}{c|}{6.88} &
  \multicolumn{1}{c|}{14.55} &
  \multicolumn{1}{c|}{9.02} &
  \multicolumn{1}{c|}{\textbf{6.59}} &
  15.52 \\ \hline
\emph{nft} &
  \multicolumn{1}{c|}{\textbf{2.50}} &
  \multicolumn{1}{c|}{4.53} &
  5.78 &
  \multicolumn{1}{c|}{6.70} &
  \multicolumn{1}{c|}{15.44} &
  \multicolumn{1}{c|}{11.48} &
  \multicolumn{1}{c|}{\textbf{6.34}} &
  15.11 \\ \hline
\textbf{SSSP} &
  \multicolumn{3}{c|}{} &
  \multicolumn{5}{c|}{} \\ \hline
\emph{lj} &
  \multicolumn{1}{c|}{\textbf{1.42}} &
  \multicolumn{1}{c|}{2.18} &
  2.36 &
  \multicolumn{1}{c|}{\textbf{2.81}} &
  \multicolumn{1}{c|}{7.23} &
  \multicolumn{1}{c|}{2.82} &
  \multicolumn{1}{l|}{2.84} &
  \multicolumn{1}{l|}{3.96} \\ \hline
\emph{dl} &
  \multicolumn{1}{c|}{\textbf{0.17}} &
  \multicolumn{1}{c|}{0.23} &
  0.26 &
  \multicolumn{1}{c|}{0.69} &
  \multicolumn{1}{c|}{0.71} &
  \multicolumn{1}{c|}{\textbf{0.34}} &
  \multicolumn{1}{l|}{0.58} &
  \multicolumn{1}{l|}{0.67} \\ \hline
\emph{ldbc} &
  \multicolumn{1}{c|}{\textbf{6.36}} &
  \multicolumn{1}{c|}{8.65} &
  9.17 &
  \multicolumn{1}{c|}{11.58} &
  \multicolumn{1}{c|}{36.42} &
  \multicolumn{1}{c|}{15.46} &
  \multicolumn{1}{l|}{\textbf{9.99}} &
  \multicolumn{1}{l|}{21.69} \\ \hline
\emph{g5} &
  \multicolumn{1}{c|}{\textbf{4.25}} &
  \multicolumn{1}{c|}{5.78} &
  10.00 &
  \multicolumn{1}{c|}{7.75} &
  \multicolumn{1}{c|}{14.68} &
  \multicolumn{1}{c|}{\textbf{6.46}} &
  \multicolumn{1}{l|}{7.50} &
  \multicolumn{1}{l|}{9.61} \\ \hline
\emph{wk} &
  \multicolumn{1}{c|}{\textbf{1.88}} &
  \multicolumn{1}{c|}{2.44} &
  3.56 &
  \multicolumn{1}{c|}{3.58} &
  \multicolumn{1}{c|}{7.71} &
  \multicolumn{1}{c|}{4.38} &
  \multicolumn{1}{l|}{\textbf{3.42}} &
  \multicolumn{1}{l|}{5.61} \\ \hline
\emph{ct} &
  \multicolumn{1}{c|}{\textbf{0.79}} &
  \multicolumn{1}{c|}{1.64} &
  1.45 &
  \multicolumn{1}{c|}{1.83} &
  \multicolumn{1}{c|}{4.65} &
  \multicolumn{1}{c|}{1.79} &
  \multicolumn{1}{l|}{\textbf{1.68}} &
  \multicolumn{1}{l|}{2.78} \\ \hline
\emph{tw} &
  \multicolumn{1}{c|}{\textbf{6.71}} &
  \multicolumn{1}{c|}{10.03} &
  11.30 &
  \multicolumn{1}{c|}{13.62} &
  \multicolumn{1}{c|}{32.49} &
  \multicolumn{1}{c|}{\textbf{12.67}} &
  \multicolumn{1}{l|}{12.09} &
  \multicolumn{1}{l|}{17.81} \\ \hline
\emph{nft} &
  \multicolumn{1}{c|}{\textbf{4.55}} &
  \multicolumn{1}{c|}{6.66} &
  8.31 &
  \multicolumn{1}{c|}{8.79} &
  \multicolumn{1}{c|}{31.42} &
  \multicolumn{1}{c|}{12.20} &
  \multicolumn{1}{l|}{\textbf{7.28}} &
  \multicolumn{1}{l|}{24.61} \\ \hline
\textbf{WCC} &
  \multicolumn{3}{c|}{} &
  \multicolumn{5}{c|}{} \\ \hline
\emph{lj} &
  \multicolumn{1}{c|}{\textbf{1.95}} &
  \multicolumn{1}{c|}{3.02} &
  3.76 &
  \multicolumn{1}{c|}{4.15} &
  \multicolumn{1}{c|}{5.82} &
  \multicolumn{1}{c|}{\textbf{4.14}} &
  \multicolumn{1}{c|}{4.76} &
  8.39 \\ \hline
\emph{dl} &
  \multicolumn{1}{c|}{\textbf{0.29}} &
  \multicolumn{1}{c|}{0.29} &
  0.42 &
  \multicolumn{1}{c|}{0.85} &
  \multicolumn{1}{c|}{0.86} &
  \multicolumn{1}{c|}{\textbf{0.45}} &
  \multicolumn{1}{c|}{0.94} &
  1.04 \\ \hline
\emph{ldbc} &
  \multicolumn{1}{c|}{\textbf{5.73}} &
  \multicolumn{1}{c|}{6.76} &
  6.93 &
  \multicolumn{1}{c|}{\textbf{9.72}} &
  \multicolumn{1}{c|}{15.03} &
  \multicolumn{1}{c|}{15.34} &
  \multicolumn{1}{c|}{10.48} &
  29.48 \\ \hline
\emph{g5} &
  \multicolumn{1}{c|}{\textbf{4.88}} &
  \multicolumn{1}{c|}{6.04} &
  10.53 &
  \multicolumn{1}{c|}{8.12} &
  \multicolumn{1}{c|}{11.10} &
  \multicolumn{1}{c|}{\textbf{6.60}} &
  \multicolumn{1}{c|}{9.95} &
  16.14 \\ \hline
\emph{wk} &
  \multicolumn{1}{c|}{\textbf{1.22}} &
  \multicolumn{1}{c|}{1.68} &
  2.76 &
  \multicolumn{1}{c|}{\textbf{2.96}} &
  \multicolumn{1}{c|}{5.61} &
  \multicolumn{1}{c|}{5.19} &
  \multicolumn{1}{c|}{3.37} &
  7.85 \\ \hline
\emph{ct} &
  \multicolumn{1}{c|}{\textbf{0.97}} &
  \multicolumn{1}{c|}{2.27} &
  2.13 &
  \multicolumn{1}{c|}{\textbf{2.45}} &
  \multicolumn{1}{c|}{3.51} &
  \multicolumn{1}{c|}{3.46} &
  \multicolumn{1}{c|}{2.84} &
  5.71 \\ \hline
\emph{tw} &
  \multicolumn{1}{c|}{\textbf{10.48}} &
  \multicolumn{1}{c|}{12.55} &
  15.31 &
  \multicolumn{1}{c|}{\textbf{17.84}} &
  \multicolumn{1}{c|}{24.15} &
  \multicolumn{1}{c|}{18.62} &
  \multicolumn{1}{c|}{20.27} &
  41.14 \\ \hline
\emph{nft} &
  \multicolumn{1}{c|}{\textbf{4.22}} &
  \multicolumn{1}{c|}{5.83} &
  7.90 &
  \multicolumn{1}{c|}{\textbf{8.77}} &
  \multicolumn{1}{c|}{17.95} &
  \multicolumn{1}{c|}{16.00} &
  \multicolumn{1}{c|}{9.70} &
  33.28 \\ \hline
\end{tabular}
\end{table}

Figure \ref{tab:performance_comparison_containers_appendix} illustrates the performance of graph containers on various graph analytic queries, including BFS, SSSP, and WCC. In all three datasets, continuous methods exhibit superior performance compared to segmented methods due to their rapid scan speed. Among the segmented methods, Teseo demonstrates the best performance among methods without optimizations. Enabling adaptive indexing significantly enhances speed, achieving a speedup of 1.1-3.4 times. As a result, Sortledton-w outperforms its counterparts in 15 out of 24 cases, despite having lower scan and search efficiency compared to Teseo. Additionally, enabling flatten optimization improves performance by 1.0-3.6 times. Nevertheless, CSR consistently achieves a speedup of 1.3-3.5 times over the best segmented methods in each scenario.

\vspace{2pt}
\noindent\textbf{Graph Operations Latency.} To compare the quality of service, we measure the graph operation latency during concurrent execution. Specifically, we execute 32 threads where four threads conduct insert operations, eight threads conduct search operations, and twenty threads conduct scan operations. We configure four insert threads because graph workloads are generally read-intensive, and competing methods have limited scalability for insert operations.

\begin{table}[h]
\caption{Mean, max and tail latency of search operations (100 operations).}
\label{tab:search_tail_latency}
\begin{tabular}{|lc|ccccc|}
\hline
\multicolumn{2}{|l|}{\multirow{2}{*}{}} &
  \multicolumn{5}{c|}{\textbf{Search}} \\ \cline{3-7} 
\multicolumn{2}{|l|}{} &
  \multicolumn{1}{c|}{\textbf{AdjLst}} &
  \multicolumn{1}{c|}{\textbf{LiveGraph}} &
  \multicolumn{1}{c|}{\textbf{Aspen}} &
  \multicolumn{1}{c|}{\textbf{Teseo}} &
  \multicolumn{1}{c|}{\textbf{Sortledton}} \\ \hline
\multicolumn{1}{|l|}{\multirow{5}{*}{\emph{lj}}} &
  p50 &
  \multicolumn{1}{c|}{57} &
  \multicolumn{1}{c|}{135} &
  \multicolumn{1}{c|}{117} &
  \multicolumn{1}{c|}{66} &
  \textbf{54} \\ \cline{2-7} 
\multicolumn{1}{|l|}{} &
  p95 &
  \multicolumn{1}{c|}{65} &
  \multicolumn{1}{c|}{173} &
  \multicolumn{1}{c|}{139} &
  \multicolumn{1}{c|}{72} &
  \textbf{61} \\ \cline{2-7} 
\multicolumn{1}{|l|}{} &
  p99 &
  \multicolumn{1}{c|}{\textbf{71}} &
  \multicolumn{1}{c|}{200} &
  \multicolumn{1}{c|}{165} &
  \multicolumn{1}{c|}{78} &
  85 \\ \cline{2-7} 
\multicolumn{1}{|l|}{} &
  Max &
  \multicolumn{1}{c|}{\textbf{105}} &
  \multicolumn{1}{c|}{280} &
  \multicolumn{1}{c|}{379} &
  \multicolumn{1}{c|}{134} &
  143 \\ \cline{2-7} 
\multicolumn{1}{|l|}{} &
  Mean &
  \multicolumn{1}{c|}{57} &
  \multicolumn{1}{c|}{138} &
  \multicolumn{1}{c|}{120} &
  \multicolumn{1}{c|}{66} &
  \textbf{55} \\ \hline
\multicolumn{1}{|l|}{\multirow{5}{*}{\emph{g5}}} &
  p50 &
  \multicolumn{1}{c|}{\textbf{99}} &
  \multicolumn{1}{c|}{1,491} &
  \multicolumn{1}{c|}{188} &
  \multicolumn{1}{c|}{122} &
  201 \\ \cline{2-7} 
\multicolumn{1}{|l|}{} &
  p95 &
  \multicolumn{1}{c|}{\textbf{106}} &
  \multicolumn{1}{c|}{2,410} &
  \multicolumn{1}{c|}{197} &
  \multicolumn{1}{c|}{130} &
  225 \\ \cline{2-7} 
\multicolumn{1}{|l|}{} &
  p99 &
  \multicolumn{1}{c|}{\textbf{113}} &
  \multicolumn{1}{c|}{3,064} &
  \multicolumn{1}{c|}{208} &
  \multicolumn{1}{c|}{134} &
  235 \\ \cline{2-7} 
\multicolumn{1}{|l|}{} &
  Max &
  \multicolumn{1}{c|}{\textbf{193}} &
  \multicolumn{1}{c|}{3,995} &
  \multicolumn{1}{c|}{458} &
  \multicolumn{1}{c|}{212} &
  362 \\ \cline{2-7} 
\multicolumn{1}{|l|}{} &
  Mean &
  \multicolumn{1}{c|}{\textbf{99}} &
  \multicolumn{1}{c|}{1,562} &
  \multicolumn{1}{c|}{189} &
  \multicolumn{1}{c|}{122} &
  202 \\ \hline
\end{tabular}%
\end{table}

\small
\begin{table}[h]
\caption{Mean, max and tail latency of insert operations (100 operations).}
\label{tab:insert_tail_latency}
\begin{tabular}{|lc|ccccc|}
\hline
\multicolumn{2}{|l|}{\multirow{2}{*}{}} &
  \multicolumn{5}{c|}{\textbf{Insert}} \\ \cline{3-7} 
\multicolumn{2}{|l|}{} &
  \multicolumn{1}{c|}{\textbf{AdjLst}} &
  \multicolumn{1}{c|}{\textbf{LiveGraph}} &
  \multicolumn{1}{c|}{\textbf{Aspen}} &
  \multicolumn{1}{c|}{\textbf{Teseo}} &
  \multicolumn{1}{c|}{\textbf{Sortledton}} \\ \hline
\multicolumn{1}{|l|}{\multirow{5}{*}{\emph{lj}}} &
  p50 &
  \multicolumn{1}{c|}{166} &
  \multicolumn{1}{c|}{\textbf{151}} &
  \multicolumn{1}{c|}{1,093} &
  \multicolumn{1}{c|}{194} &
  174 \\ \cline{2-7} 
\multicolumn{1}{|l|}{} &
  p95 &
  \multicolumn{1}{c|}{206} &
  \multicolumn{1}{c|}{\textbf{168}} &
  \multicolumn{1}{c|}{2,060} &
  \multicolumn{1}{c|}{217} &
  204 \\ \cline{2-7} 
\multicolumn{1}{|l|}{} &
  p99 &
  \multicolumn{1}{c|}{220} &
  \multicolumn{1}{c|}{\textbf{177}} &
  \multicolumn{1}{c|}{2,825} &
  \multicolumn{1}{c|}{228} &
  245 \\ \cline{2-7} 
\multicolumn{1}{|l|}{} &
  Max &
  \multicolumn{1}{c|}{\textbf{475}} &
  \multicolumn{1}{c|}{746} &
  \multicolumn{1}{c|}{17,835} &
  \multicolumn{1}{c|}{709} &
  757 \\ \cline{2-7} 
\multicolumn{1}{|l|}{} &
  Mean &
  \multicolumn{1}{c|}{166} &
  \multicolumn{1}{c|}{\textbf{152}} &
  \multicolumn{1}{c|}{1,184} &
  \multicolumn{1}{c|}{190} &
  175 \\ \hline
\multicolumn{1}{|l|}{\multirow{5}{*}{\emph{g5}}} &
  p50 &
  \multicolumn{1}{c|}{1,397} &
  \multicolumn{1}{c|}{1,007} &
  \multicolumn{1}{c|}{1,688} &
  \multicolumn{1}{c|}{\textbf{281}} &
  304 \\ \cline{2-7} 
\multicolumn{1}{|l|}{} &
  p95 &
  \multicolumn{1}{c|}{2,979} &
  \multicolumn{1}{c|}{2,295} &
  \multicolumn{1}{c|}{2,822} &
  \multicolumn{1}{c|}{\textbf{316}} &
  368 \\ \cline{2-7} 
\multicolumn{1}{|l|}{} &
  p99 &
  \multicolumn{1}{c|}{3,595} &
  \multicolumn{1}{c|}{2,796} &
  \multicolumn{1}{c|}{3,189} &
  \multicolumn{1}{c|}{\textbf{367}} &
  414 \\ \cline{2-7} 
\multicolumn{1}{|l|}{} &
  Max &
  \multicolumn{1}{c|}{6,545} &
  \multicolumn{1}{c|}{4,732} &
  \multicolumn{1}{c|}{7,853} &
  \multicolumn{1}{c|}{8,749} &
  \textbf{2,625} \\ \cline{2-7} 
\multicolumn{1}{|l|}{} &
  Mean &
  \multicolumn{1}{c|}{1,467} &
  \multicolumn{1}{c|}{1,080} &
  \multicolumn{1}{c|}{1,763} &
  \multicolumn{1}{c|}{\textbf{275}} &
  291 \\ \hline
\end{tabular}%
\end{table}

\small
\begin{table}[h]

\caption{Mean, max and tail latency of scan operations (100 operations).}
\label{tab:scan_tail_latency}
\begin{tabular}{|lc|ccccc|}
\hline
\multicolumn{2}{|l|}{\multirow{2}{*}{}} &
  \multicolumn{5}{c|}{\textbf{Scan}} \\ \cline{3-7} 
\multicolumn{2}{|l|}{} &
  \multicolumn{1}{c|}{\textbf{AdjLst}} &
  \multicolumn{1}{c|}{\textbf{LiveGraph}} &
  \multicolumn{1}{c|}{\textbf{Aspen}} &
  \multicolumn{1}{c|}{\textbf{Teseo}} &
  \multicolumn{1}{c|}{\textbf{Sortledton}} \\ \hline
\multicolumn{1}{|l|}{\multirow{5}{*}{\emph{lj}}} &
  p50 &
  \multicolumn{1}{c|}{161} &
  \multicolumn{1}{c|}{225} &
  \multicolumn{1}{c|}{228} &
  \multicolumn{1}{c|}{\textbf{133}} &
  248 \\ \cline{2-7} 
\multicolumn{1}{|l|}{} &
  p95 &
  \multicolumn{1}{c|}{257} &
  \multicolumn{1}{c|}{458} &
  \multicolumn{1}{c|}{392} &
  \multicolumn{1}{c|}{\textbf{241}} &
  543 \\ \cline{2-7} 
\multicolumn{1}{|l|}{} &
  p99 &
  \multicolumn{1}{c|}{299} &
  \multicolumn{1}{c|}{895} &
  \multicolumn{1}{c|}{435} &
  \multicolumn{1}{c|}{\textbf{297}} &
  594 \\ \cline{2-7} 
\multicolumn{1}{|l|}{} &
  Max &
  \multicolumn{1}{c|}{12,798} &
  \multicolumn{1}{c|}{1,038} &
  \multicolumn{1}{c|}{498} &
  \multicolumn{1}{c|}{\textbf{312}} &
  703 \\ \cline{2-7} 
\multicolumn{1}{|l|}{} &
  Mean &
  \multicolumn{1}{c|}{185} &
  \multicolumn{1}{c|}{263} &
  \multicolumn{1}{c|}{249} &
  \multicolumn{1}{c|}{\textbf{146}} &
  287 \\ \hline
\multicolumn{1}{|l|}{\multirow{5}{*}{\emph{g5}}} &
  p50 &
  \multicolumn{1}{c|}{\textbf{3,336}} &
  \multicolumn{1}{c|}{9,228} &
  \multicolumn{1}{c|}{5,879} &
  \multicolumn{1}{c|}{5,404} &
  9,149 \\ \cline{2-7} 
\multicolumn{1}{|l|}{} &
  p95 &
  \multicolumn{1}{c|}{\textbf{5,162}} &
  \multicolumn{1}{c|}{15,354} &
  \multicolumn{1}{c|}{8,701} &
  \multicolumn{1}{c|}{8,423} &
  13,722 \\ \cline{2-7} 
\multicolumn{1}{|l|}{} &
  p99 &
  \multicolumn{1}{c|}{\textbf{6,834}} &
  \multicolumn{1}{c|}{19,553} &
  \multicolumn{1}{c|}{10,422} &
  \multicolumn{1}{c|}{10,486} &
  17,961 \\ \cline{2-7} 
\multicolumn{1}{|l|}{} &
  Max &
  \multicolumn{1}{c|}{11,332} &
  \multicolumn{1}{c|}{21,107} &
  \multicolumn{1}{c|}{\textbf{10,569}} &
  \multicolumn{1}{c|}{10,800} &
  18,533 \\ \cline{2-7} 
\multicolumn{1}{|l|}{} &
  Mean &
  \multicolumn{1}{c|}{\textbf{3,425}} &
  \multicolumn{1}{c|}{9,707} &
  \multicolumn{1}{c|}{5,936} &
  \multicolumn{1}{c|}{5,499} &
  9,249 \\ \hline
\end{tabular}%
\end{table}

Tables \ref{tab:search_tail_latency}, \ref{tab:scan_tail_latency}, and \ref{tab:insert_tail_latency} present the experimental results on latency. 
For search latency, AdjLst outperforms all other methods on both datasets. Sortledton demonstrates strong performance on the \emph{lj} dataset, which features a smaller average degree, thereby benefiting from Sortledton's adaptive indexing. LiveGraph exhibits the worst performance on both datasets due to its requirement to scan neighbors.

For insert latency, Teseo generally outperforms other methods in most scenarios, as it often has empty slots available for new elements. However, its needing for periodic rebalancing leads to relatively high maximum latency. Sortledton exhibits lower maximum latency on the \emph{g5} dataset because its split operations are confined to a single block. LiveGraph performs well on the \emph{lj} dataset but poorly on the \emph{g5} dataset, as the bloom filter fails to be effective in large neighborhoods. Aspen's limitation of allowing only a single writer results in the worst insert latency.For scan latency, both AdjLst and Teseo exhibit strong performance due to their contiguous storage structure. In contrast, LiveGraph incurs relatively high latency as it scans from the end of the array.

\begin{figure}[h]
	\setlength{\abovecaptionskip}{0pt}
	\setlength{\belowcaptionskip}{0pt}
		\captionsetup[subfigure]{aboveskip=0pt,belowskip=0pt}
	\centering
    \includegraphics[width=0.60\textwidth]{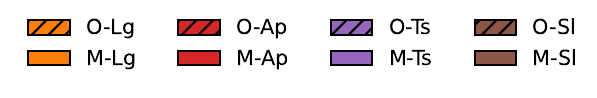}\\
    \begin{subfigure}[t]{0.20\textwidth}
    		\centering
    		\includegraphics[width=\textwidth]{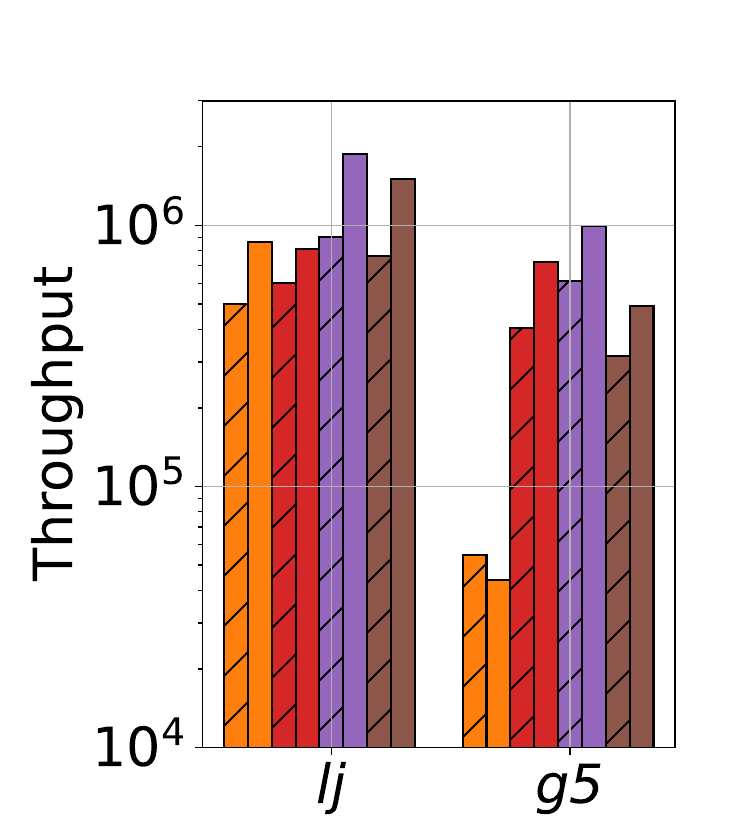}
    		\caption{\textsc{SearchEdge}}
    		\label{fig:comparison_search}
    \end{subfigure}
    \begin{subfigure}[t]{0.20\textwidth}
    		\centering
    		\includegraphics[width=\textwidth]{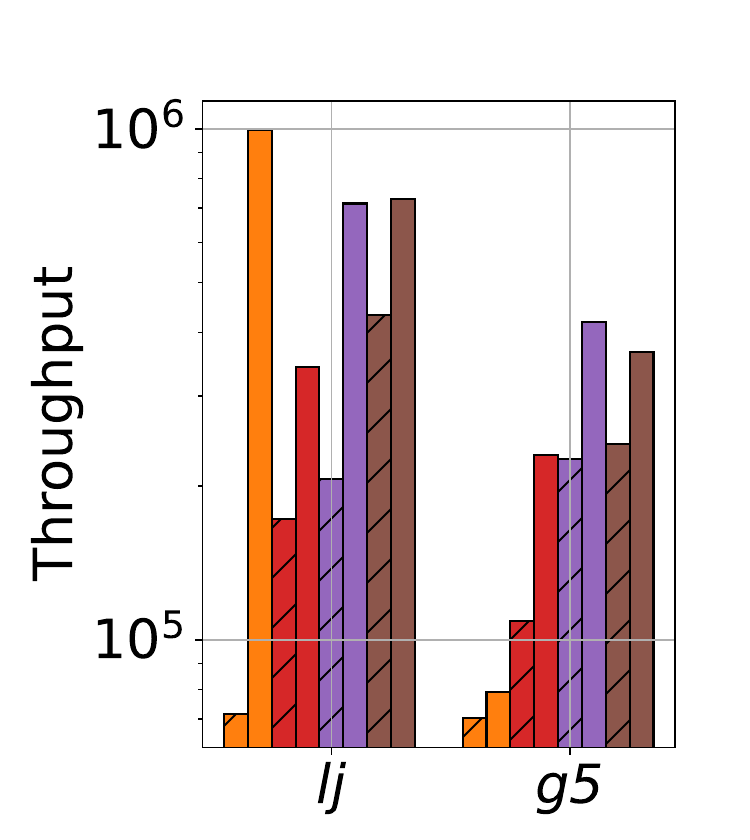}
    		\caption{\textsc{InsertEdge}}
    		\label{fig:comparison_insert}
    \end{subfigure}
    \begin{subfigure}[t]{0.20\textwidth}
    		\centering
    		\includegraphics[width=\textwidth]{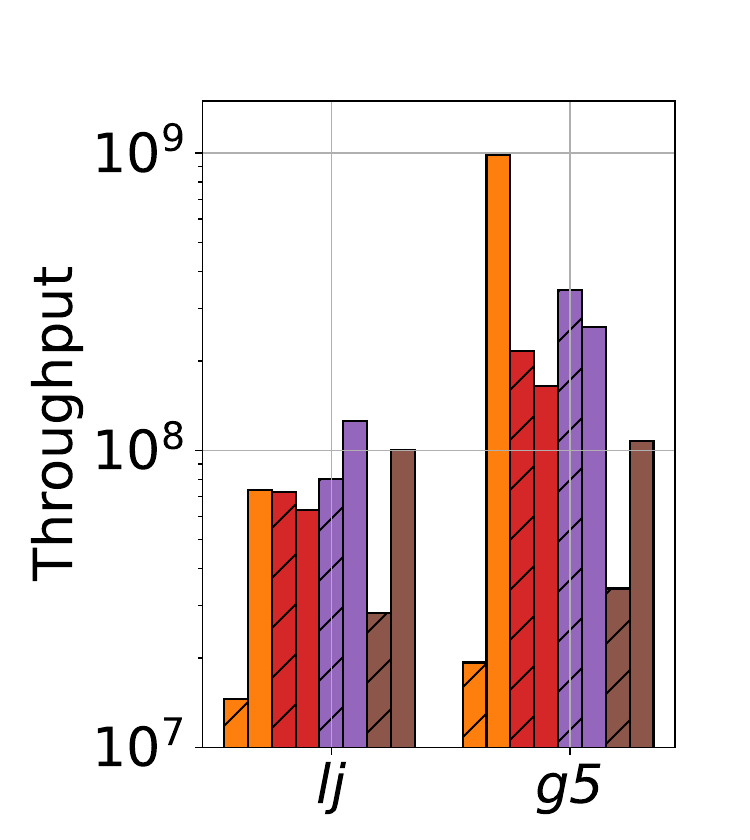}
    		\caption{\textsc{ScanNbr}}
    		\label{fig:comparison_scan}
    \end{subfigure}
    \caption{Comparison of efficiency with original implementations. Unshadowed denotes our implementation, and shadowed denotes the origin ones. Lg, Ap, Ts and Sl denote LiveGraph, Aspen, Teseo and Sortledton, respectively.}
    \label{fig:comparison_of_efficiency_with_origin}
\end{figure}

\vspace{2pt}
\noindent\textbf{Comparison of Efficiency with Original Implementation.} 
Figure \ref{fig:comparison_of_efficiency_with_origin} illustrates the comparison of search, insert, and scan efficiencies between our implementations and the original ones. Our implementations demonstrate either comparable or superior efficiency across the two datasets and three operations.
\small
\begin{table}[htbp]
\caption{Comparison of memory consumption with original implementation. 'M' denotes our implementation and 'O' denotes the original implementation. Lg, Ap, Ts and Sl denote LiveGraph, Aspen, Teseo and Sortledton, respectively.}
\label{tab:memory_comparison_with_origin}
\begin{tabular}{|l|l|ll|ll|ll|ll|}
\hline
  \textbf{Dataset}        & \textbf{CSR}  & \textbf{M-Lg} & \textbf{O-Lg}  & \textbf{M-Ap}  & \textbf{O-Ap}  & \textbf{M-Ts}  & \textbf{O-Ts}  & \textbf{M-Sl}  & \textbf{O-Sl}  \\ \hline
\emph{lj}   & 0.67 & 6.24  & 6.98  & 3.58  & 28.25 & 30.98 & 5.25  & 4.31  & 3.66  \\ \hline
\emph{dl}   & 0.76 & 3.71  & 6.35  & 2.52  & 27.82 & 6.87  & 4.73  & 3.11  & 4.10  \\ \hline
\emph{ldbc}   & 2.84 & 28.95 & 32.48 & 17.96 & 33.88 & OOM  & 20.15 & 20.19 & 19.00 \\ \hline
\emph{g5} & 3.95 & 24.62 & 40.46 & 15.33 & 33.02 & 78.42 & 19.34 & 19.55 & 19.16 \\ \hline
\emph{wk}   & 0.98 & 13.09 & 11.20 & 8.18  & 29.41 & 86.08 & 8.31  & 7.12  & 6.56  \\ \hline
\emph{ct}   & 0.27 & 3.32  & 3.49  & 2.06  & 27.60 & 22.81 & 3.14  & 2.14  & 1.80  \\ \hline
\emph{tw}   & 4.11 & 31.98 & 37.56 & 20.20 & 34.43 & OOM  & 22.03 & 22.04 & 23.53 \\ \hline
\emph{nft}  & 1.38 & 23.01 & 17.30 & 14.94 & 31.57 & OOM  & 11.77 & 12.27 & 11.12 \\ \hline
\end{tabular}%
\end{table}

\small
\begin{table}[htbp]
\caption{Comparison of graph analytic queries with original implementations  (in seconds). 'M' denotes our implementation and 'O' denotes the original implementation. Lg, Ap, Ts and Sl denote LiveGraph, Aspen, Teseo and Sortledton, respectively.}
\label{tab:performance_comparison_of_analytic_queries_appendix}
\centering
\begin{tabular}{|c|cccccccc|}
    \hline
    \textbf{Dataset} &
      \textbf{M-Lg} &
      \multicolumn{1}{c|}{\textbf{O-Lg}} &
      \textbf{M-As} &
      \multicolumn{1}{c|}{\textbf{O-As}} &
      \textbf{M-Ts} &
      \multicolumn{1}{c|}{\textbf{O-Ts}} &
      \textbf{M-Sl} &
      \textbf{O-Sl} \\ \hline
    \multicolumn{1}{|c|}{\textbf{BFS}} &
      \multicolumn{8}{l|}{} \\ \hline
    \emph{lj} &
      2.20 &
      \multicolumn{1}{c|}{12.19} &
      3.56 &
      \multicolumn{1}{c|}{3.98} &
      2.81 &
      \multicolumn{1}{c|}{4.09} &
      2.19 &
      4.45 \\ \hline
    \emph{dl} &
      0.03 &
      \multicolumn{1}{c|}{0.65} &
      0.04 &
      \multicolumn{1}{c|}{0.03} &
      0.03 &
      \multicolumn{1}{c|}{0.07} &
      0.04 &
      0.10 \\ \hline
    \emph{ldbc} &
      9.04 &
      \multicolumn{1}{c|}{30.70} &
      13.20 &
      \multicolumn{1}{c|}{23.86} &
      OOM &
      \multicolumn{1}{c|}{13.70} &
      10.03 &
      27.78 \\ \hline
    \emph{g5} &
      5.38 &
      \multicolumn{1}{c|}{17.79} &
      6.77 &
      \multicolumn{1}{c|}{8.69} &
      5.80 &
      \multicolumn{1}{c|}{5.91} &
      6.19 &
      9.76 \\ \hline
    \emph{wk} &
      3.95 &
      \multicolumn{1}{c|}{15.46} &
      6.24 &
      \multicolumn{1}{c|}{10.12} &
      4.83 &
      \multicolumn{1}{c|}{6.17} &
      4.11 &
      10.60 \\ \hline
    \emph{ct} &
      1.99 &
      \multicolumn{1}{c|}{8.23} &
      3.04 &
      \multicolumn{1}{c|}{4.33} &
      2.81 &
      \multicolumn{1}{c|}{3.60} &
      1.91 &
      5.44 \\ \hline
    \emph{tw} &
      8.77 &
      \multicolumn{1}{c|}{36.17} &
      12.58 &
      \multicolumn{1}{c|}{22.31} &
      OOM &
      \multicolumn{1}{c|}{16.14} &
      9.29 &
      24.09 \\ \hline
    \emph{nft} &
      9.15 &
      \multicolumn{1}{c|}{37.89} &
      14.66 &
      \multicolumn{1}{c|}{27.18} &
      OOM &
      \multicolumn{1}{c|}{18.88} &
      9.59 &
      24.92 \\ \hline
    \multicolumn{1}{|c|}{\textbf{SSSP}} &
      \multicolumn{8}{l|}{} \\ \hline
    \emph{lj} &
      2.87 &
      \multicolumn{1}{c|}{6.76} &
      7.31 &
      \multicolumn{1}{c|}{6.83} &
      3.34 &
      \multicolumn{1}{c|}{5.84} &
      2.79 &
      3.93 \\ \hline
    \emph{dl} &
      0.55 &
      \multicolumn{1}{c|}{0.75} &
      0.66 &
      \multicolumn{1}{c|}{0.37} &
      0.50 &
      \multicolumn{1}{c|}{1.83} &
      1.07 &
      1.94 \\ \hline
    \emph{ldbc} &
      12.18 &
      \multicolumn{1}{c|}{31.76} &
      32.70 &
      \multicolumn{1}{c|}{37.97} &
      OOM &
      \multicolumn{1}{c|}{31.81} &
      15.19 &
      26.91 \\ \hline
    \emph{g5} &
      7.79 &
      \multicolumn{1}{c|}{17.31} &
      14.81 &
      \multicolumn{1}{c|}{14.67} &
      9.79 &
      \multicolumn{1}{c|}{16.21} &
      10.25 &
      10.60 \\ \hline
    \emph{wk} &
      4.46 &
      \multicolumn{1}{c|}{12.04} &
      7.99 &
      \multicolumn{1}{c|}{9.31} &
      4.32 &
      \multicolumn{1}{c|}{7.84} &
      4.34 &
      9.04 \\ \hline
    \emph{ct} &
      1.71 &
      \multicolumn{1}{c|}{4.17} &
      4.68 &
      \multicolumn{1}{c|}{4.04} &
      2.00 &
      \multicolumn{1}{c|}{4.83} &
      1.74 &
      3.41 \\ \hline
    \emph{tw} &
      12.79 &
      \multicolumn{1}{c|}{29.91} &
      32.78 &
      \multicolumn{1}{c|}{35.06} &
      OOM &
      \multicolumn{1}{c|}{30.34} &
      15.05 &
      26.72 \\ \hline
    \emph{nft} &
      10.79 &
      \multicolumn{1}{c|}{31.73} &
      30.02 &
      \multicolumn{1}{c|}{37.59} &
      OOM &
      \multicolumn{1}{c|}{27.62} &
      10.37 &
      20.51 \\ \hline
    \multicolumn{1}{|c|}{\textbf{PR}} &
      \multicolumn{8}{l|}{} \\ \hline
    \emph{lj} &
      17.98 &
      \multicolumn{1}{c|}{79.78} &
      26.21 &
      \multicolumn{1}{c|}{33.99} &
      20.24 &
      \multicolumn{1}{c|}{35.15} &
      18.46 &
      36.48 \\ \hline
    \emph{dl} &
      3.29 &
      \multicolumn{1}{c|}{12.87} &
      4.84 &
      \multicolumn{1}{c|}{4.29} &
      3.40 &
      \multicolumn{1}{c|}{12.41} &
      8.92 &
      21.77 \\ \hline
    \emph{ldbc} &
      67.68 &
      \multicolumn{1}{c|}{286.22} &
      104.58 &
      \multicolumn{1}{c|}{186.50} &
      OOM &
      \multicolumn{1}{c|}{117.54} &
      101.31 &
      239.40 \\ \hline
    \emph{g5} &
      50.54 &
      \multicolumn{1}{c|}{202.23} &
      66.53 &
      \multicolumn{1}{c|}{88.30} &
      53.01 &
      \multicolumn{1}{c|}{162.39} &
      82.31 &
      104.32 \\ \hline
    \emph{wk} &
      32.39 &
      \multicolumn{1}{c|}{139.40} &
      49.75 &
      \multicolumn{1}{c|}{82.51} &
      40.58 &
      \multicolumn{1}{c|}{45.86} &
      38.53 &
      92.28 \\ \hline
    \emph{ct} &
      10.59 &
      \multicolumn{1}{c|}{52.23} &
      14.22 &
      \multicolumn{1}{c|}{22.73} &
      13.35 &
      \multicolumn{1}{c|}{15.92} &
      10.04 &
      28.21 \\ \hline
    \emph{tw} &
      75.95 &
      \multicolumn{1}{c|}{331.24} &
      105.64 &
      \multicolumn{1}{c|}{174.88} &
      OOM &
      \multicolumn{1}{c|}{138.30} &
      110.80 &
      256.53 \\ \hline
    \emph{nft} &
      59.74 &
      \multicolumn{1}{c|}{285.08} &
      93.48 &
      \multicolumn{1}{c|}{180.99} &
      OOM &
      \multicolumn{1}{c|}{91.07} &
      67.43 &
      173.85 \\ \hline
    \multicolumn{1}{|c|}{\textbf{WCC}} &
      \multicolumn{8}{l|}{} \\ \hline
    \emph{lj} &
      4.36 &
      \multicolumn{1}{c|}{13.66} &
      5.81 &
      \multicolumn{1}{c|}{7.75} &
      4.81 &
      \multicolumn{1}{c|}{7.00} &
      4.14 &
      7.34 \\ \hline
    \emph{dl} &
      0.66 &
      \multicolumn{1}{c|}{1.40} &
      0.99 &
      \multicolumn{1}{c|}{0.69} &
      0.79 &
      \multicolumn{1}{c|}{0.75} &
      1.62 &
      4.18 \\ \hline
    \emph{ldbc} &
      9.85 &
      \multicolumn{1}{c|}{36.71} &
      14.93 &
      \multicolumn{1}{c|}{26.24} &
      OOM &
      \multicolumn{1}{c|}{25.05} &
      17.03 &
      35.54 \\ \hline
    \emph{g5} &
      9.03 &
      \multicolumn{1}{c|}{28.01} &
      10.93 &
      \multicolumn{1}{c|}{14.31} &
      10.88 &
      \multicolumn{1}{c|}{16.31} &
      15.60 &
      15.89 \\ \hline
    \emph{wk} &
      3.96 &
      \multicolumn{1}{c|}{18.08} &
      6.86 &
      \multicolumn{1}{c|}{10.50} &
      5.08 &
      \multicolumn{1}{c|}{8.35} &
      5.08 &
      12.47 \\ \hline
    \emph{ct} &
      2.64 &
      \multicolumn{1}{c|}{8.65} &
      3.55 &
      \multicolumn{1}{c|}{5.04} &
      2.85 &
      \multicolumn{1}{c|}{4.58} &
      2.29 &
      5.54 \\ \hline
    \emph{tw} &
      18.91 &
      \multicolumn{1}{c|}{65.94} &
      23.71 &
      \multicolumn{1}{c|}{40.73} &
      OOM &
      \multicolumn{1}{c|}{40.92} &
      29.66 &
      61.61 \\ \hline
    \emph{nft} &
      10.65 &
      \multicolumn{1}{c|}{53.80} &
      16.85 &
      \multicolumn{1}{c|}{34.33} &
      OOM &
      \multicolumn{1}{c|}{26.52} &
      13.42 &
      33.40 \\ \hline
\end{tabular}

\end{table}

Table \ref{tab:performance_comparison_of_analytic_queries_appendix} illustrates the performance comparison with original methods on graph analytic queries.
Our implementations shows similar or superior efficiency in all cases. Although LiveGraph utilizes sequential storage, which offers significant advantages for scan operations, it does not store the degree of each vertex. Instead, it computes the degree dynamically through traversal each time. As a result, in the GAPBS benchmark implementation, LiveGraph does not exhibit a performance advantage. To optimize this, the degrees of vertices could be precomputed and stored within snapshots.

\noindent\textbf{Comparison of Memory Consumption with Original Implementation.} Table \ref{tab:memory_comparison_with_origin} presents a comparison of memory consumption between our implementations and the original versions. Our implementations of LiveGraph, Aspen, and Sortledton exhibit comparable memory usage to the original versions. However, Teseo optimizes memory storage by storing neighbor sets in a compact manner, allowing multiple neighbor sets to reside within the same PMA leaf. We allocate a PMA leaf for each vertex to enhance efficiency, which results in higher memory overhead.

\clearpage

\end{document}